\documentclass[a4paper,11pt]{article}
\usepackage{aaskaiid}

\usepackage{upgreek}
\usepackage{xcolor}
\usepackage{amsmath}
\usepackage{enumerate}
\usepackage{subcaption}
\usepackage{orcidlink}

\newcommand{\arcsec}{^{\prime\prime}}
\def\Rm{\rm Rm}

\def\Rey{{\rm Re}}
\def\Pm{\rm Pm}

\newcommand{\sigmarm}{\sigma_{\rm RM}}

\newcommand{\kk}{\mbox{\boldmath $k$} {}}
\newcommand{\FF}{\mbox{\boldmath $F$} {}}
\newcommand{\cmn}{\,{\rm cm^{-3}}}
\newcommand{\gcm}{\,{\rm gm\,cm^{-3}}}

\newcommand{\kpc}{\,{\rm kpc}}

\newcommand{\kms}{\,{\rm km\,s^{-1}}}

\newcommand{\yr}{\,{\rm yr}}

\newcommand{\mkG}{\,\upmu{\rm G}}

\newcommand{\radm}{\,{\rm rad\,m^{-2}}}
\newcommand{\mhz}{\,{\rm MHz}}
\newcommand{\ghz}{\,{\rm GHz}}
\newcommand{\urms}{{u_{\rm rms}}}
\newcommand{\mrms}{\mathcal{M}_{\rm rms}}
\newcommand{\brms}{{b_{\rm rms}}}
\newcommand{\pmean}{{\langle p_{\rm f} \rangle}}
\newcommand{\lf}{{l_{\rm f}}}
\newcommand{\kf}{{k_{\rm f}}}
\newcommand{\pf}{{p_{\rm f}}}
\newcommand{\EQ}{\begin{equation}}
\newcommand{\EN}{\end{equation}}
\newcommand{\EQA}{\begin{eqnarray}}
\newcommand{\ENA}{\end{eqnarray}}

\title{Unraveling the Imprints of Fluctuation-dynamo on the Intracluster Medium with the SKA}
\ShortTitle{ICM magnetic fields}

\author[1,2]{Aritra Basu\orcidlink{0000-0003-2030-3394}}
\ShortName{Aritra Basu et al.} 
\author[3]{Sharanya Sur\orcidlink{0000-0003-4286-8476}}
\author[4,1]{Nadia Biava\orcidlink{0000-0001-7947-6447}}
\author[4]{Marisa Brienza\orcidlink{0000-0003-4120-9970}}

\affiliation[1]{Th\"uringer Landessternwarte, Sternwarte 5, 07778 Tautenburg, Germany}
\affiliation[2]{Max-Planck-Institut f\"ur Radioastronomie, Auf dem H\"ugel 69, 53121 Bonn, Germany}
\affiliation[3]{Indian Institute of Astrophysics, 2nd Block, 100 Feet Road, Koramangala, Bangalore 560034, India}
\affiliation[4]{Istituto Nazionale di Astrofisica – Istituto di Radioastronomia, Via Gobetti 101, 40129 Bologna, Italy}
\emailAdd{abasu@tls-tautenburg.de}
\emailAdd{sharanya.sur@iiap.res.in}
\emailAdd{nadia.biava@inaf.it}
\emailAdd{marisa.brienza@inaf.it}

\abstract{Studying the morphology and coherence scale of the magnetic fields in synchrotron emitting halos of galaxy clusters through detection of polarized synchrotron emission is important in order to understand how they distribute relativistic plasma, their contribution to pressure balance, and how they may affect gas content in galaxies and set up initial conditions during non-linear collapse to eventually form galaxies in the intracluster medium (ICM). Using synthetic maps over broad-bandwidths, generated from high resolution magnetohydrodynamic simulations of fluctuation dynamo, we study the efficacy of SKA-Low and Mid in Array Assembly AA4 in our quest for detecting polarized synchrotron emission from the intracluster medium (ICM). Fluctuation dynamo action in the ICM are expected to generate ubiquitous filamentary and sheet-like magnetic field structures. The associated synchrotron emission projected in the plane of the sky appear highly filamentary that can span hundreds of kiloparsecs. Such filaments can be robustly identified and quantified through high angular resolution ($\lesssim 1\arcsec$), sensitive observations (about $\rm 0.25\textrm{--}1\,\upmu Jy\,beam^{-1}$) above about 3\,GHz, a niche for the SKA-Mid in Band\,5a. Statistical properties of polarized emission and morphological measures, such as, Minkowski functionals, can then be used to infer the turbulence driving scales in the ICM with deep observations of galaxy clusters using SKA-Mid. Availability of a Band\,4 receiver covering 2--4\,GHz would be a major boost in studying the polarization properties of the ICM.}

\begin{document}
\maketitle

\section{Introduction}

Magnetic fields play a key dynamical role in the evolution of astrophysical objects on a wide range of scales. On sub-parsec to tens of parsec scales, it regulates the collapse of gas clouds to form stars in galaxies, launching of jets from accretion disks of supermassive black holes in active galactic nuclei (AGN); on hundreds of parsec scale it determines the small-scale structures in galaxies and regulates the multiphase interstellar medium (ISM); on kiloparsec scales, magnetic fields shape up the large-scale structure in galaxies by regulating the transport of cosmic rays and how gas is accreted onto them. On several tens to hundreds of kiloparsec scales, magnetic fields determine the dynamics of cosmic ray electrons in lobes of AGN and in the largest gravitationally bound structures in the Universe, the intracluster medium (ICM) of galaxy clusters, and megaparsec-scale shocks in radio relics emanating from cluster mergers.

Observationally, little is known about the dynamical importance of magnetic fields and their morphology in halos of galaxy clusters. So far, microgauss ($\upmu$G) strength magnetic fields in halos have been indirectly inferred via Faraday rotation measure (RM) estimated towards polarized sources located in the background \citep{bonaf10, osinga2022, osinga2025}, or from depolarization studies of radio relics, remnants of cluster collisions \citep{kierdorf+17, rajpurohit2022, digennaro2023, deRubeis2024}. Synchrotron intensity gradient has been used to trace magnetic fields in relics and halo \citep{hu2020, hu2024}. However, a direct detection of polarized halo emission remains elusive and is a major science driver for the Square Kilometre Array (SKA).
The main open questions in the context of magnetic fields in the ICM are --- (1) what drives turbulence and on what scales? (2) what are the strengths and morphology of magnetic fields? (3) what role do they play in the distribution and (re)acceleration of cosmic ray electrons (CREs)? and (4) what are the properties of, yet elusive, polarized emission in the ICM?

Massive merging clusters often show diffuse radio emission (radio halo) originating from relativistic
electrons, thought to be accelerated by turbulence in the ICM \citep{Brunetti2014}, that are illuminated via synchrotron mechanism by the cluster magnetic
fields 
\citep{Fer+12, weere19}. This emission is expected to be partially polarized and its measurement provides insight into the statistical properties of the magnetic field structure
in the ICM \citep{vazza+18, Domin+19, loi+19b, sur21, bs21}. In the absence of large-scale rotation in galaxy clusters, the magnetic fields in the ICM are expected to be amplified from insignificant seed-fields and maintained by the action of \textit{fluctuation dynamo} \citep[e.g.,][]{K68, zeldovich1990, scheko+04,r19,ss21}. Assuming large turbulent length scales $l_{0} \sim 100\kpc$ and turbulent velocities $v_{0} \sim 300\kms$, fluctuation dynamos can efficiently amplify magnetic fields by random stretching of the field by turbulent eddies \citep{SSH06, CR09, vazza+18} on timescales 
$\tau = l_{0}/v_{0} \sim 10^{8}\yr$, much shorter than the age of the cluster.

A characteristic signature of fluctuation dynamo action is that the generated field components are non-Gaussian, and, along with density and velocity fields, they are spatially intermittent \citep[e.g.,][]{hbd04, scheko+04, seta+20, sur21}. Since the Stokes parameters are non-linearly related to the magnetic field components, they are more sensitive to the morphology of these fields in the ICM \citep{WSE09}.
However, in contrast, observational studies of RM and depolarization in the ICM are often compared to numerical models derived by assuming Gaussian random magnetic fields having a simple power-law power spectrum \citep[e.g.,][]{murgia+04, laing+08, vacca+10, osinga2022}. It is thus imperative to extract information about the RM, synchrotron emissivity, and polarization signals directly from numerical magnetohydrodynamic (MHD) simulations of fluctuation dynamos in order to make a meaningful comparison to observations.
With the advent of advanced cosmological simulations in recent years, synthetic observations from them have been analyzed \citep{vazza+18, loi+19b, vacca+2024}.
But, cosmological MHD simulations of galaxy clusters, while indispensable for capturing large-scale structure formation, are unable to resolve the enormous separation between the driving, viscous, and resistive scales relevant for fluctuation dynamos. Moreover, their effective Reynolds and magnetic Prandtl numbers are determined by numerical dissipation rather than physical transport coefficients. Idealised simulations of fluctuation dynamos in non-ideal ICM-like settings are therefore essential to isolate the underlying amplification physics, determine saturation levels and magnetic-field morphology and thus provide a complimentary approach, however, for a smaller representative volume compared to the cosmological simulations.

In this chapter, we use MHD simulations of fluctuation dynamo operating in the ICM to explore the statistical properties of polarized emission and the observational advancements SKAO will make towards our understanding of the magnetic fields and turbulence driving in the ICM. With additional information from high quality X-ray spectroscopy from XRISM, we can glean direct information on the velocity fields and power spectrum of kinetic energy in galaxy clusters \citep[see][]{XGRISM2025coma}. The synergy between SKAO and XRISM will lead to a paradigm shift in our understanding of turbulence in the ICM.

\section{MHD simulations of fluctuation dynamo and the SKAO} \label{sec:simulation}

To explore the efficacy of detecting and characterizing the properties of total and polarized synchrotron emission from the ICM, we made use of non-ideal MHD simulations of fluctuation dynamo to generate synthetic maps of Stokes parameters ($I, Q$ and $U$) using the  \texttt{COSMIC} polarization transfer package \citep{basu19b}. 
We briefly summarize the key aspects of the simulations. A complete description of the numerical setup is provided in \citet{bs21}. The simulations solve the full set of MHD equations in dimensionless units using the {\texttt{FLASH}} code (\texttt{version 4.2})\footnote{\url{https://flash.rochester.edu/site/flashcode/}}
 on a uniform, periodic grid with a resolution of $512^{3}$. An isothermal equation of state is adopted, along with explicit viscosity and magnetic resistivity. The initial conditions consist of a uniform density and sound speed, both set to unity, and zero initial velocity. We emphasize that the inclusion of non-ideal effects is crucial for fluctuation dynamo studies, as the dynamo process is inherently governed by dynamics occurring near the resistive scales. Starting from an initially weak seed magnetic field with an initial plasma beta $\beta_{\rm init} \sim 10^{6}$, fluctuation dynamos drive amplification of the field assisted by an artificial driving of turbulence as a stochastic Ornstein-Uhlenbeck (OU) process \citep{Fry+00,Benzi+08,EP88} with a finite time correlation. The amplitude of the driving was adjusted such that in the steady state, the rms Mach number $\mrms = \urms/c_{\rm s}\approx 0.2$, consistent with the typical subsonic, turbulent velocities in the core regions of the ICM \citep{hitomi+16, Hitomi+18b}. Here, $\urms$ is the turbulent velocity and 
$c_{\rm s}$ is the isothermal sound speed. Recent cosmological simulations of hierarchical structure formation, including the formation of galaxy clusters, show that turbulence in the cluster core is dominated by solenoidal modes \citep[e.g.,][]{miniati15,vazza+17,wittor+17}. Thus, to maximize the efficiency of the dynamo and to explore the impact of different turbulent driving scales in our simulations, turbulence was driven with solenoidal modes\footnote{This is achieved by 
decomposing the acceleration field into solenoidal and compressive 
components using a projection operator in Fourier space. In index 
notation the operator is $\mathcal{P}^{\zeta}_{ij}(\kk) = \zeta\mathcal{P}^{\perp}_{ij} + (1-\zeta)\mathcal{P}^{\parallel}_{ij}$, where $\mathcal{P}^{\perp}_{ij}$ and $\mathcal{P}^{\parallel}_{ij}$ are the solenoidal and compressive projection operators, respectively and $\zeta \in [0,1]$ is an adjustable parameter which controls the solenoidal contribution.
We chose $\zeta = 1$ for purely solenoidal driving.} (i.e., $\nabla\cdot\FF = 0$, where $\FF$ is the forcing term in the momentum equation) over three characteristic range of wave-numbers ($\kk$) --- (i) $1 \leq |\kk|L/2\uppi \leq 3$, (ii) $4 \leq |\kk|L/2\uppi \leq 6$, and (iii) $7 \leq |\kk|L/2\uppi \leq 9$. Here, $L$ is the size of the domain. This 
implies that the average forcing wave numbers are $\kf = 2, 5$ and $8$. 
In all three simulations, the magnetic Prandtl number is fixed at 
$\Pm = \Rm/\Rey = 1$, with Reynolds and magnetic Reynolds numbers of 
$\Rey = \Rm = 1080, 1450$, and $1425$ for $\kf = 2, 5$, and $8$, respectively. In each case, the magnetic field is evolved from the kinematic regime to the saturated state over several eddy turnover times, thereby sampling many statistically independent realizations of the dynamo in both phases. For all subsequent analyses presented in this work, we made use of multiple realizations drawn from the saturated state of the dynamo.

\paragraph{Normalization of physical units:} 
We normalize the length of the simulation domain to $L = 512$\,kpc, implying a resolution of 1\,kpc. Thus, for the average forcing wave-numbers used in the simulations, the turbulent driving scales ($\lf = 2\uppi/\kf$) correspond to $L/2 = 256$\,kpc, $L/5 = 102.4$\,kpc, and $L/8 = 64$\,kpc. It is important to note that, the scales, and therefore the results can be generalized to a sample of clusters by comparing $\lf$ to their core radius $r_{\rm c}$.\footnote{We will also represent a length scale associated with the simulations to the corresponding scaling with $r_{\rm c}$ for generalization to other clusters.} For example, $\lf = 256$\,kpc roughly corresponds to $r_{\rm c} \approx 300$\,kpc of the Coma cluster \citep{BHB92, bonaf10}, while $\lf = 64$\,kpc is similar to the scale height of the cluster core. In other words, the different $\lf$s can be equivalently interpreted as---
\begin{enumerate}[(i)]
    \item $\lf = 256\,{\rm kpc} \equiv \lf/2\,r_{\rm c} \gtrsim 1/2$ corresponds to turbulence driven by the cascade of vortical motions generated in oblique accretion shocks and instabilities during cluster formation on megaparsec scales \citep{SSH06, RKCD08, miniati15}.
    \item $\lf = 102.4\,{\rm kpc} \equiv \lf/2\,r_{\rm c} \approx 1/20\textrm{--}1/4$ corresponds to feedback from AGN in the ICM \citep{fabian2012, bourne2017,ehlert2021}, or turbulent heating \citep{zhuravleva2014}.
    \item $\lf = 64\,{\rm kpc} \equiv \lf/2\,r_{\rm c} \lesssim 1/30$ corresponds to turbulent energy input on galactic scales, e.g., driven by gas accretion and/or star formation driven feedback from galaxies \citep{donnert2009, dubois2012, pakmor2016, wiene17}.
\end{enumerate}
In order to facilitate computation of the observable quantities in physical units, we assume a typical ICM electron density $\langle n_{\rm e} \rangle = 10^{-3}\cmn$ and rms sound speed $c_{\rm s} = 10^{3}\kms$ \citep{Sar88}. Thus, rms $\mathcal{M} \approx 0.2$ in our simulations implies rms velocity $\urms \approx 200\kms$, consistent with measurements of pressure fluctuations and spectral lines observed in the X-ray \citep[e.g.,][]{sanders2013, hitomi18a, zhurav+19, XGRISM2025centaurus, XGRISM2025coma}. Therefore, for fully ionized gas in  the ICM at $\mathcal{O}(10^8\,{\rm K})$ temperatures, this implies a mean gas mass density $\langle \rho\rangle = \langle n_{\rm e}\rangle\,\mu_{\rm e}\,m_{\rm p} \approx 1.97\times 10^{-27}\gcm$, and rms field strengths ($\brms$) between 1.3 and $1.7\mkG$.
Here, $\mu_{\rm e} = 1.18$ is the mean molecular weight per free electron and $m_{\rm p}$ is the proton mass. For details, please see \citet{bs21}.

\paragraph{Synthetic observations:} 
After converting the output of the MHD simulations to physical units, \texttt{COSMIC} package \citep{basu19b} was used to compute broad-bandwidth synthetic maps of the total and polarized synchrotron emission. For the purpose of our analyses, we chose $x$- and $y$-axes to be in the plane of the sky, i.e., $B_\perp = \sqrt{B_x^2 + B_y^2}$ contributes towards the synchrotron emission, while $z$-axis was chosen to be along the line of sight (LOS), so that $B_\| = B_z$ contributes to Faraday rotation. Fluctuations in $B_\perp$ and $B_\|$, both in the plane of the sky and along the LOS, cause depolarization. Because our simulations are devoid of cosmic rays, we assume a uniform number density of CREs ($n_{\rm CRE}$) in each mesh point. CREs follow a power-law energy spectrum of the form $n_{\rm CRE}(E) = n_0\,E^\gamma$ with a constant energy index $\gamma = -3$. Furthermore, the normalization $n_0$ is chosen such that the total synchrotron flux density at 1\,GHz over the entire synthetic map is 1\,Jy, i.e., the frequency spectra is given as, $I(\nu) = I_0\,\left[\nu/1\ghz\right]^\alpha$, where $I_0 = 1$\,Jy, and spectral index $\alpha = -1$. In this chapter, we will focus mainly on using fractional polarization ($\pf$) as it is independent of the fiducial choice of $I_0$, and appropriately scale $I_0$ to investigate the requisite sensitivity with the SKA. With these assumptions, synthetic 2-D images 
of Stokes $I, Q$ and $U$ were generated between 0.5 and 8\,GHz. This choice of frequency setting is sufficient to glean insights into what is expected to be observed with the SKA-Mid in Bands\,1, 2 and 5a.

\paragraph{On the assumption of constant CRE density:} Because our simulations do not consider cosmic rays as mentioned above, to compute synchrotron emission, we assumed a constant $n_{\rm CRE}$ throughout the simulated domain. Furthermore, the energy index $\gamma = -3$ was chosen so that the synthetic total intensity maps have a spectral slope of $\alpha = -1$, typically observed in galaxy clusters \citep{Fer+12}. We note that, varying CRE acceleration efficiency and subsequent energy losses via synchrotron and inverse-Compton cooling could give rise to spatially varying $\alpha$ and $n_{\rm CRE}$. This requires a quantitative analysis of CRE transport and solving the diffuse-loss equation. As we are interested in the statistical properties of the synchrotron emission from the ICM and not the evolution over cosmic time, we have not included such a treatment. 
However, as discussed in \citet{sur21}, for clusters at redshift $z$ with 
$\brms < 3.25\,\upmu{\rm G}\,(1+z)^2$, inverse-Compton losses would dominate over synchrotron cooling, and in the presence of nearly uniform CMB energy density on cluster scales, variation of synchrotron spectrum due to local magnetic fields would be low. Furthermore,
the cooling timescale for CREs emitting at gigahertz frequencies in $\upmu$G-strength magnetic fields is $\mathcal{O}(10^8\,\textrm{yr})$ \citep[e.g.,][]{longa11}, within which, under Bohm diffusion \citep{Drury83, Bagchi+02}, the CREs diffuse away from the injection regions to at most $\mathcal{O}({\rm kpc})$ \citep{brunetti+jones2014}.
This indicates that the CREs are likely re-accelerated away, by the same turbulence that is driving the fluctuation dynamo, from such small scales to Mpc scales associated with radio emission from cluster halos.
We therefore expect, for cluster-wide turbulence, the variation of energy going into CREs vary on hundreds of kpc scale, and thus a nearly constant $n_{\rm CRE}$ over the scale of turbulent eddies is plausible.
Any further spatial variations of $n_{\rm CRE}$ caused by cooling could be smoothed by the same turbulence through efficient mixing of CREs due to large turbulent diffusivity. Therefore, we believe that the assumption of constant $n_{\rm CRE}(E)$ would provide reasonable insights into the properties of synchrotron emission from the ICM. However, we note that any global trends caused by stratification, e.g., radial profile of $n_{\rm e}$, $n_{\rm CRE}$, $b_{\rm rms}$ and $\alpha$ are not captured in our isothermal simulations. Hence, the results discussed in this chapter are representative of a localized volume of the ICM and not its global property.

\section{Characteristics of synchrotron emission induced by fluctuation dynamo} \label{sec:synthmaps}

Here, we briefly summarize the salient features of total ($I_{\rm sync}$) and polarized ($PI$) synchrotron emission from ICM originating due to the action of fluctuation dynamo. For detailed discussions on the analyses, see \citet{sur21, bs21} and \citet{dutta2024}.

\begin{figure}
\centering
\begin{tabular}{ccc}
{\large $l_{\rm f}=256\kpc$} & {\large $l_{\rm f}=102.4\kpc$} & {\large $l_{\rm f}=64\kpc$} \\
\includegraphics[width=5.2cm, trim=10mm 15mm 12mm 15mm, clip]{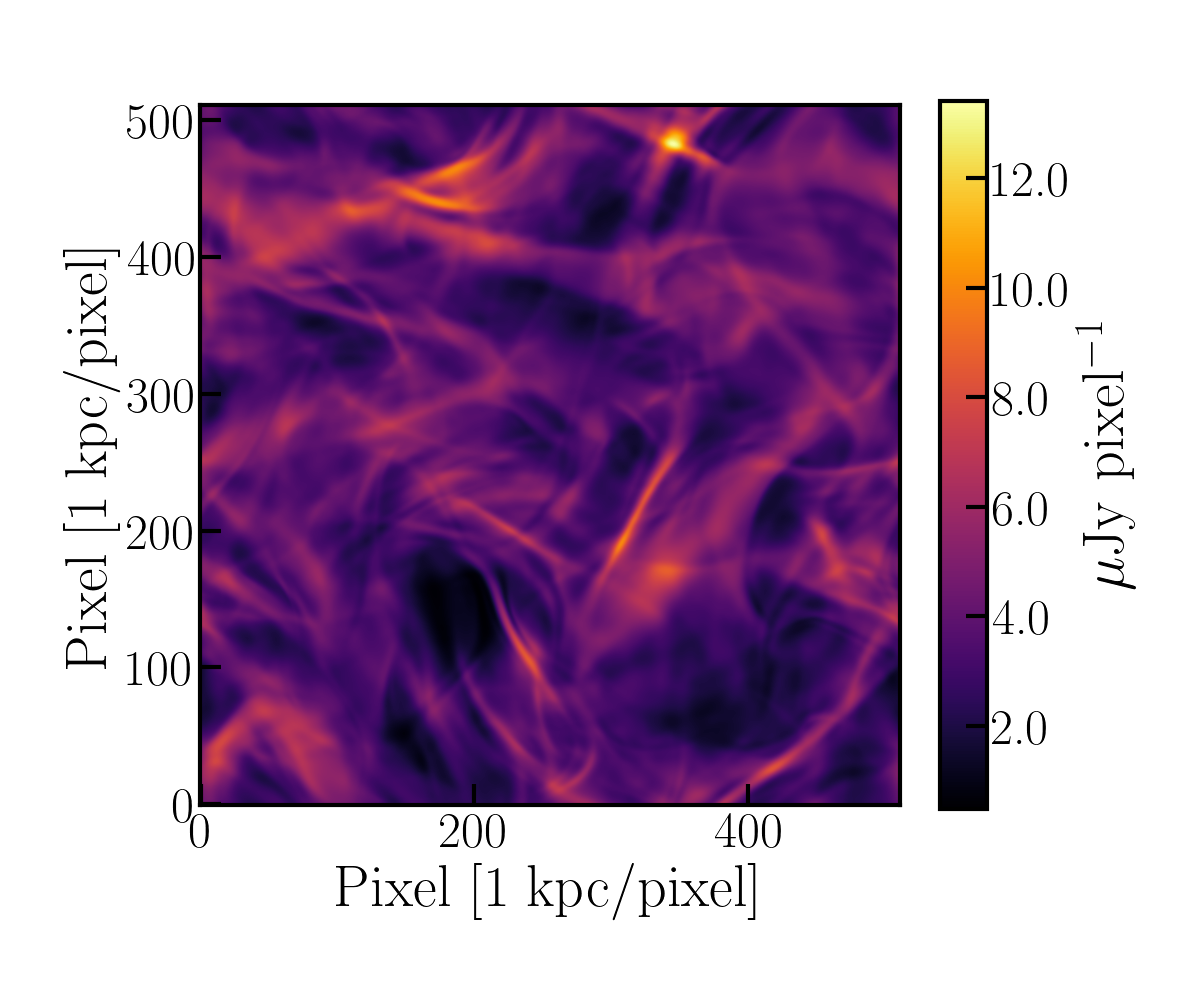} &
\includegraphics[width=5.2cm, trim=10mm 15mm 12mm 15mm, clip]{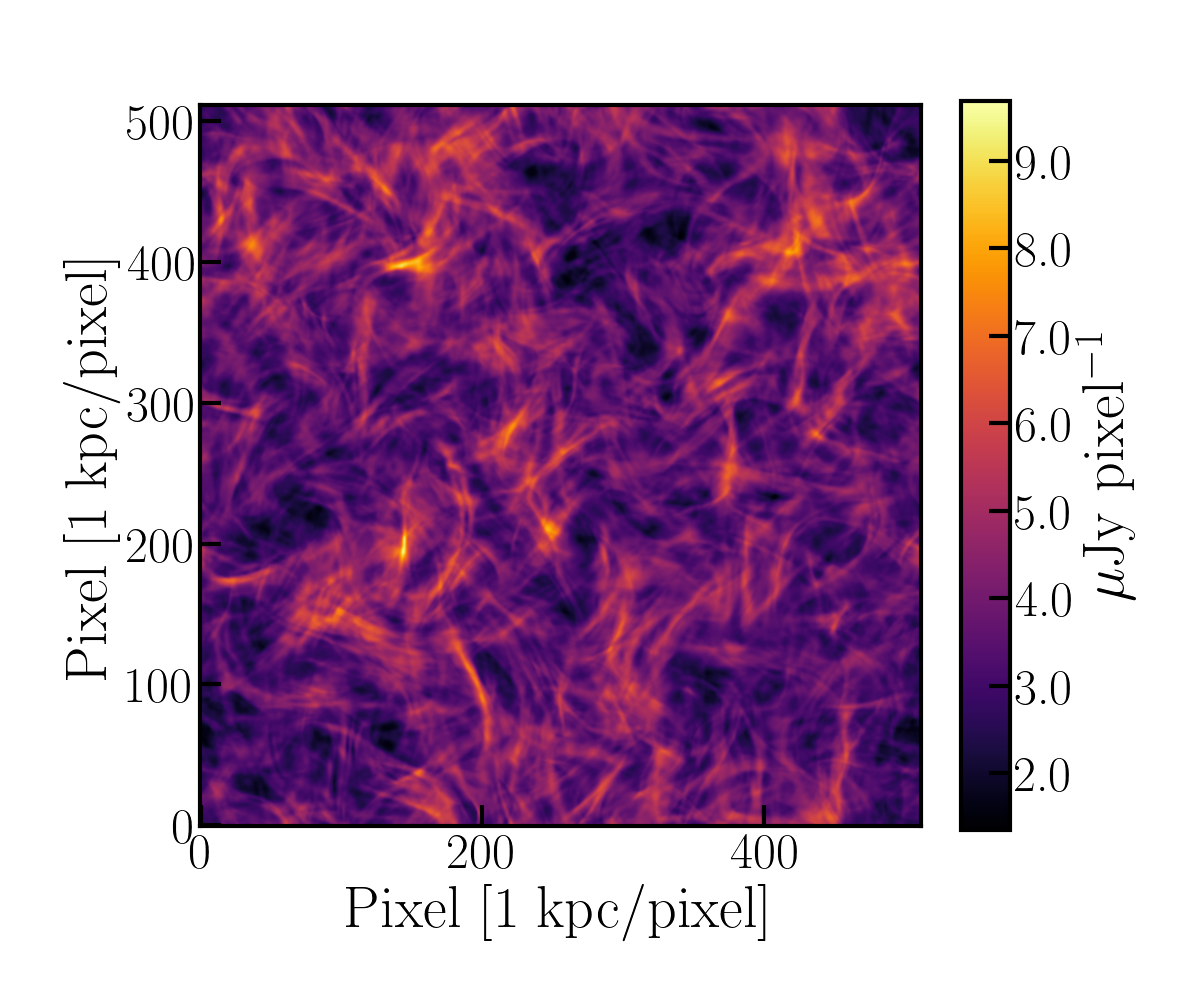} &
\includegraphics[width=5.2cm, trim=10mm 15mm 12mm 15mm, clip]{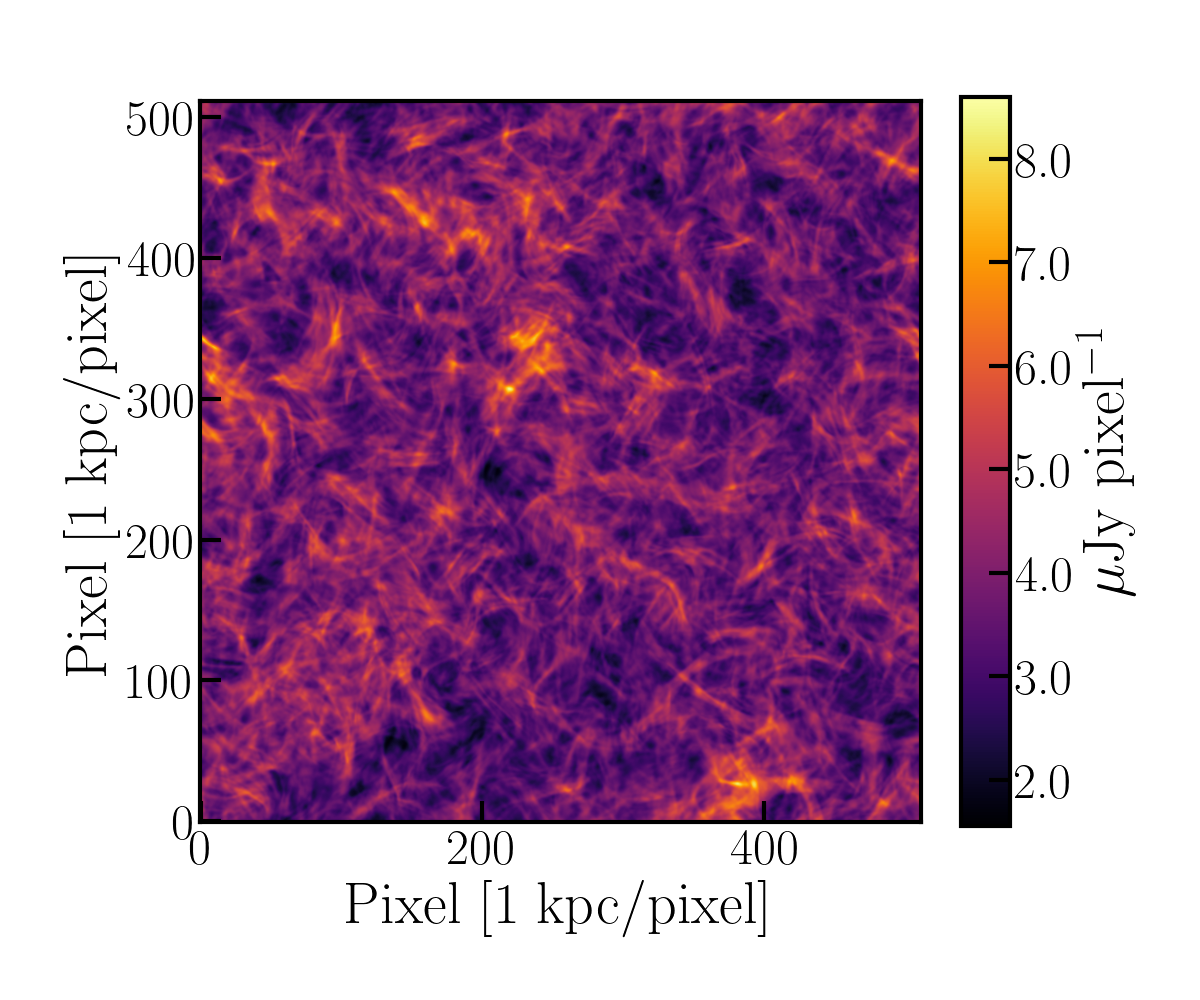} \\
\includegraphics[width=5.2cm, trim=10mm 15mm 12mm 15mm, clip]{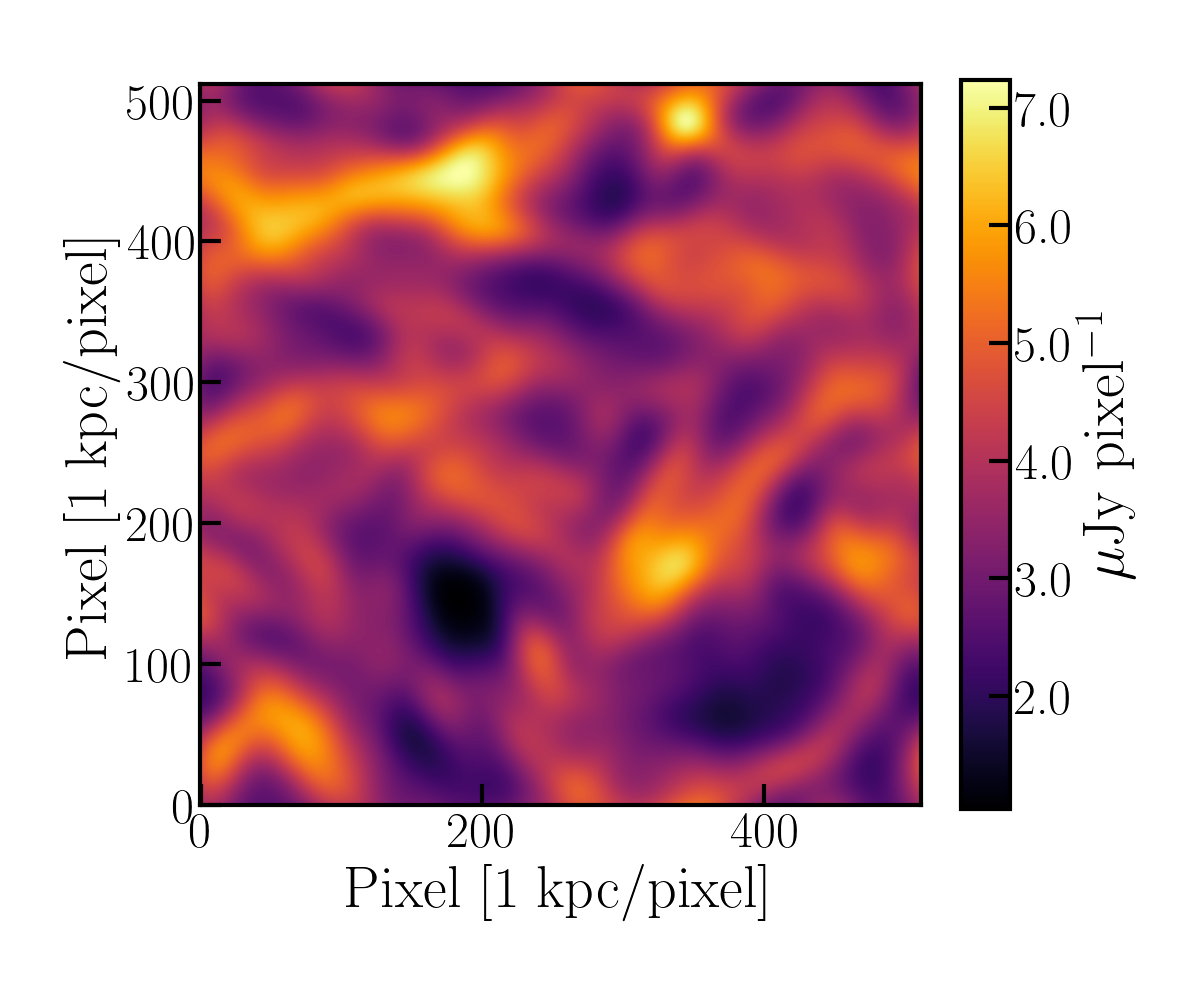} &
\includegraphics[width=5.2cm, trim=10mm 15mm 12mm 15mm, clip]{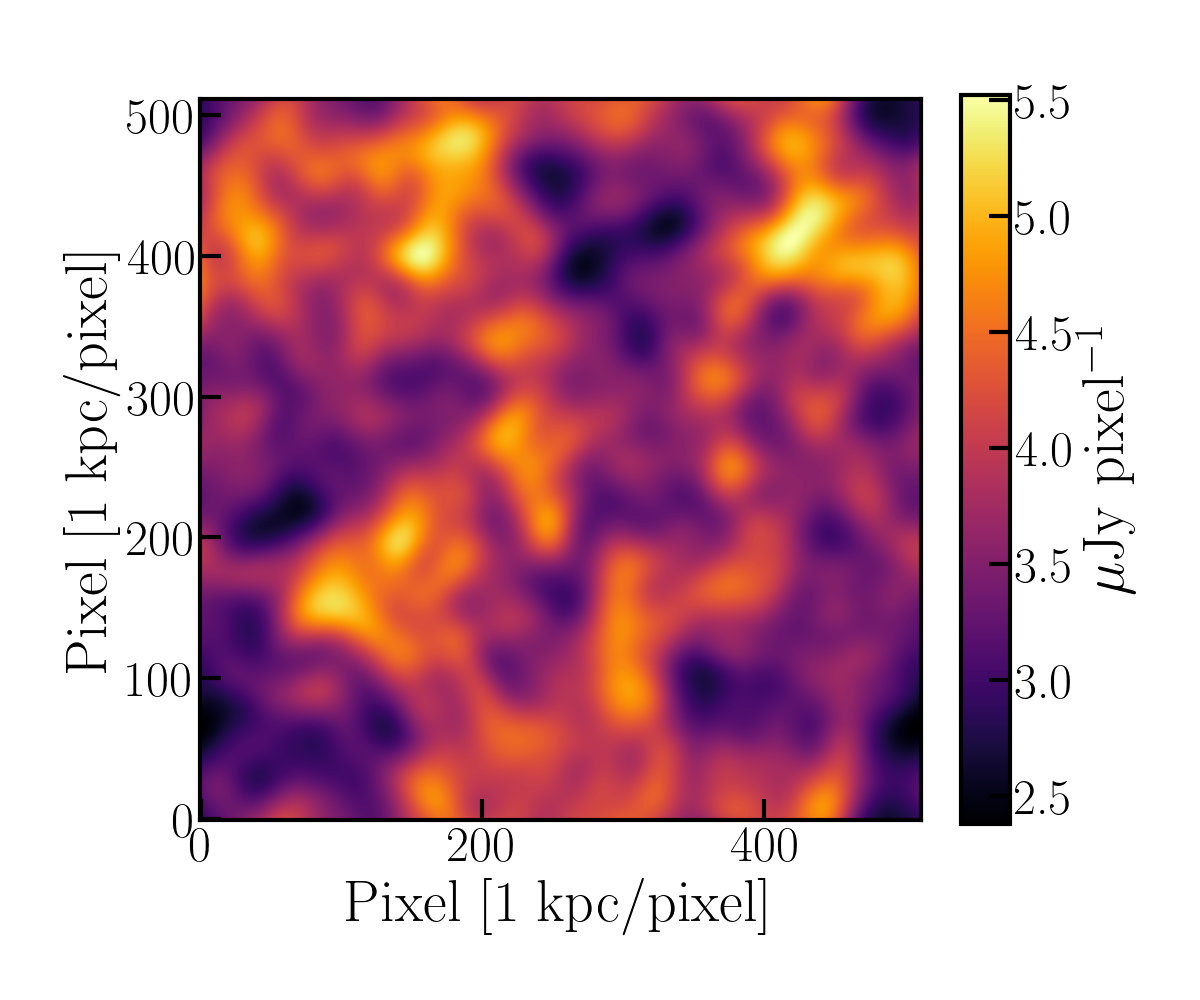} &
\includegraphics[width=5.2cm, trim=10mm 15mm 12mm 15mm, clip]{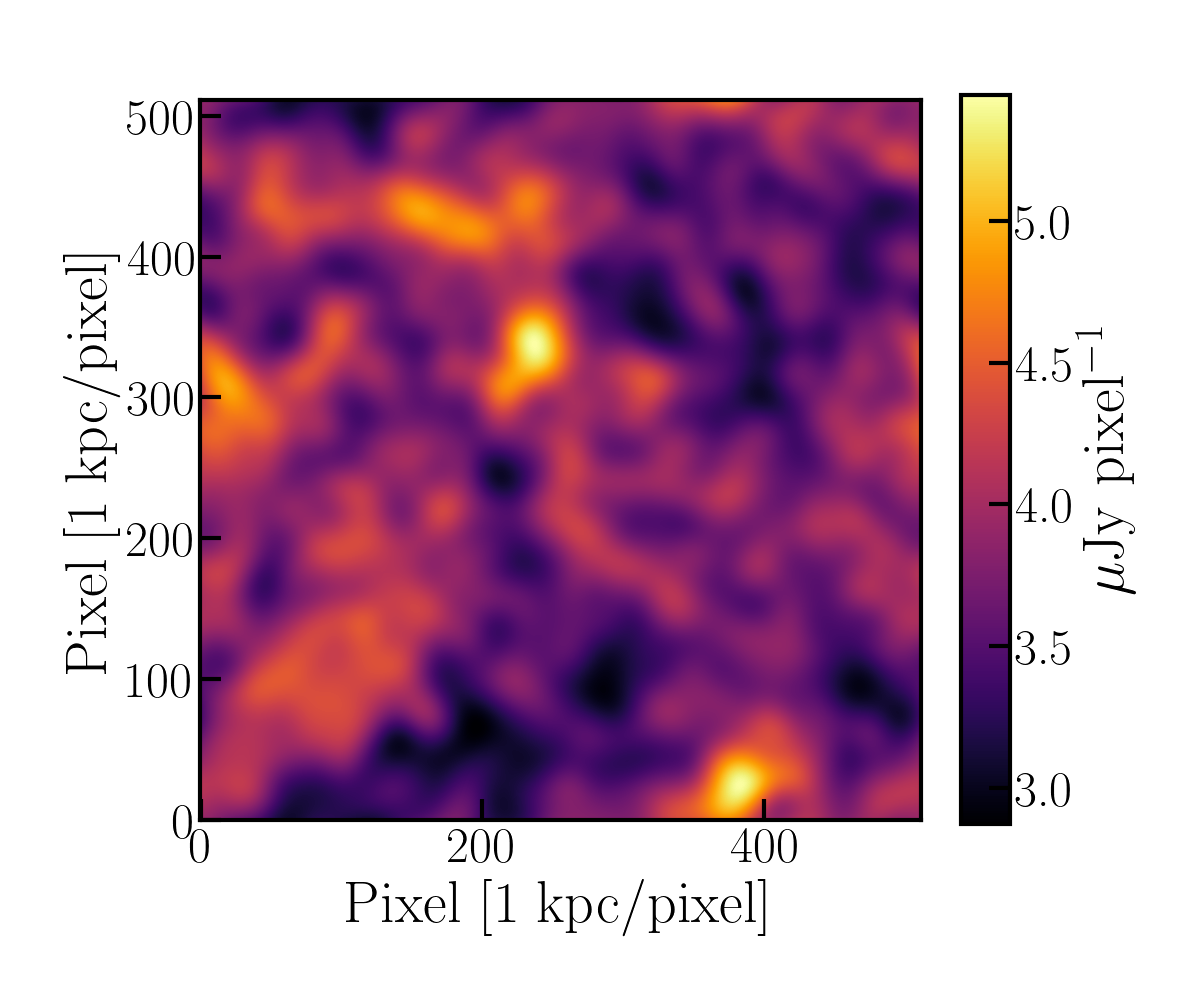} \\
\end{tabular}
\caption{\textbf{\textit{Top row:}} Synthetic maps of the total intensity synchrotron surface brightness ($I_{\rm sync}$) at the native 1\,kpc resolution of the 
simulations. \textbf{\textit{Bottom row:}} Maps of $I_{\rm sync}$
in the top row smoothed by a Gaussian kernel with a FWHM of $30\kpc$.
The \textbf{\textit{Left}}, \textbf{\textit{middle}}, and 
\textbf{\textit{right}} columns are for $l_{\rm f}=256$, $102.4$, and $64\kpc$.}
\label{fig:totI}
\end{figure}

\subsection{Filamentary synchrotron emission in the ICM} \label{sec:totalI}

Fig.~\ref{fig:totI} shows the 2-D synthetic maps of $I_{\rm sync}$ at 1\,GHz for the three different $\lf$. At the native 1-kpc resolution of the simulations shown in the top panels, it is evident that synchrotron emission has pervasive, highly filamentary structures that fill up more volume for smaller $\lf$. The typical length scale associated with $I_{\rm sync}$, characterized by the integral length ($L_{\rm int}$), increases linearly from $\sim80\textbf{--}90$\,kpc to $\sim220\textrm{--}230$\,kpc, corresponding to $L_{\rm int}/2\,r_{\rm c} \approx 1/10$ to $1/2$, as $\lf$ increases from 64 to 256\,kpc \citep[see][]{bs21}. Such filaments are characteristic signatures of the action of fluctuation dynamo, and a proper description of them in observations is necessary to advance our knowledge on ICM dynamics. Quantitative analyses of morphology, e.g., through statistical measures like power spectrum, structure functions, Minkowski functionals, etc., and the probability density function of fluctuations of $I_{\rm sync}$ can provide direct insights into the properties of the turbulence driving mechanism in ICM \citep[see e.g.,][]{hu2019, bs21, dutta2024}. In fact, filamentary synchrotron emission have been also reported in synthetic maps generated from cosmological MHD simulations \cite[][]{loi+19b}. Such filamentary structures have a typical width of few kiloparsecs and span several tens of kiloparsecs. Hence, a spatial resolution similar to that used in the simulations, $\mathcal{O}(\rm 1\,kpc)$, is necessary to resolve them.

In the bottom panels of Fig.~\ref{fig:totI}, we show the $I_{\rm sync}$ maps smoothed by a $30\kpc$ Gaussian kernel, equivalent to observing a cluster at $z\approx 0.2\,(0.1)$ with a commonly used $10\arcsec\,(15\arcsec)$ resolution provided by current telescopes. It is apparent that the strong intensity contrasts are washed out, especially for smaller $\lf$ where the emission is more volume-filling. Often, due to low surface brightness of radio halos, data are further smoothed to even lower angular resolutions to improve sensitivity. Therefore, it is likely that these characteristic substructures in ICM are smoothed out, making radio halos appear spatially smooth.

\begin{figure}
\centering
\begin{tabular}{ccc}
{\large $l_{\rm f}=256\kpc$} & {\large $l_{\rm f}=102.4\kpc$} & {\large $l_{\rm f}=64\kpc$} \\
 & \large{Polarized intensity} &  \\
\includegraphics[width=5.2cm, trim=10mm 15mm 12mm 15mm, clip]{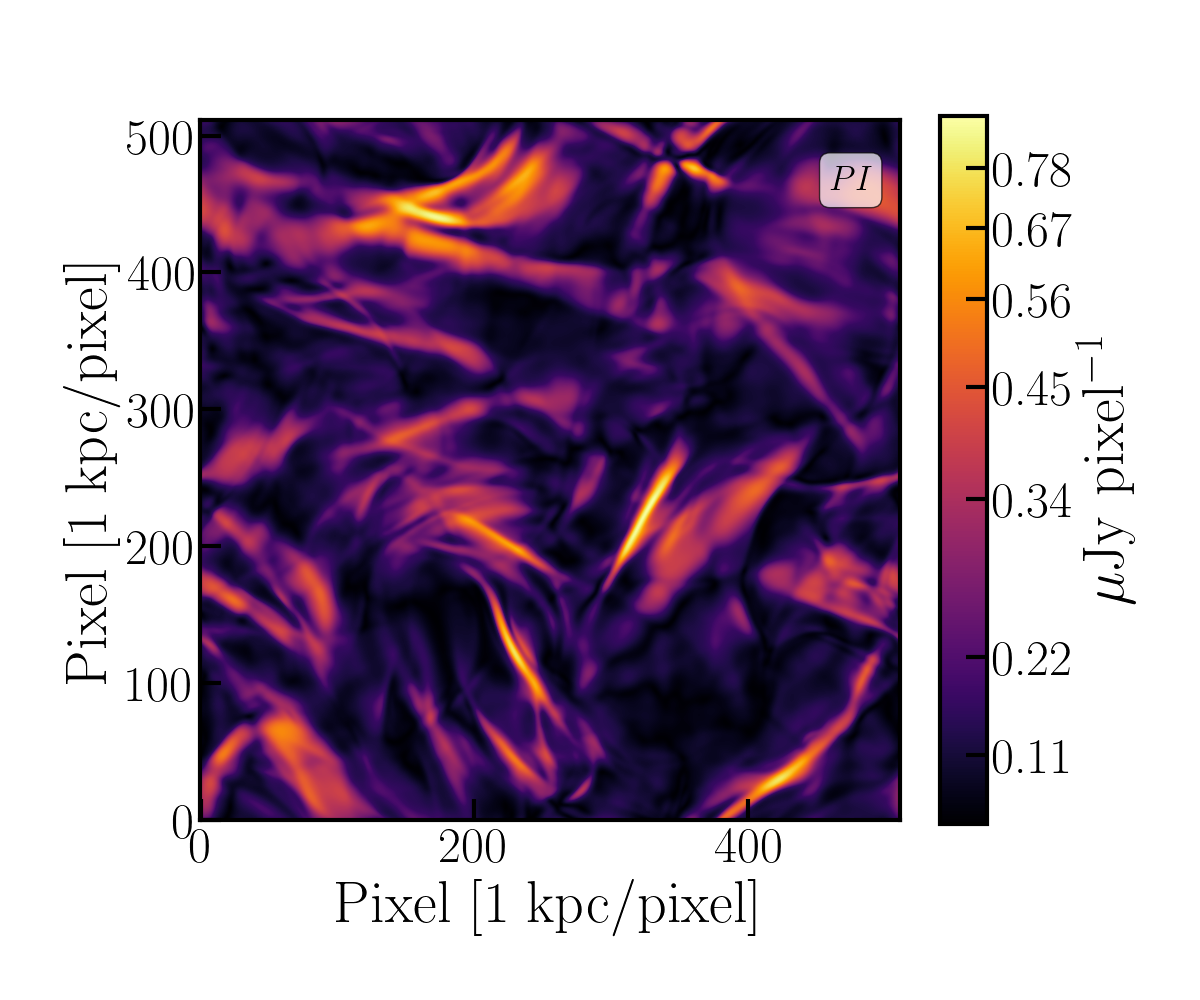} &
\includegraphics[width=5.2cm, trim=10mm 15mm 12mm 15mm, clip]{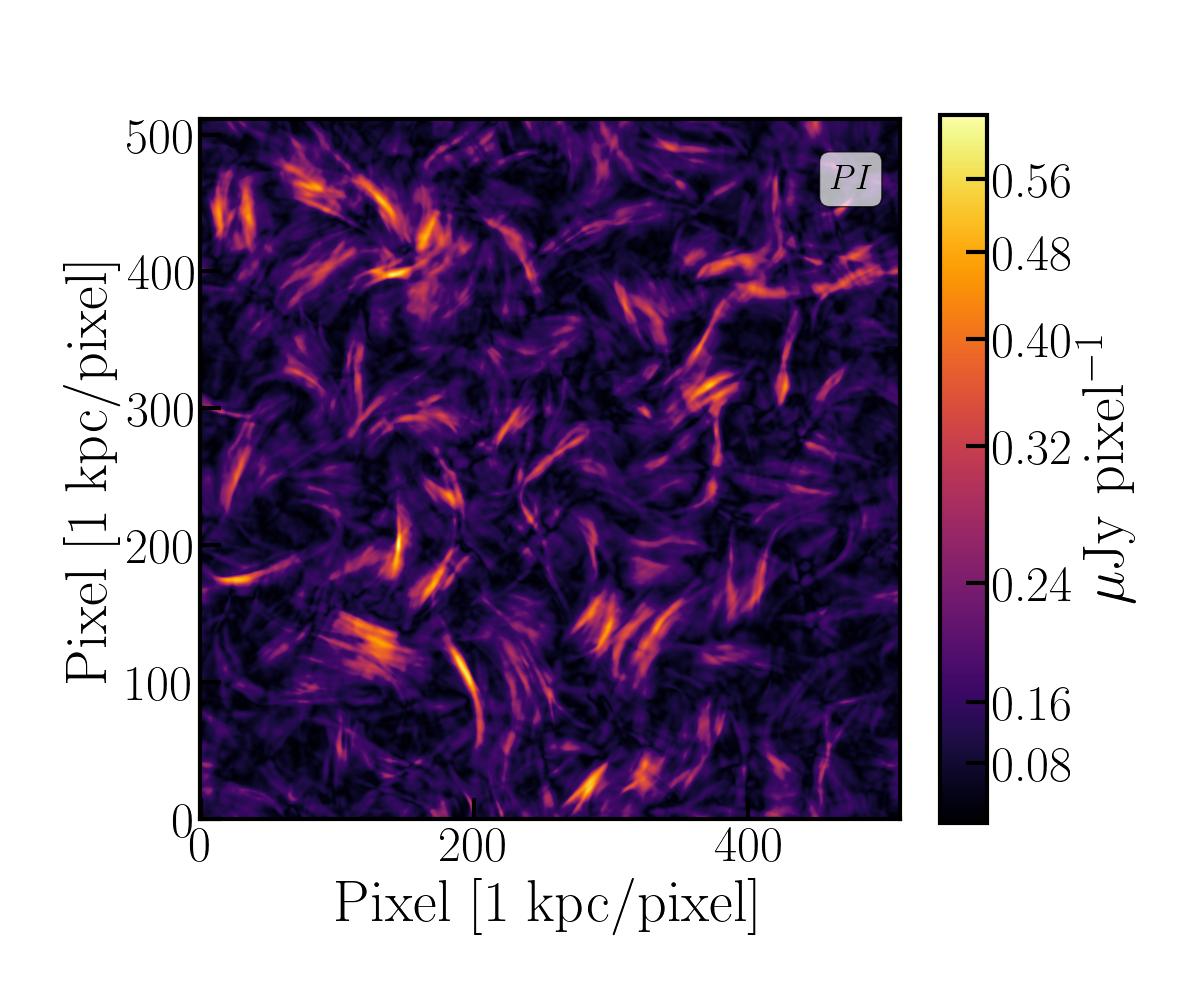} &
\includegraphics[width=5.2cm, trim=10mm 15mm 12mm 15mm, clip]{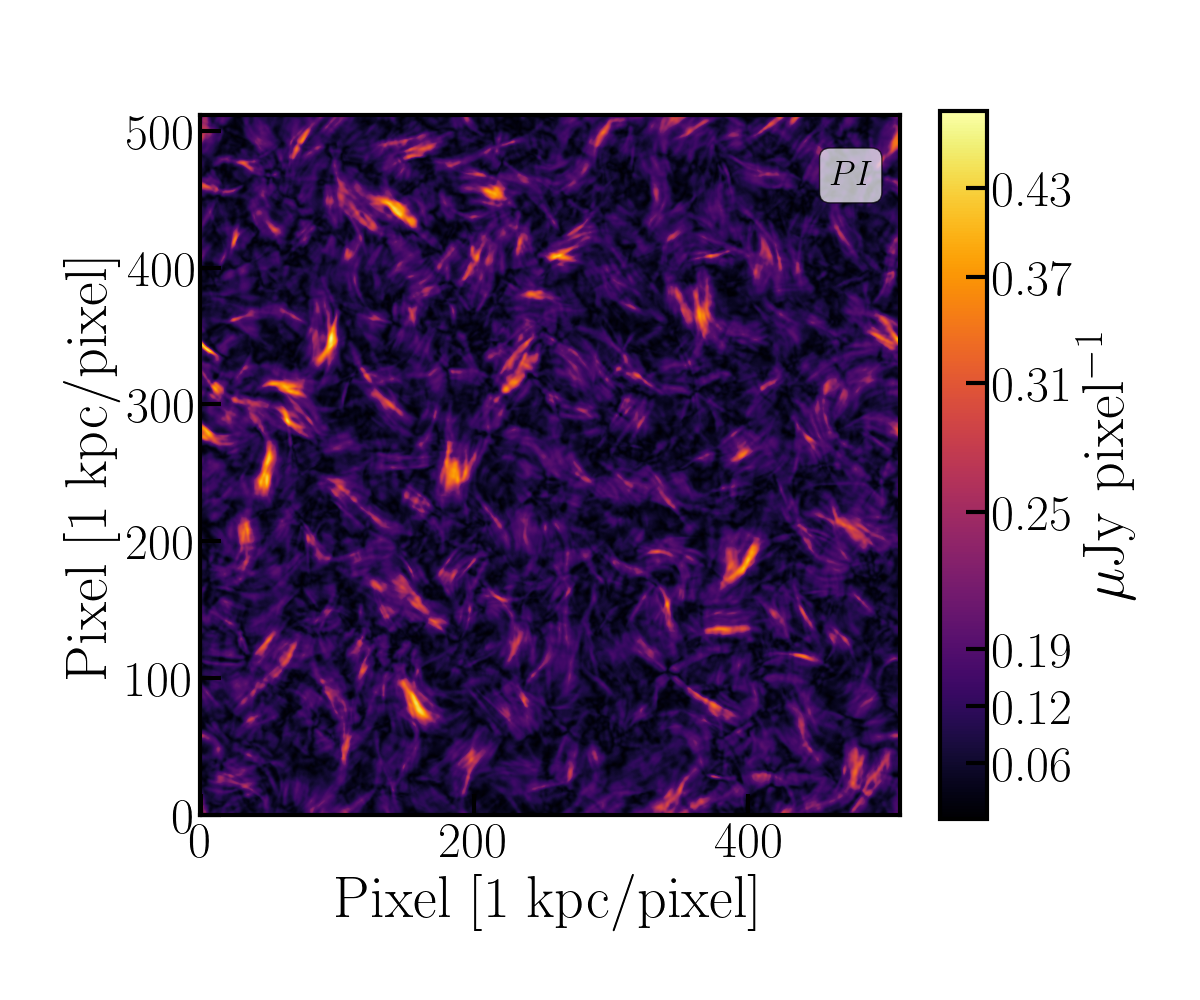} \\
 \multicolumn{3}{c}{\large{Fractional polarization (1\,kpc)}}\\ 
\includegraphics[width=5.2cm, trim=10mm 15mm 12mm 15mm, clip]{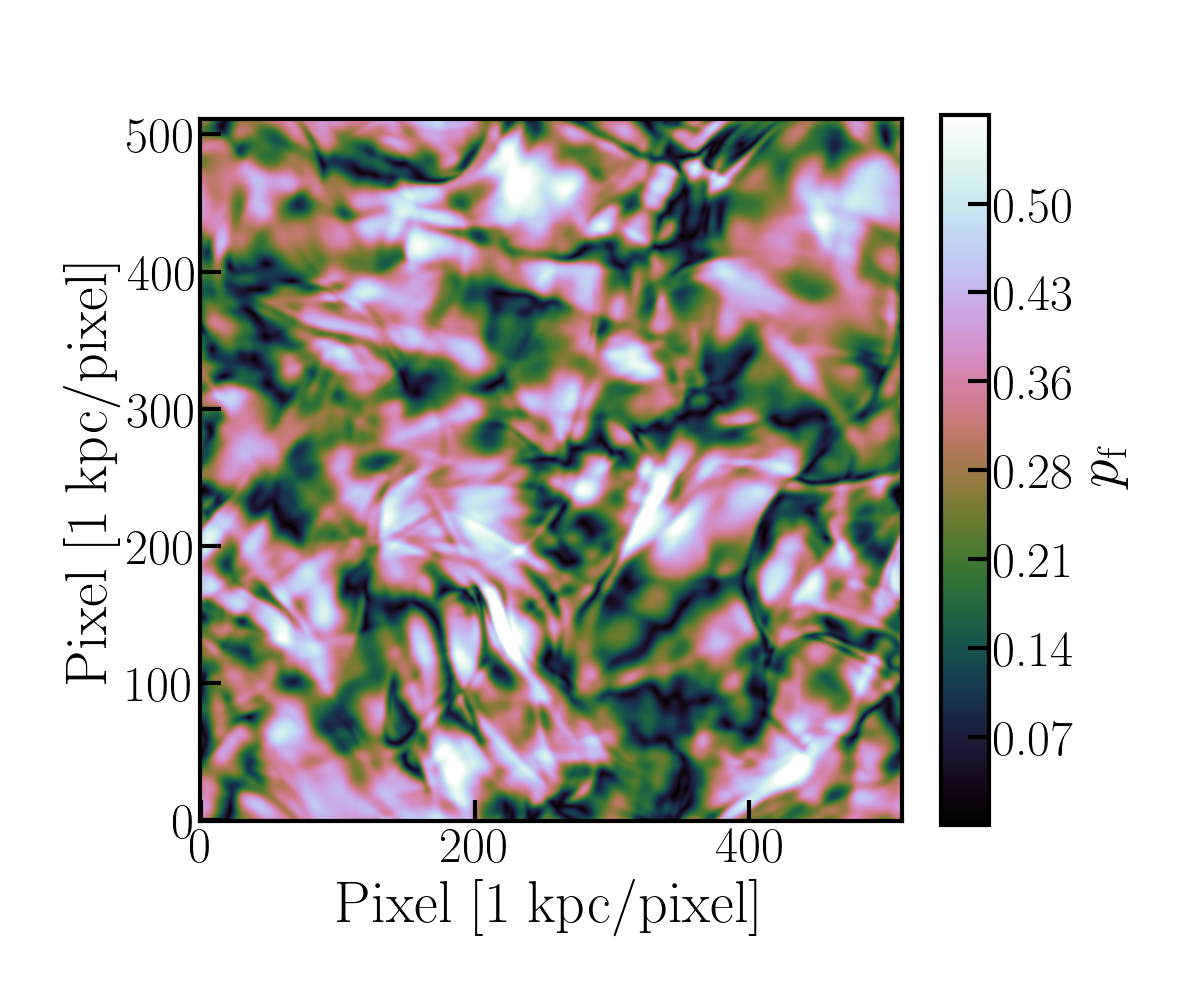} &
\includegraphics[width=5.2cm, trim=10mm 15mm 12mm 15mm, clip]{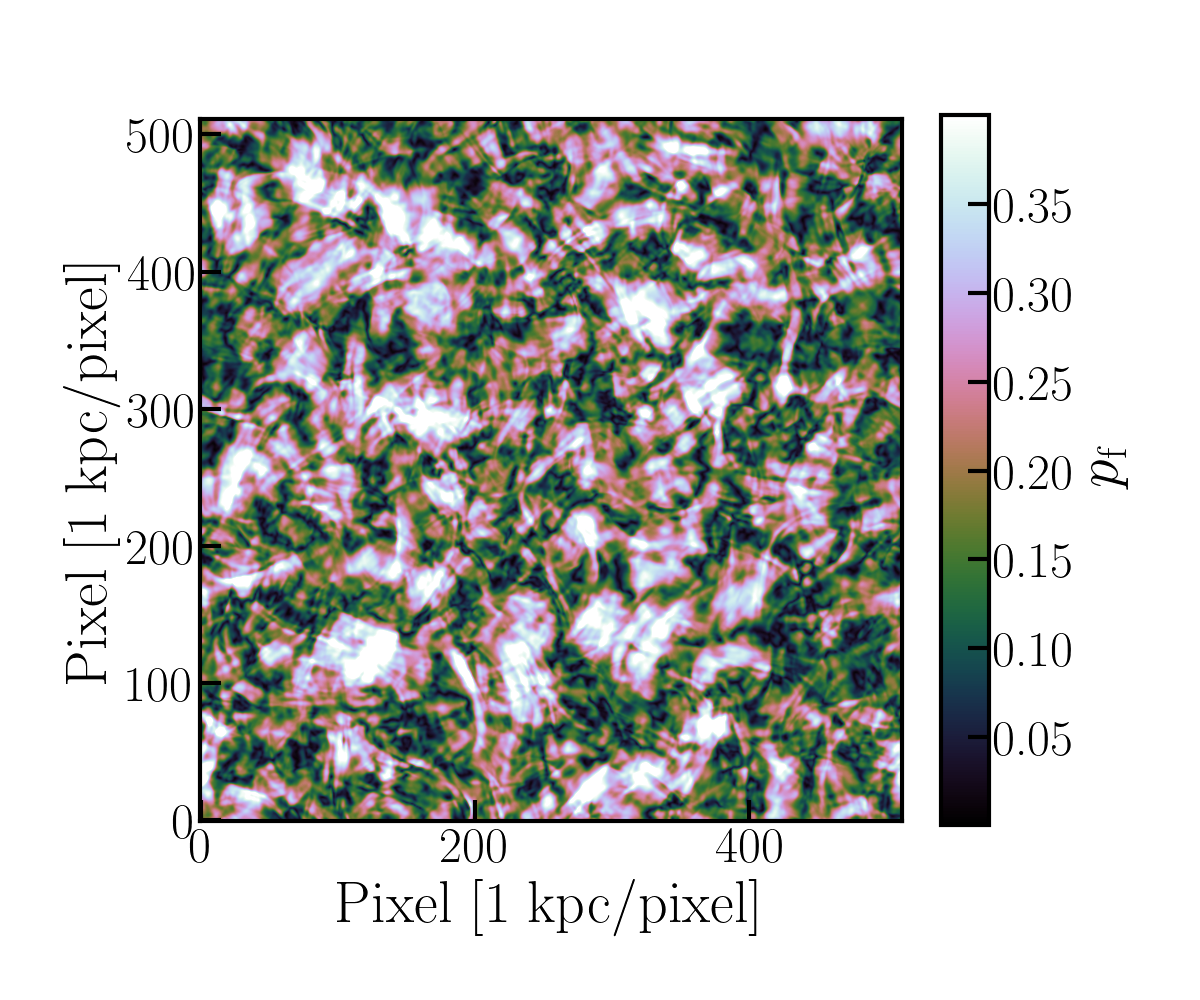} &
\includegraphics[width=5.2cm, trim=10mm 15mm 12mm 15mm, clip]{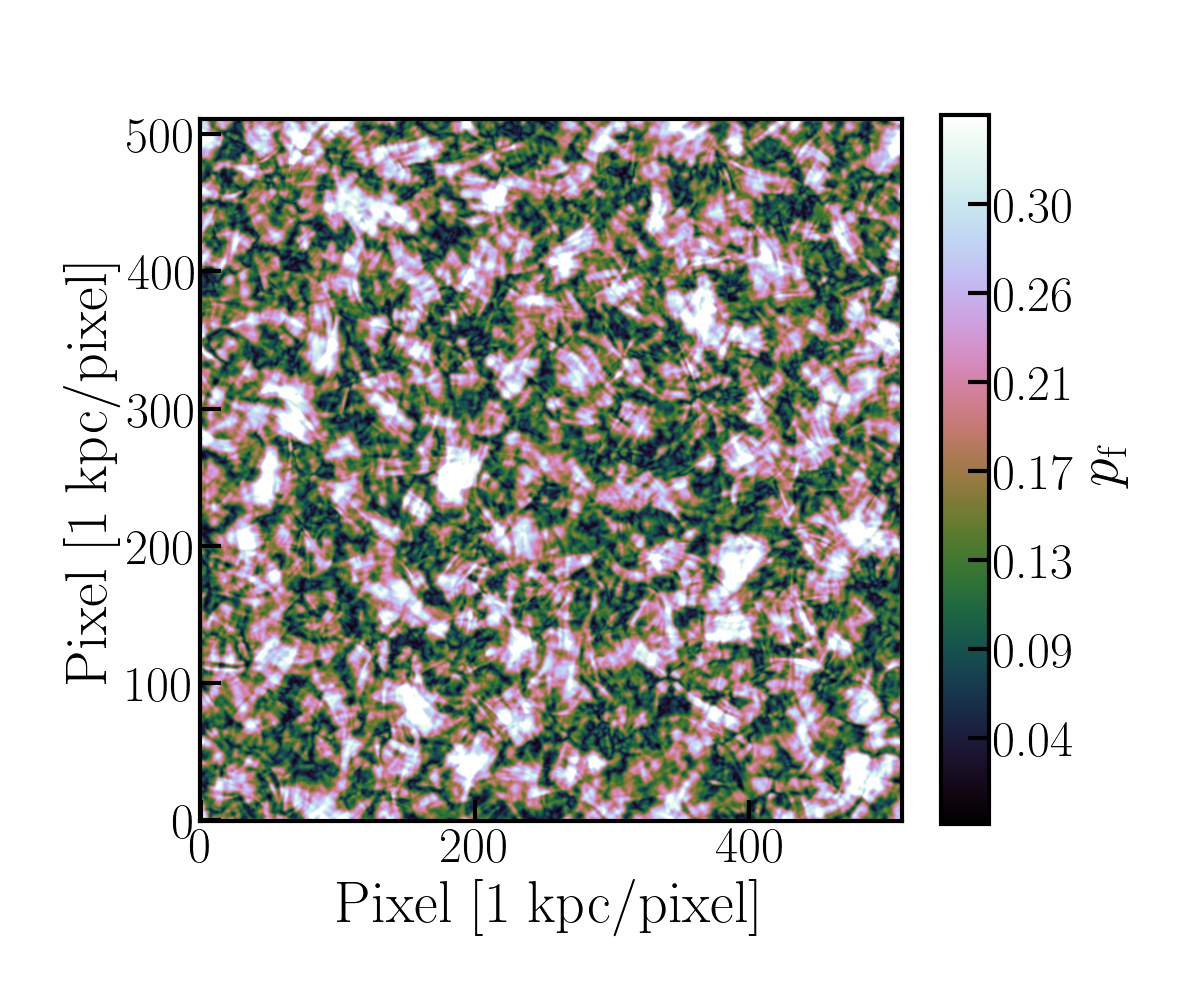} \\
 \multicolumn{3}{c}{\large{Fractional polarization (smoothed 10\,kpc)}}\\ 
\includegraphics[width=5.2cm, trim=10mm 15mm 12mm 15mm, clip]{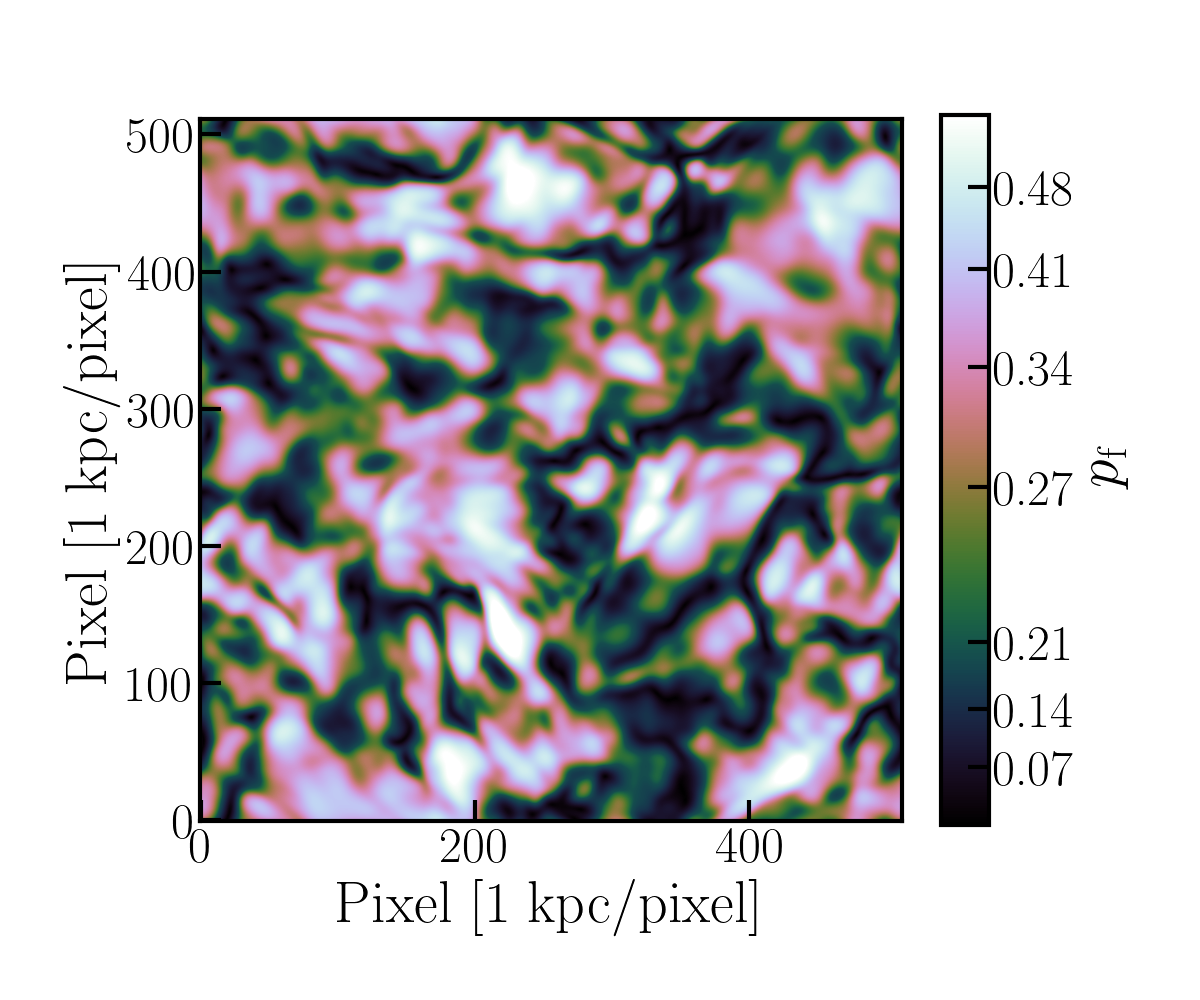} &
\includegraphics[width=5.2cm, trim=10mm 15mm 12mm 15mm, clip]{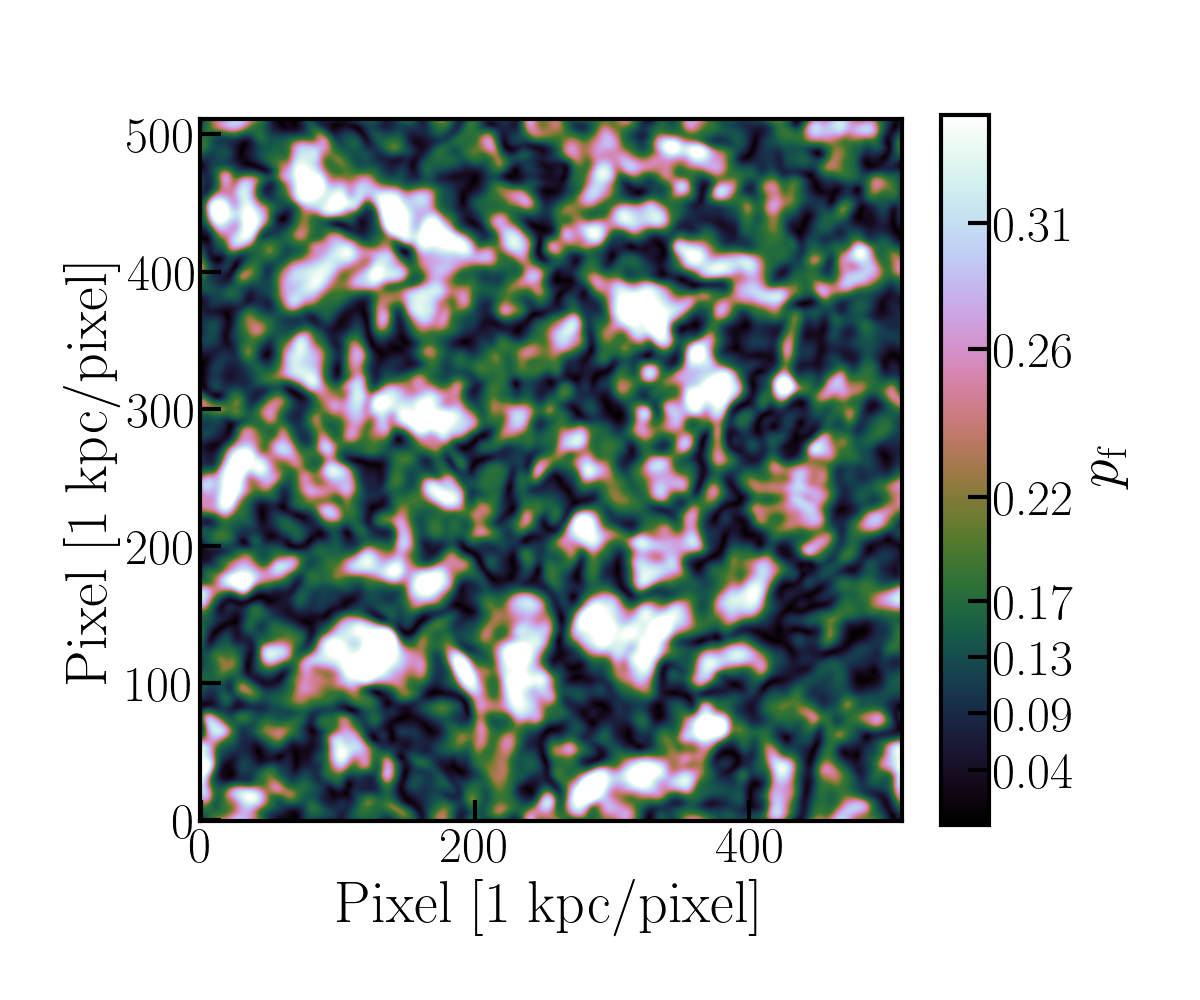} &
\includegraphics[width=5.2cm, trim=10mm 15mm 12mm 15mm, clip]{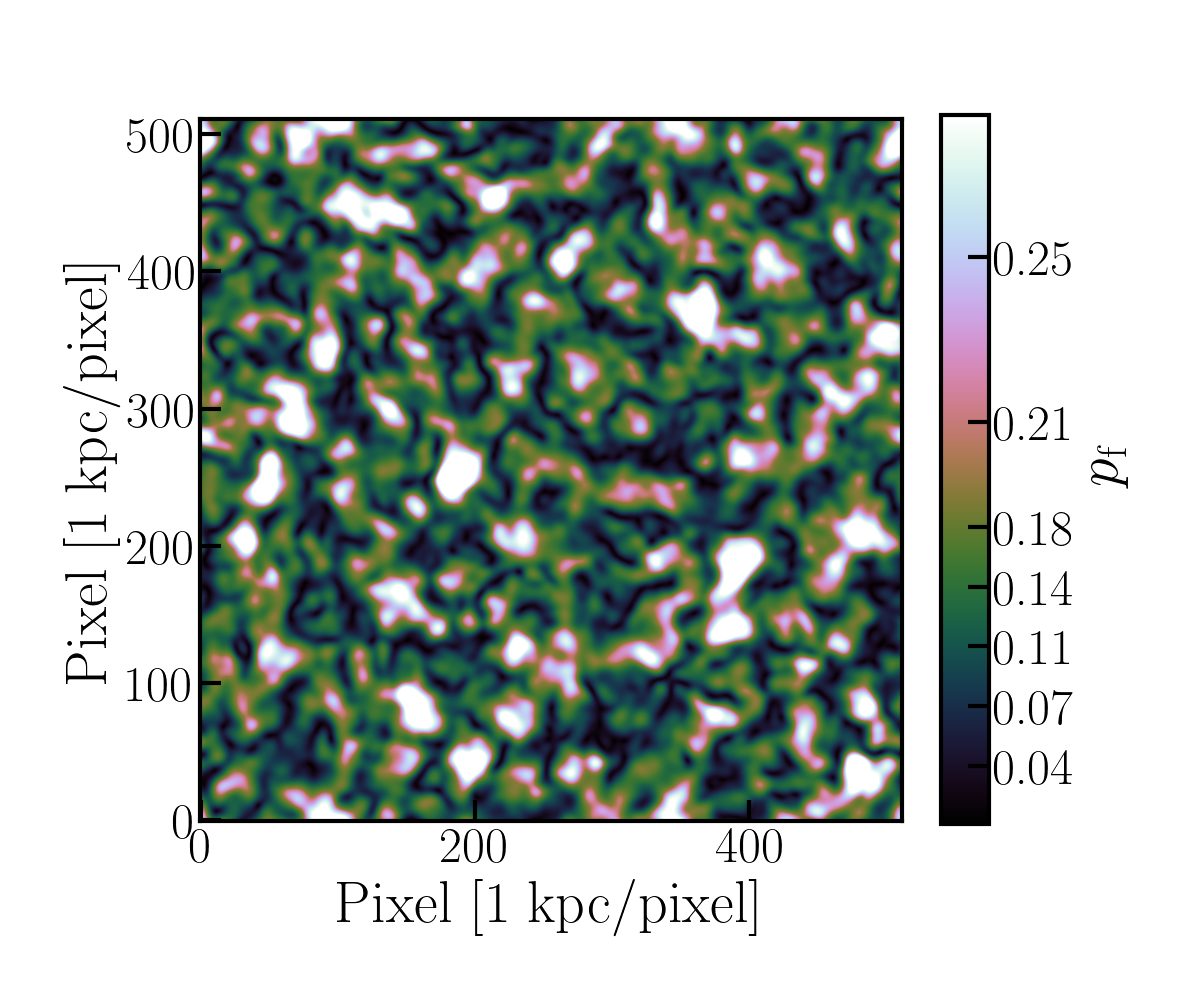} \\
\end{tabular}
\caption{\textbf{\textit{Top row:}} Polarized intensity maps at 6\,GHz in units of Jy~pixel$^{-2}$ for the three $\lf$ at 1-kpc resolution.  \textbf{\textit{Middle row:}} Fractional polarization ($\pf$) maps at 6\,GHz at 1-kpc resolution. \textbf{\textit{Bottom row:}}
$\pf$ after smoothing by a Gaussian kernel with FWHM 10\,kpc ($\equiv 10$\,pixels).}
\label{fig:polI_6GHz}
\end{figure}

\begin{figure}
\centering
\begin{tabular}{ccc}
{\large $l_{\rm f}=256\kpc$} & {\large $l_{\rm f}=102.4\kpc$} & {\large $l_{\rm f}=64\kpc$} \\
 & \large{Polarized intensity} &  \\
\includegraphics[width=5.2cm, trim=10mm 15mm 12mm 15mm, clip]{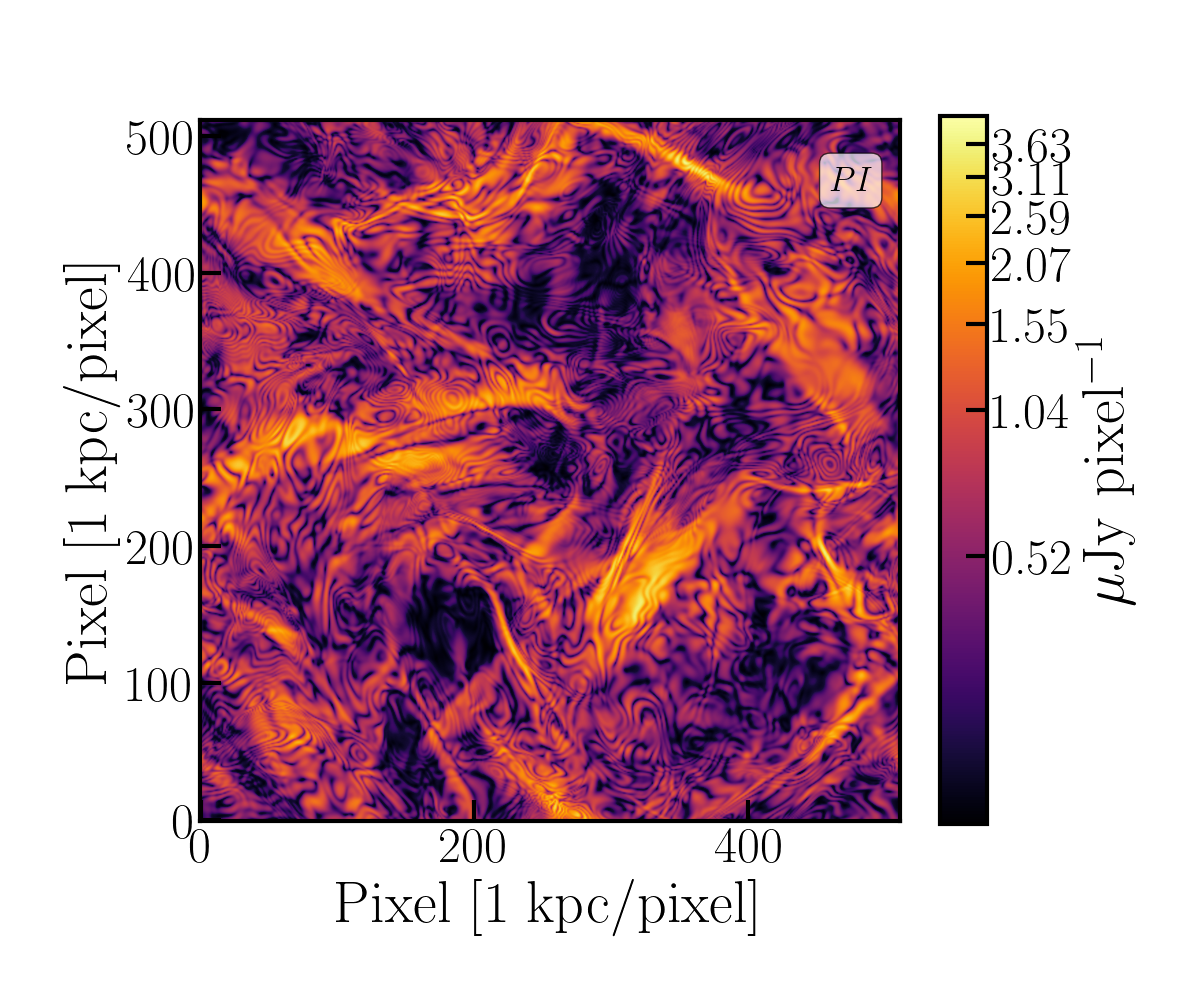} &
\includegraphics[width=5.2cm, trim=10mm 15mm 12mm 15mm, clip]{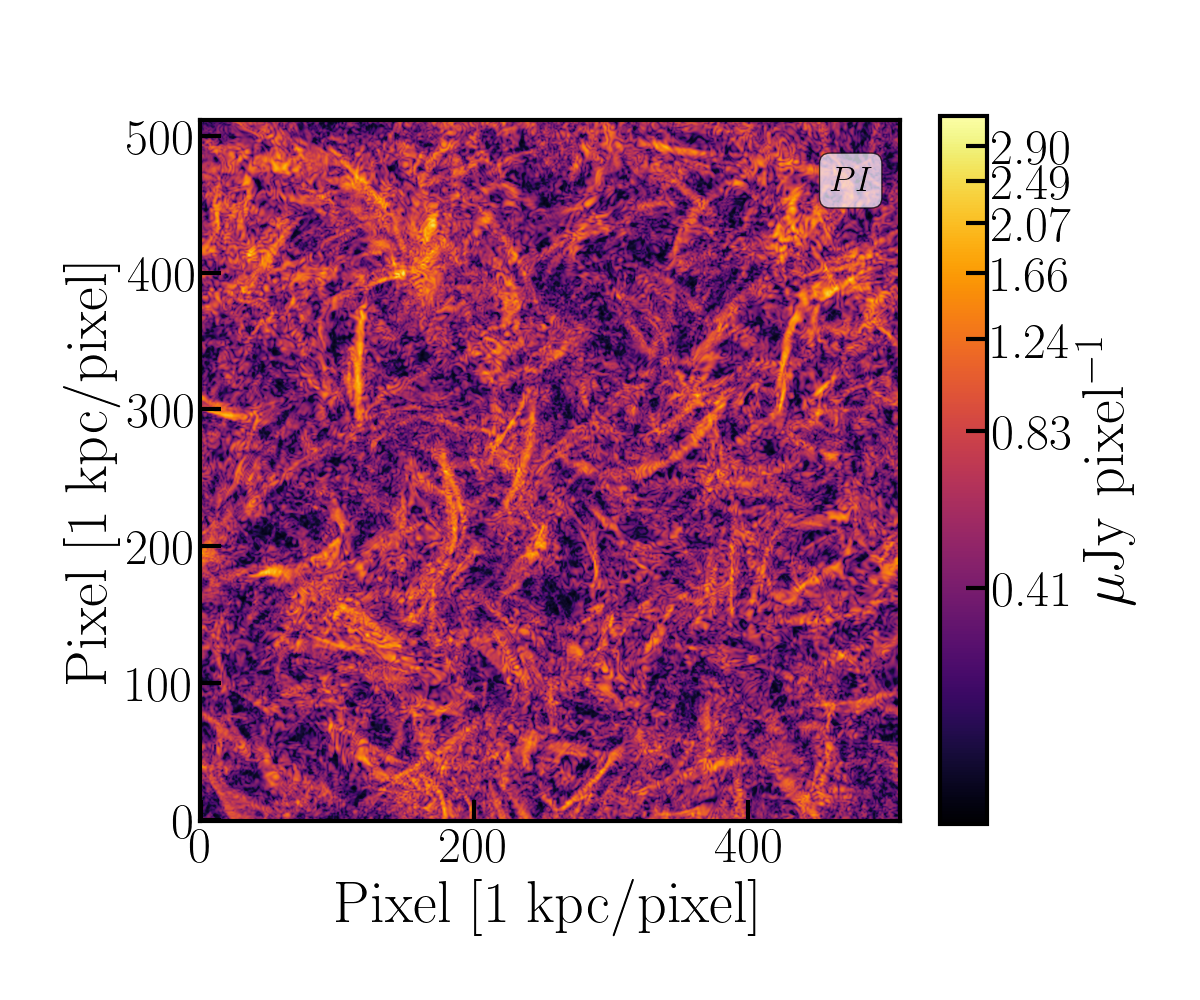} &
\includegraphics[width=5.2cm, trim=10mm 15mm 12mm 15mm, clip]{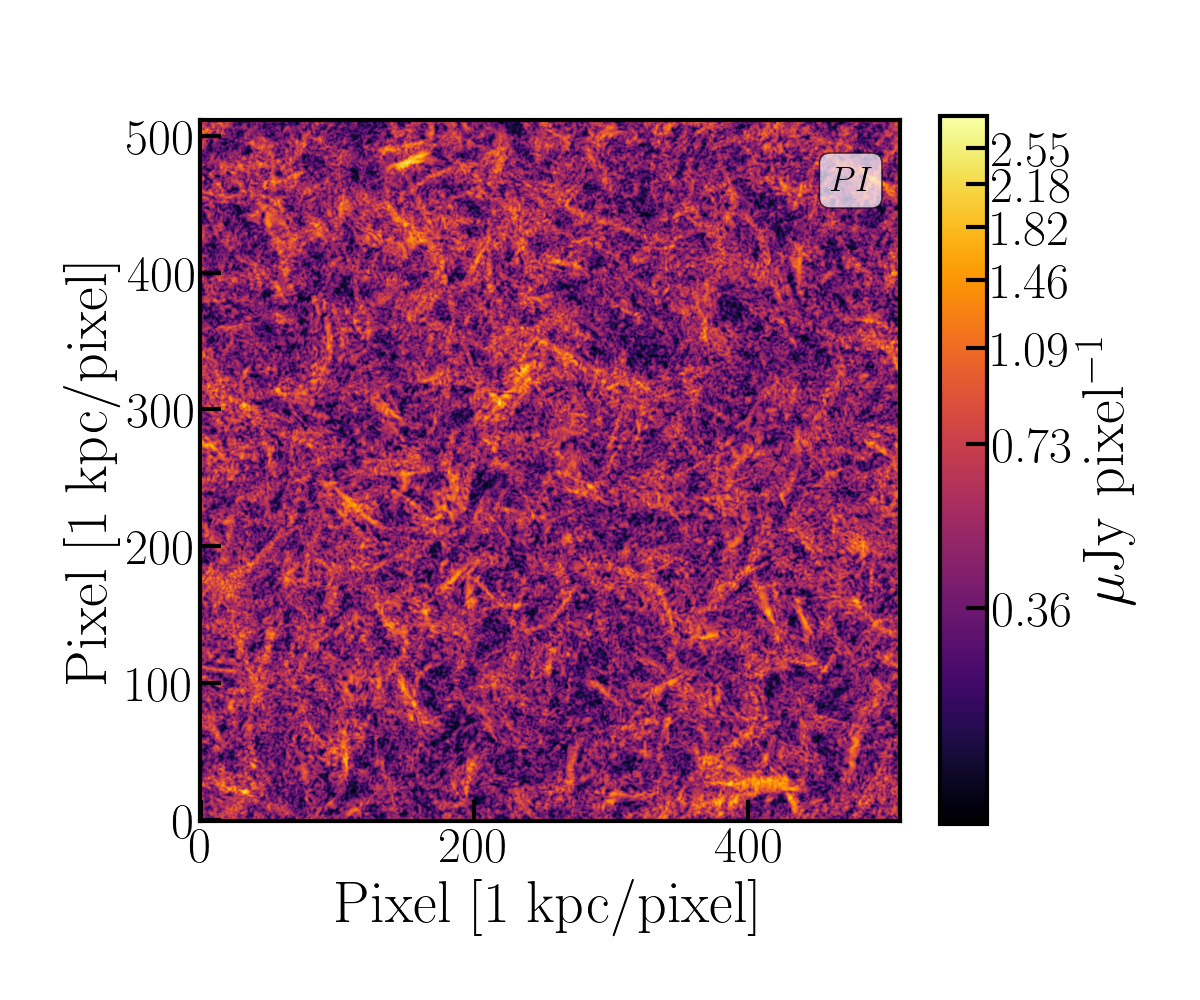} \\
\multicolumn{3}{c}{\large{Fractional polarization (1\,kpc)}}\\ 
\includegraphics[width=5.2cm, trim=10mm 15mm 12mm 15mm, clip]{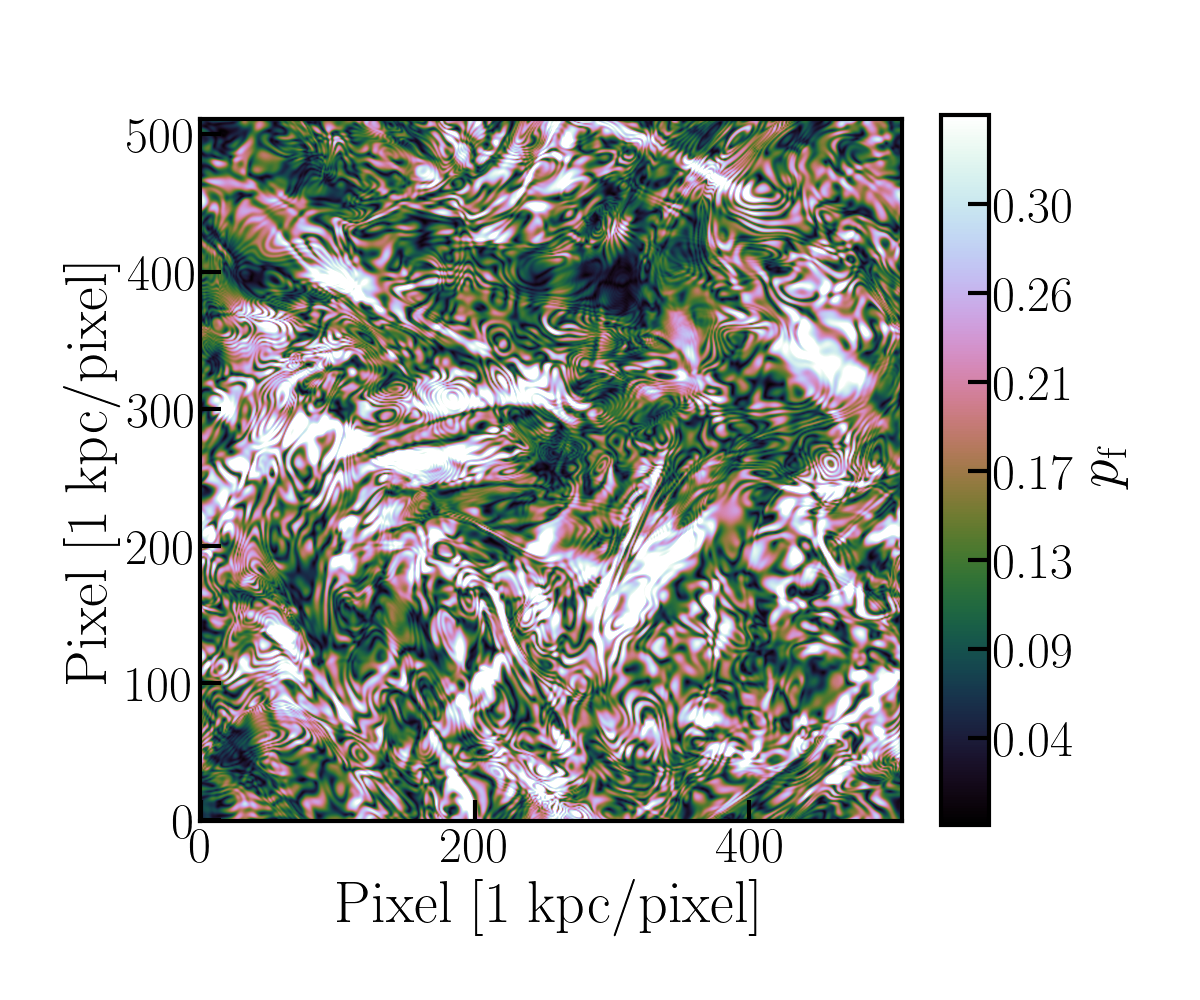} &
\includegraphics[width=5.2cm, trim=10mm 15mm 12mm 15mm, clip]{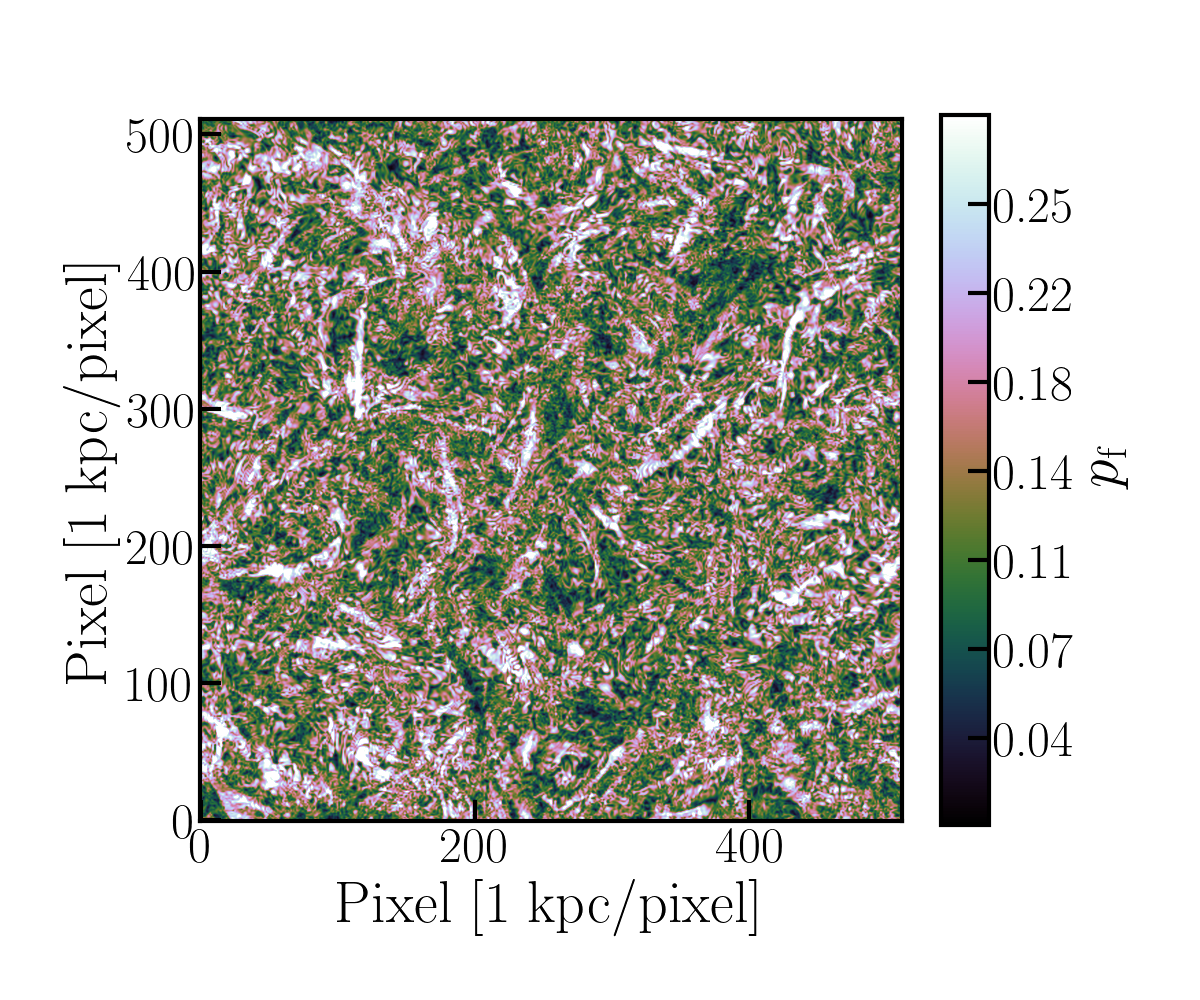} &
\includegraphics[width=5.2cm, trim=10mm 15mm 12mm 15mm, clip]{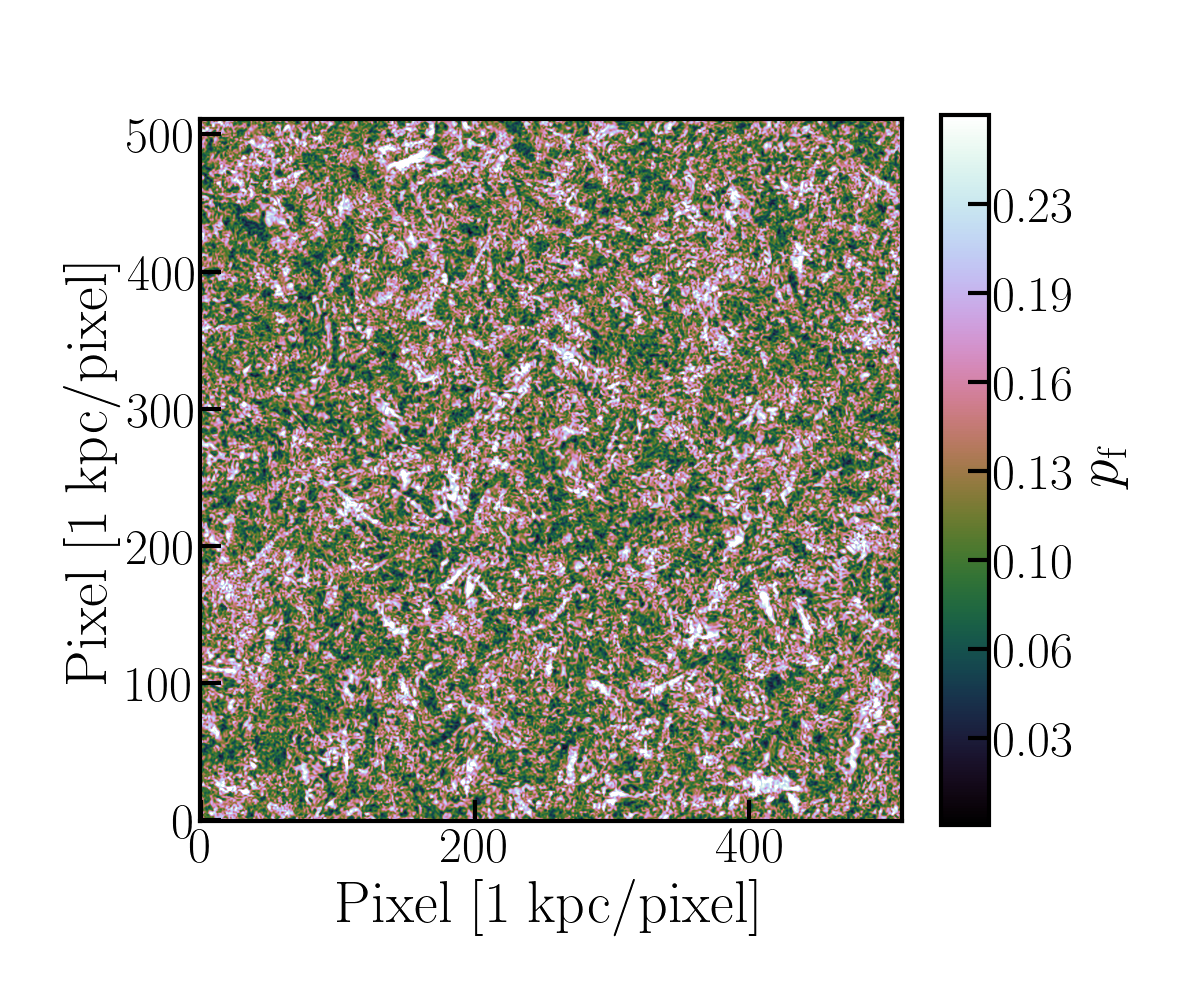} \\
 \multicolumn{3}{c}{\large{Fractional polarization (smoothed 10\,kpc)}}\\ 
\includegraphics[width=5.2cm, trim=10mm 15mm 12mm 15mm, clip]{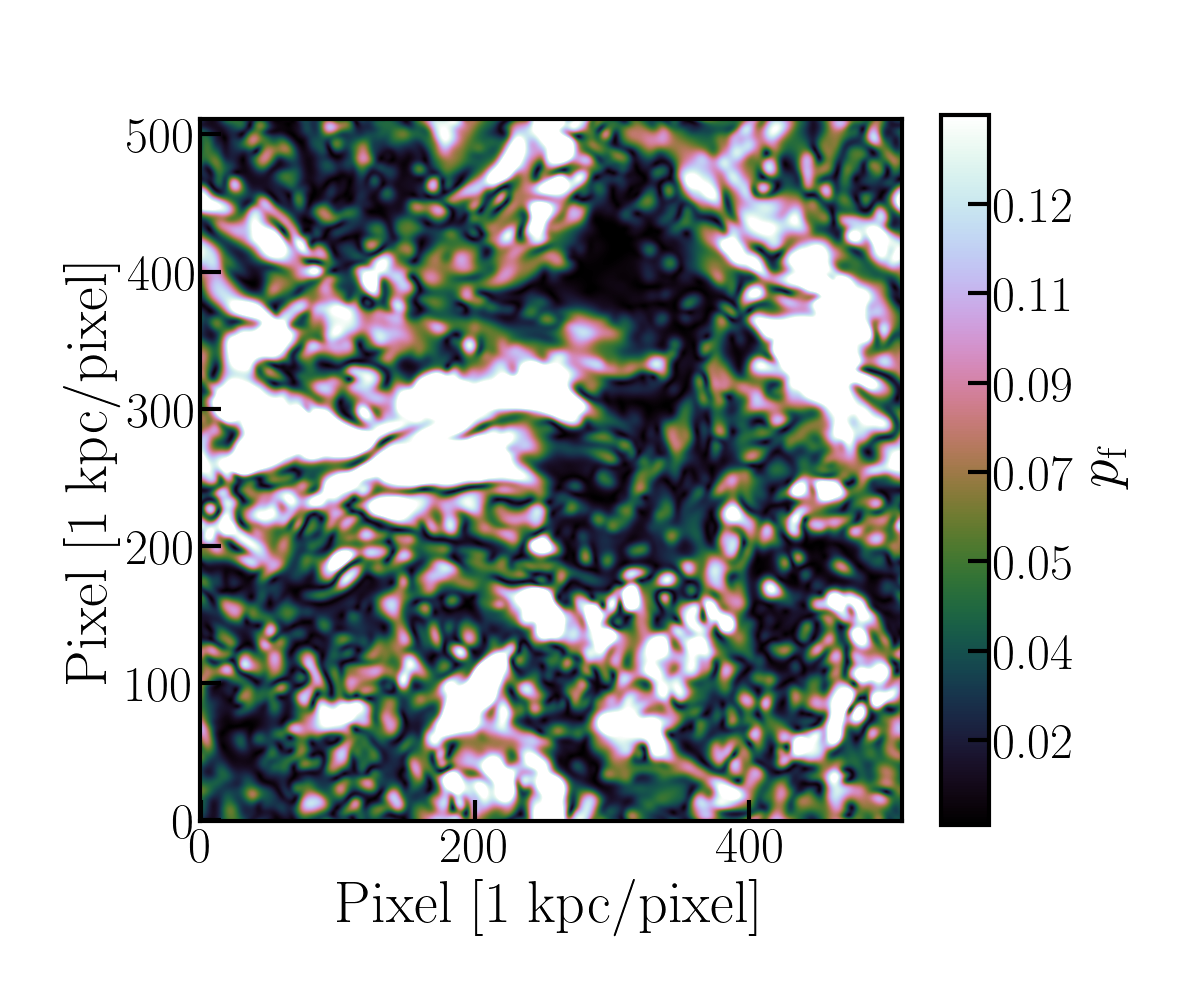} &
\includegraphics[width=5.2cm, trim=10mm 15mm 12mm 15mm, clip]{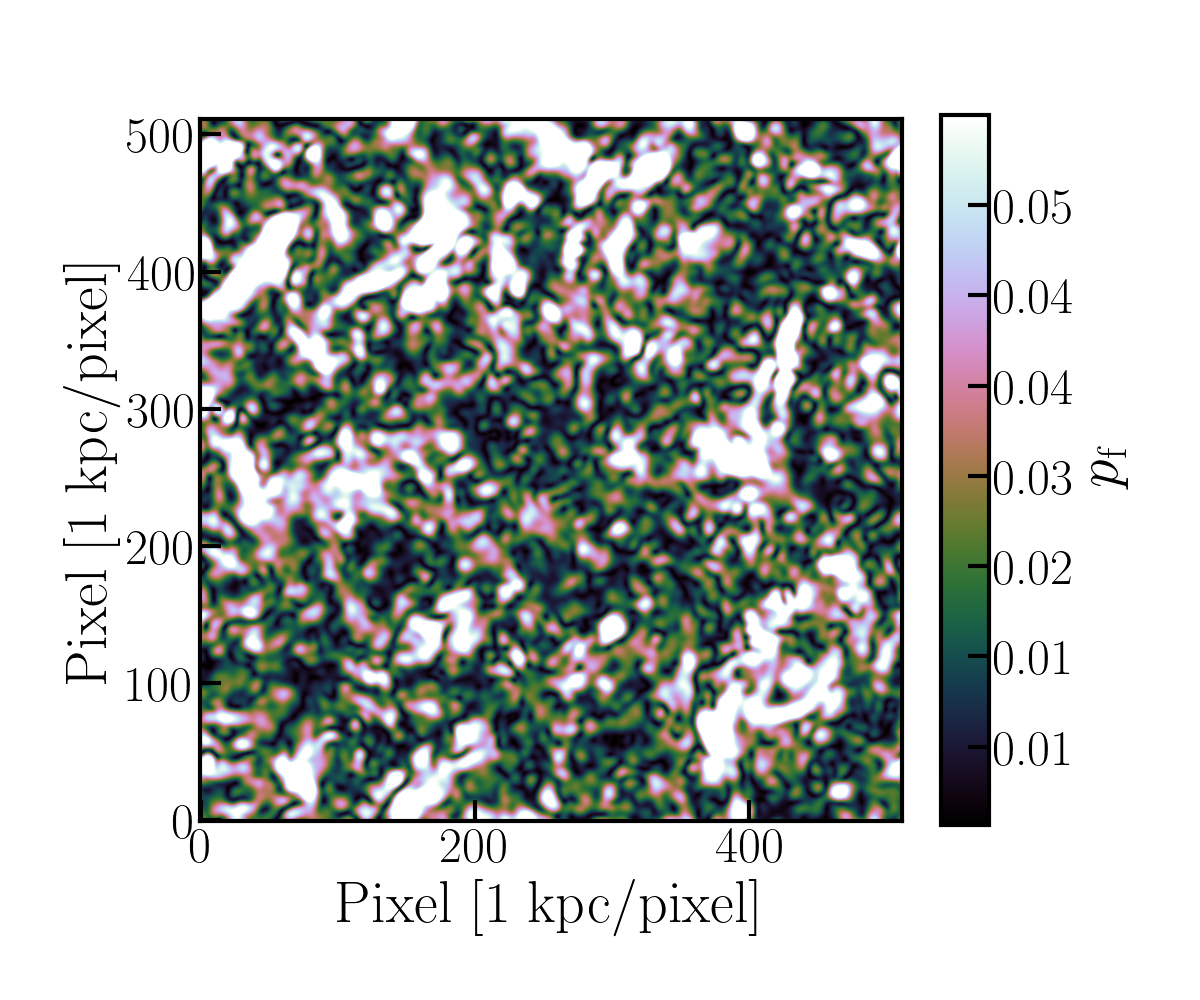} &
\includegraphics[width=5.2cm, trim=10mm 15mm 12mm 15mm, clip]{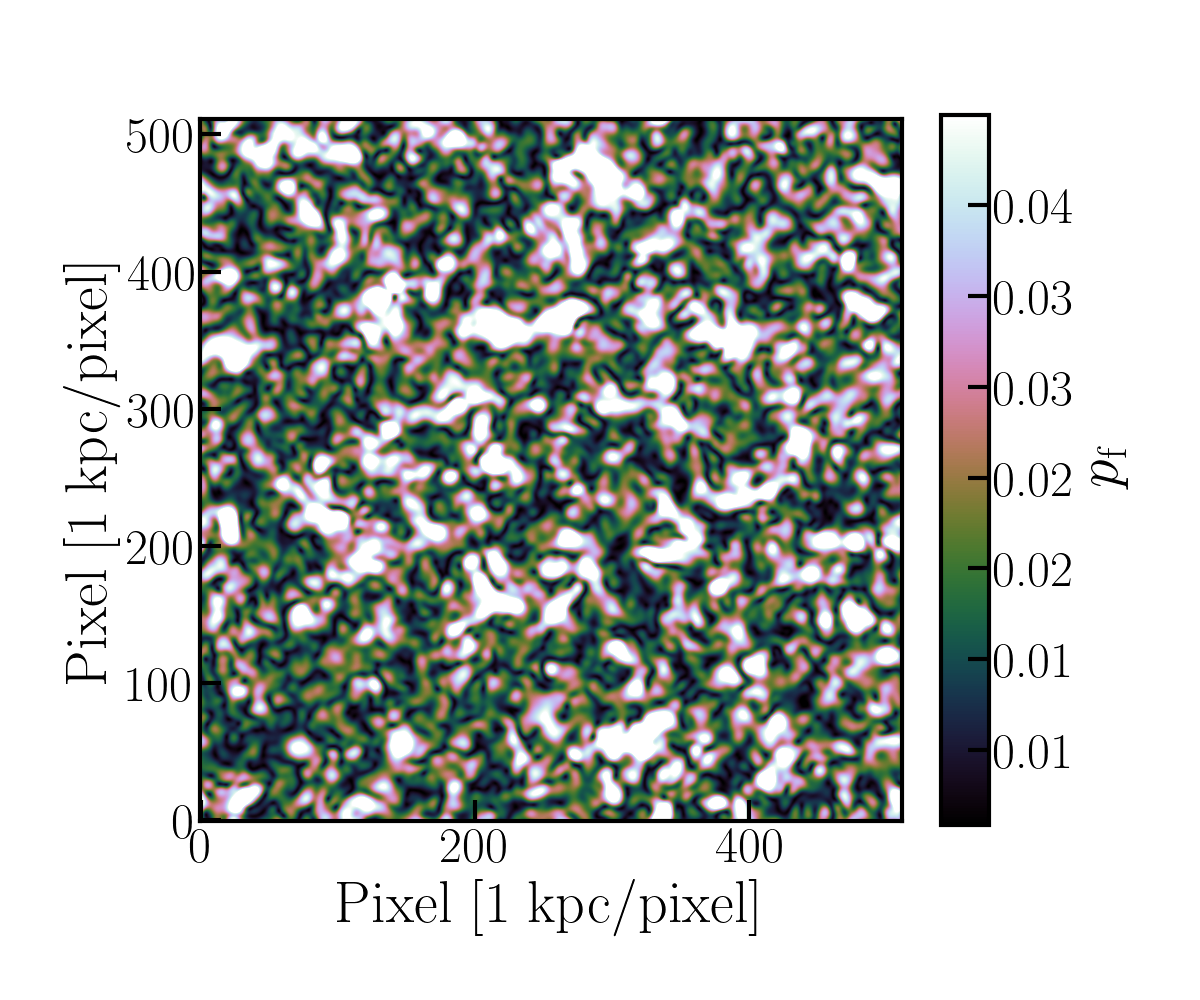} \\
\end{tabular}
\caption{Same as Fig.~\ref{fig:polI_6GHz}, but at 1\,GHz.}
\label{fig:polI_1GHz}
\end{figure}

\subsection{Severe Faraday depolarization in the ICM} \label{sec:depol}

Similar to $I_{\rm sync}$, the polarized emission ($PI$) in ICM also exhibit filamentary morphology, as shown in Figs.~\ref{fig:polI_6GHz} and \ref{fig:polI_1GHz}, and the intrinsic fractional polarization ($p_{\rm f,int}$)\footnote{The intrinsic fractional polarization, $p_{\rm f,int} = I_{\rm sync}/PI$ is the $\pf$ measured in the absence of frequency-dependent Faraday rotation and depolarization, equivalent of observing at wavelength $\lambda \to 0$.} is expected to be substantially high, between $\sim0.2$ to $0.35$ for different $\lf$s, at the native 1-kpc resolution of the simulations \citep[see][]{sur21, bs21}. However, at decimeter wavelengths, frequency-dependent Faraday depolarization, typically characterized by the dispersion of RM ($\sigmarm$), reduces the observed $p_{\rm f,\nu} = I_{\rm sync}(\nu)/PI(\nu)$ by more than a factor of 2. These simulations yield $\sigmarm \approx 100\,\radm$, which is consistent with $\sigma_{\rm RM}$ obtained through `RM-grid'-- statistical measurement of the dispersion of RM of polarized sources lying within or behind galaxy clusters \citep{bonaf10, bohringer16, osinga2025}. 

Note that, under Gaussian random approximation, the depolarization ($DP$), a factor by which $p_{\rm f,int}$ is reduced at a given wavelength ($\lambda$), is given by, $DP \approx \exp{[-2\,\sigmarm^2\,\lambda^4]}$ \citep{Sokoloff1998}. This would lead to negligible polarized emission at $\lambda \gtrsim 10$\,cm ($\nu \lesssim 3\ghz$). In contrast, because fluctuation dynamo arranges magnetic fields in twists and folds, significant level of polarized emission with $\pf$ up to $\sim0.2$ is expected, even at 1\,GHz, as seen in Fig.~\ref{fig:polI_1GHz}. However, on a cursory look at the top panels of Figs.~\ref{fig:polI_6GHz} and \ref{fig:polI_1GHz}, the morphology of the structures are significantly modified so that, at 1\,GHz, $PI$ is fragmented into smaller/finer-scale structures compared to those at 6\,GHz. We emphasize that, at $\nu \gtrsim 3\ghz$, the morphology of $PI$ in the ICM is little affected by Faraday depolarization, and the maps shown at 6\,GHz (top and middle rows in Fig.~\ref{fig:polI_6GHz}) closely resemble the intrinsic polarized emission. Therefore, robust measurement of polarization in the ICM above $\sim 3\ghz$, i.e., higher than the frequencies at which galaxy clusters are traditionally observed, is necessary to properly study ICM magnetic fields \cite[see][]{bs21, dutta2024}.

\subsection{Need for high-resolution, high-frequency observations of galaxy clusters} \label{sec:obsreq}

As highlighted above, in order to glean insights into the properties of magnetic fields and underlying synchrotron emission in radio halos, it is imperative to have high angular resolution observations that could probe ICM on a few kiloparsec physical scales. This will allow us to resolve the filamentary structures, if present, and characterize their structural properties. For this, angular resolutions in the range $0.3\arcsec$ to $1\arcsec$ will provide the requisite physical resolution to observe clusters in the redshift range 0.02 to 0.6 \citep{Fer+12, cassano2015, weere19}. 

The surface brightness contrast of filamentary structures emanating due to the action of fluctuation dynamo depends on the phase of the dynamo amplification, i.e.,
kinematic, intermediate, and saturated phase. They are directly related to the merger history of a galaxy cluster. Hence, to establish the efficacy of the fluctuation dynamo in shaping up the nonthermal phase in the ICM, detailed observations of a large sample of galaxy clusters, covering a wide range of merger state, are necessary.

Furthermore, in addition to finer scale structures introduced by Faraday depolarization, polarized emission is further affected heavily by additional beam depolarization. In the bottom panels of Figs.~\ref{fig:polI_6GHz} and \ref{fig:polI_1GHz}, we show the maps of $\pf$ when smoothed by a 10\,kpc kernel. Clearly, near 1\,GHz, the typical $\pf$ is drastically reduced from $\sim0.15\textrm{--}0.3$ at the native 1-kpc resolution to well below $0.15$ after smoothing. In \citet{bs21}, it was shown that the mean $\pf$ as a function of smoothing length ($l$) is given by, $\langle \pf(l)\rangle = p_{\rm f, int}\,(1 + l/l_{1/2})^{-1}$, where $l_{1/2}$ is the smoothing scale at which $p_{\rm f,int}$ reduces by half. Below $\sim2$\,GHz, $l_{1/2} \approx 5$\,kpc ($l/2\,r_{\rm c} \approx 1/600 \textrm{--} 1/300$) is sufficient to severely depolarize emission in the ICM, especially for smaller $\lf$. Interestingly, at high frequencies ($\gtrsim3\ghz$), polarized emission is only mildly affected by beam smoothing, where $l_{1/2} \gtrsim 30$\,kpc ($l/2\,r_{\rm c} \gtrsim 1/200$).

In light of the discussions above, a scenario emerges, wherein high-resolution observations of galaxy clusters at frequencies $\gtrsim3\ghz$ are necessary for a deep understanding of the properties of magneto-ionic medium in the ICM. The SKA-Mid in the array assembly AA4 and AA* will play a game-changing role in unraveling the nature of turbulence and its driver in the ICM, and overcome the demanding observational challenges faced by present day telescopes to meet these requirements.

\section{Role of the SKAO}

As discussed above, fluctuation dynamos give rise to ubiquitous filamentary synchrotron emission that spans a wide range of scales determined by the turbulent driving scale. However, given the significantly larger propagation timescale ($\gg 10^8$\,yr) of CREs in the ICM compared to their typical radiative timescale, $\mathcal{O}(\rm 10^8\,yr)$, the total intensity radio continuum spectrum is expected to be significantly steeper than those assumed in Sec.~\ref{sec:simulation}. While at low frequencies ($\lesssim150\mhz$) the surface brightness is high due to the steep spectrum, achieving high angular resolution ($\sim1\arcsec$) is limited, especially in the Southern-sky.\footnote{The best possible angular resolution of $\approx5\arcsec\textrm{--}7\arcsec$ that SKA-Low will provide is insufficient for resolving the filamentary structures and the sensitivity will be limited by confusion noise.} At higher frequencies ($\gtrsim500\mhz$), although SKA-Mid in Band\,2 will provide sub-arcsec angular resolutions to discern fine-scale structures, a long integration time ($\gtrsim10$\,h) per galaxy cluster will be required to achieve adequate surface brightness sensitivity. However, this issue is somewhat alleviated when the ICM magnetic fields are illuminated by additional influx of CREs that are seeded from various astrophysical sources present in the galaxy cluster environment, e.g., remnant/dying AGN lobes.

\begin{figure*}
\centering
\begin{tabular}{ccc}
\includegraphics[width=5.0cm]{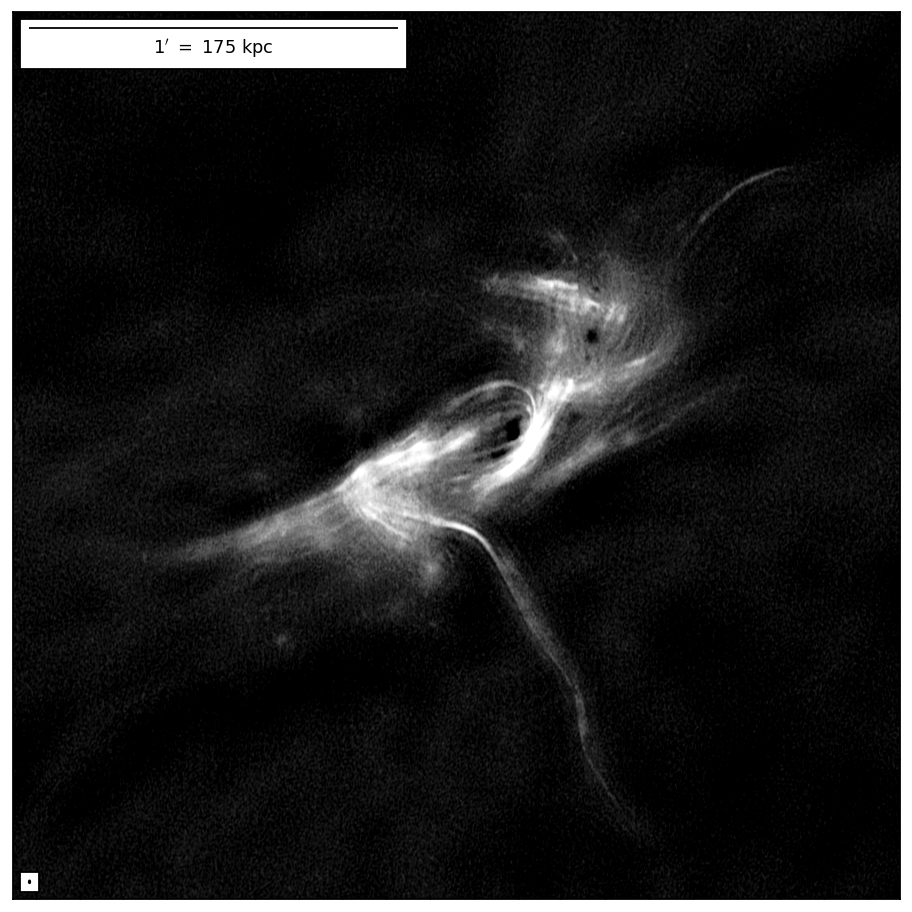} &
\includegraphics[width=5.0cm]{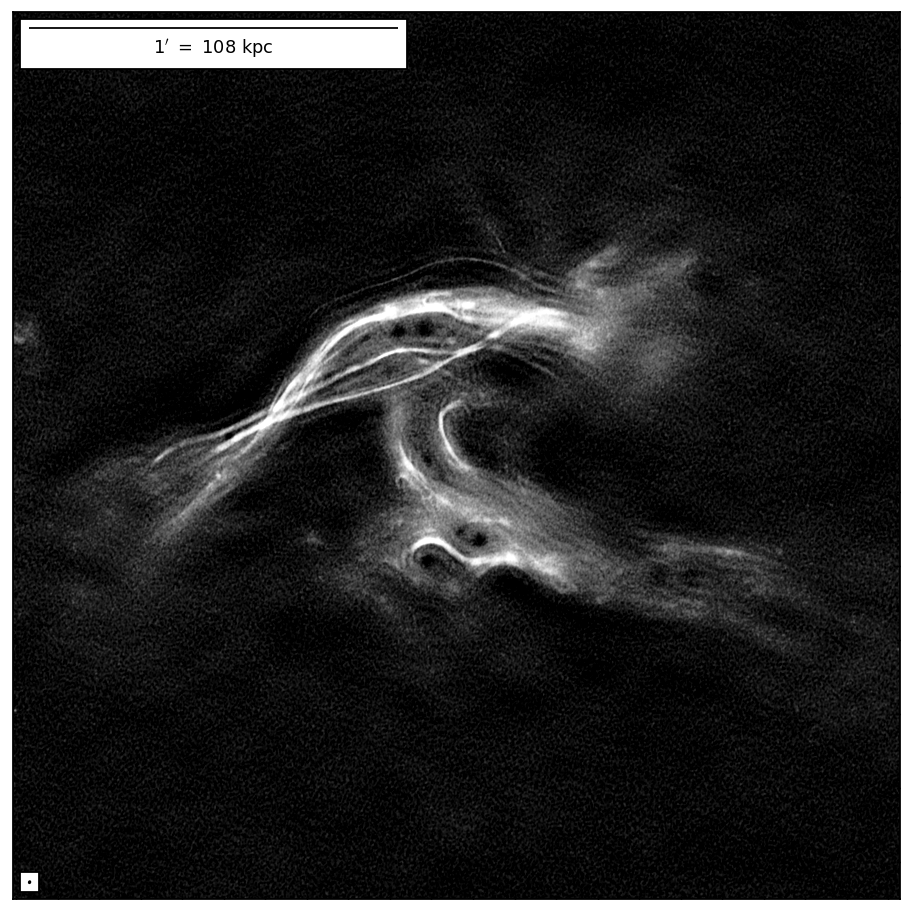} &
\includegraphics[width=4.9cm, height=4.9cm]{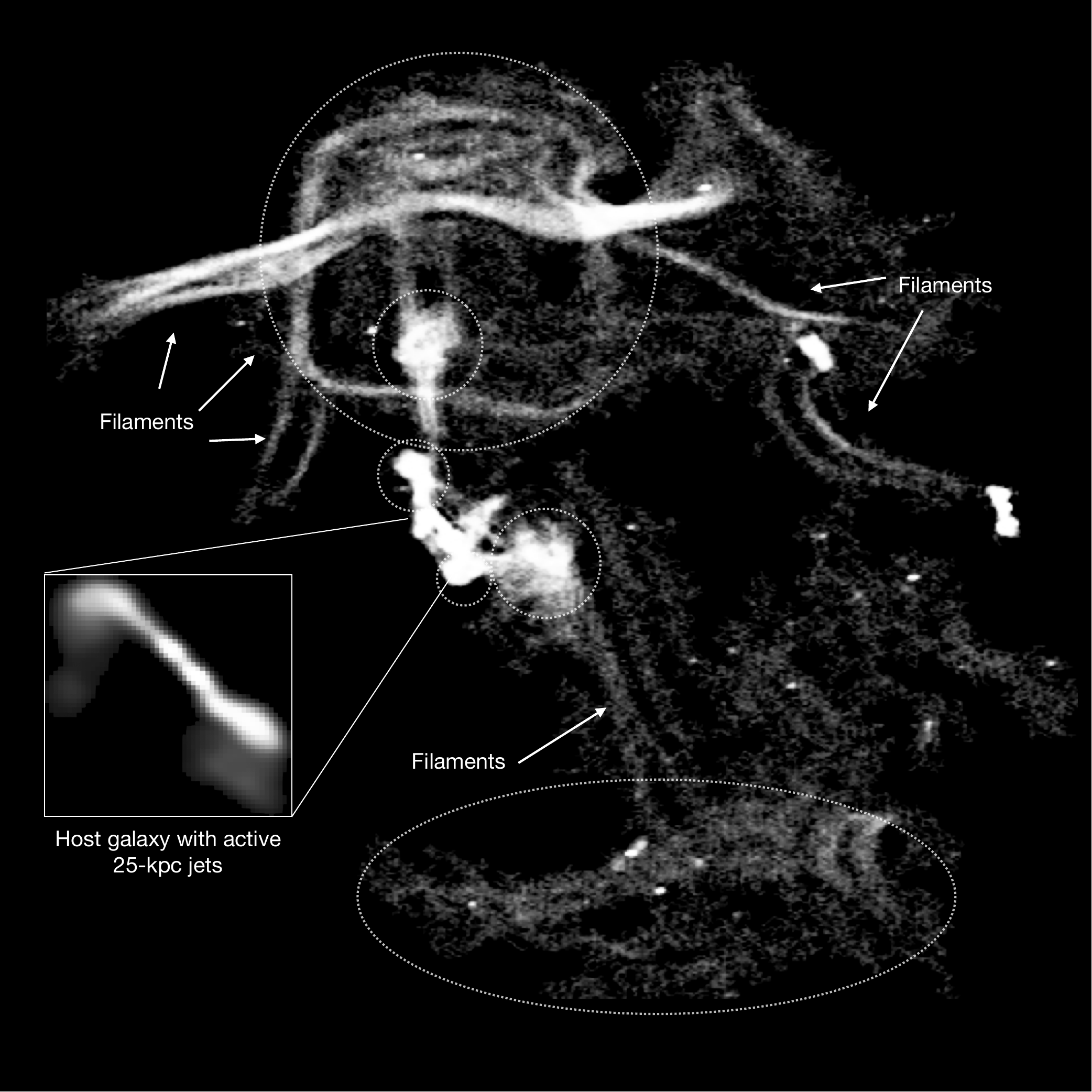} \\
\end{tabular}
\caption{Filamentary emission on $\sim100$\,kpc scales likely originated from ICM magnetic fields illuminated by CREs from AGN. \textbf{\textit{Left}} and \textbf{\textit{middle panels}} show LOFAR 144\,MHz maps of radio phoenixes in the galaxy cluster Abell\,1914 and Abell\,566. Sensitive ($\sim 40\,\upmu{\rm Jy\,beam^{-1}}$), sub-arcsec ($\sim 0.3\arcsec$) observations reveals highly filamentary morphology in radio phoenixes (Biava et al., in preparation). \textbf{\textit{Right-hand panel}:} Radio continuum morphology of Nest200047 in a galaxy group showing filamentary nonthermal emission on a wide range of scales \citep[][]{brienza2025}.}
\label{fig:obsfilaments}
\end{figure*}

\subsection{Radio phoenixes as the tracers of ICM magnetic fields} \label{sec:phoenix}

Along with diffuse, spatially smooth, synchrotron emission in the ICM, sensitive low-frequency ($\sim 150$\,MHz) observations have revealed the presence of a new type of extended emission with irregular morphology characterized by steep spectrum ($\alpha \lesssim -1.5$), known as {\textit{radio phoenixes}} \citep{slee2001, mandal2019, raja2024}. Phoenixes are of special interest because they are thought to trace old plasma from formerly jetted AGN that has been re-energized by adiabatic compression after the passage of a shock wave \citep{ensslin2001}. However, clear evidence linking phoenixes to shocks is missing.

Interestingly, high angular resolution observations that probe physical scales of $\sim1$\,kpc, have revealed intricate filamentary structures associated with phoenixes or similar objects shown in Fig.~\ref{fig:obsfilaments}. These filaments span several hundred kiloparsec in projected length, yet have widths of up to a kiloparsec, similar to the filaments seen in MHD simulations (Sec.~\ref{sec:synthmaps}).

\paragraph{Magnetic fields and CREs in phoenixes:} 
Filamentary emission has been observed in AGN lobes \citep[e.g.,][]{owen2000, ramatsoku2020, mahatma2023, derubeis2025}, thus, it is conceivable that phoenixes originate from remnant radio plasma of AGN lobes.
However, associated AGN are often not observed as seen in Fig.~\ref{fig:obsfilaments}, and the intensity variation along filaments does not show enhanced emission at any particular location that could be indicative of CREs being supplied by AGN at specific regions. In the absence of active injection of turbulent energy, i.e., switched off AGN jets, the associated filaments will unavoidably decay and diffuse out to larger distances via inverse cascade of energy during the decaying phase of turbulence \citep[see][]{SSH06, Reppin2017, sur19} and/or undergo adiabatic expansion. Therefore, an initial observed $\sim1$\,kpc width of the filaments associated with AGN \citep[][]{ramatsoku2020, rudnick2022} would grow to 10\,kpc, or even larger in the presence of adiabatic expansion within the radiative lifetime ($\sim 10^8$\,yr) of phoenixes. This is significantly larger than the observed widths of filaments in phoenixes (Biava et al., in preparation).

Furthermore, the relatively high surface brightness, compared to the ambient radio halo, suggests that the CREs in them are contributed externally, so that the ICM magnetic fields are illuminated by CREs in relic plasma of an old, adiabatically expanding, radio lobe of an, now dead and/or switched-off, AGN. In fact, in Fig.~\ref{fig:obsfilaments}, a similar scenario in which an AGN enriches its immediate environment has been observed in a galaxy group, Nest200047 \citep{brienza2021, brienza2025}. Thus, it is plausible that the filaments associated with phoenixes, like in Abell\,1914 and 566, are maintained by active turbulent energy input in the ICM on larger scales, and therefore their emission is tracing the underlying magnetic fields in the ICM (Biava et al., in preparation). 

Recently, \citet{churazov2026} put forth the idea that such filaments could arise from AGN-inflated bubbles, and survive in high plasma-$\beta$ environment, i.e., weakly magnetized ambient ICM.
Whether radio emission in phoenixes is a direct tracer of underlying ICM magnetic fields, or has other physical origin, is an open question. Therefore, studying their spatially resolved broadband spectrum and polarization properties in detail with SKA is imperative and may hold the key to unraveling magnetic fields in the ICM.

\subsubsection{Observing radio phoenixes with SKA-Low and Mid} \label{sec:obs_phoenix}

\begin{figure}
    \centering
    \begin{tabular}{cc}
    \includegraphics[width=0.49\columnwidth]{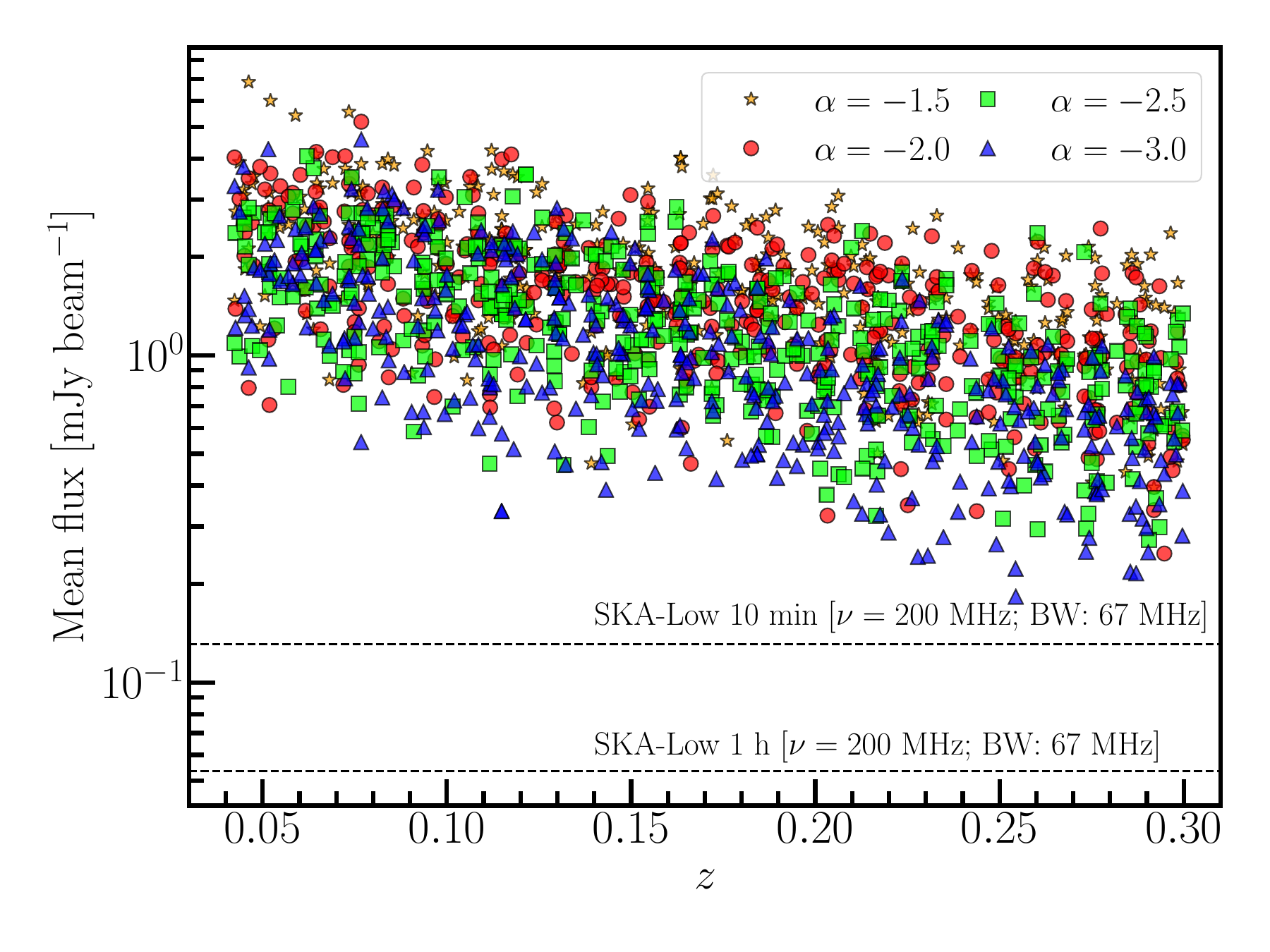} &
    \includegraphics[width=0.49\columnwidth]{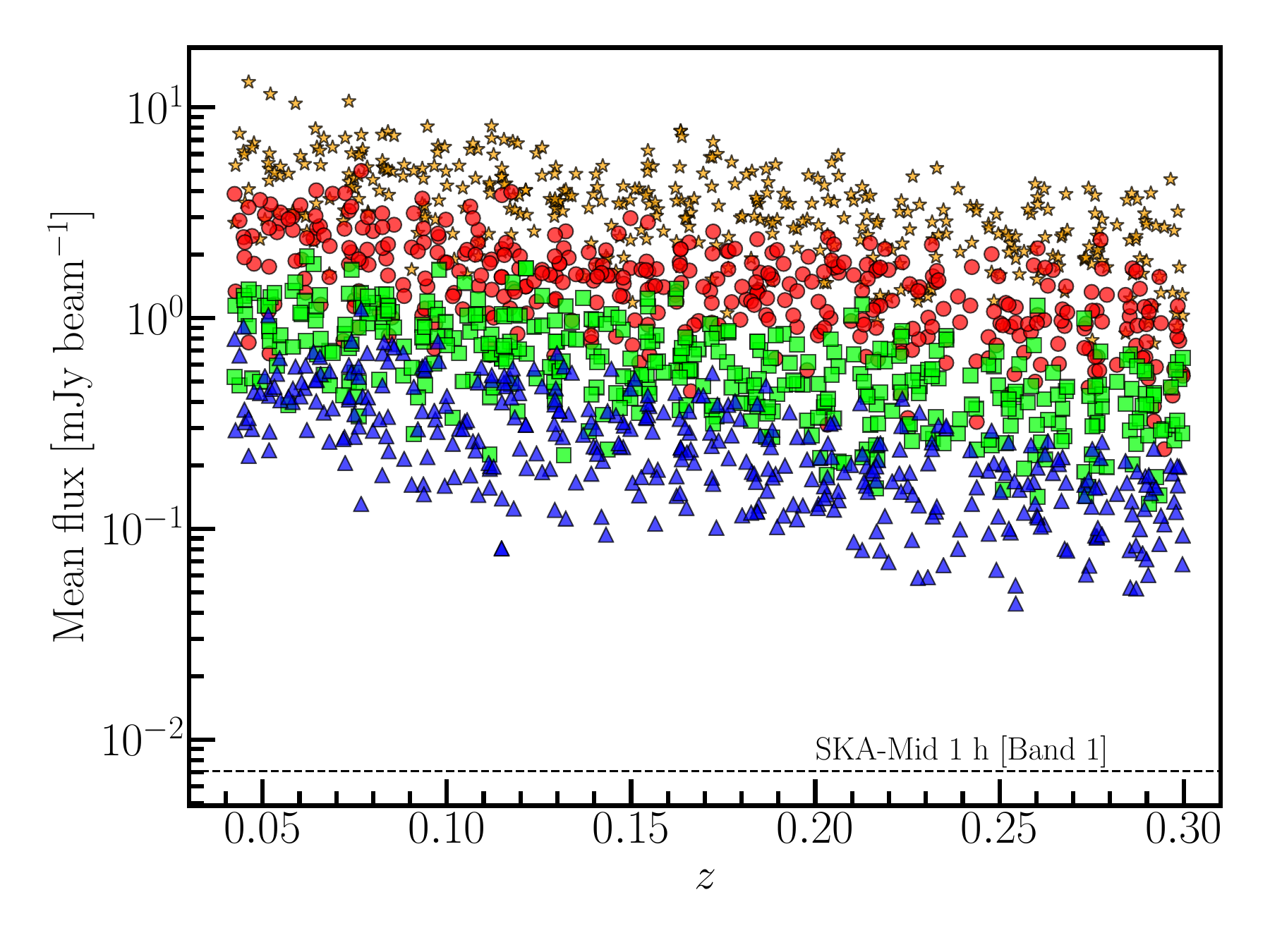}\\
    \includegraphics[width=0.49\columnwidth]{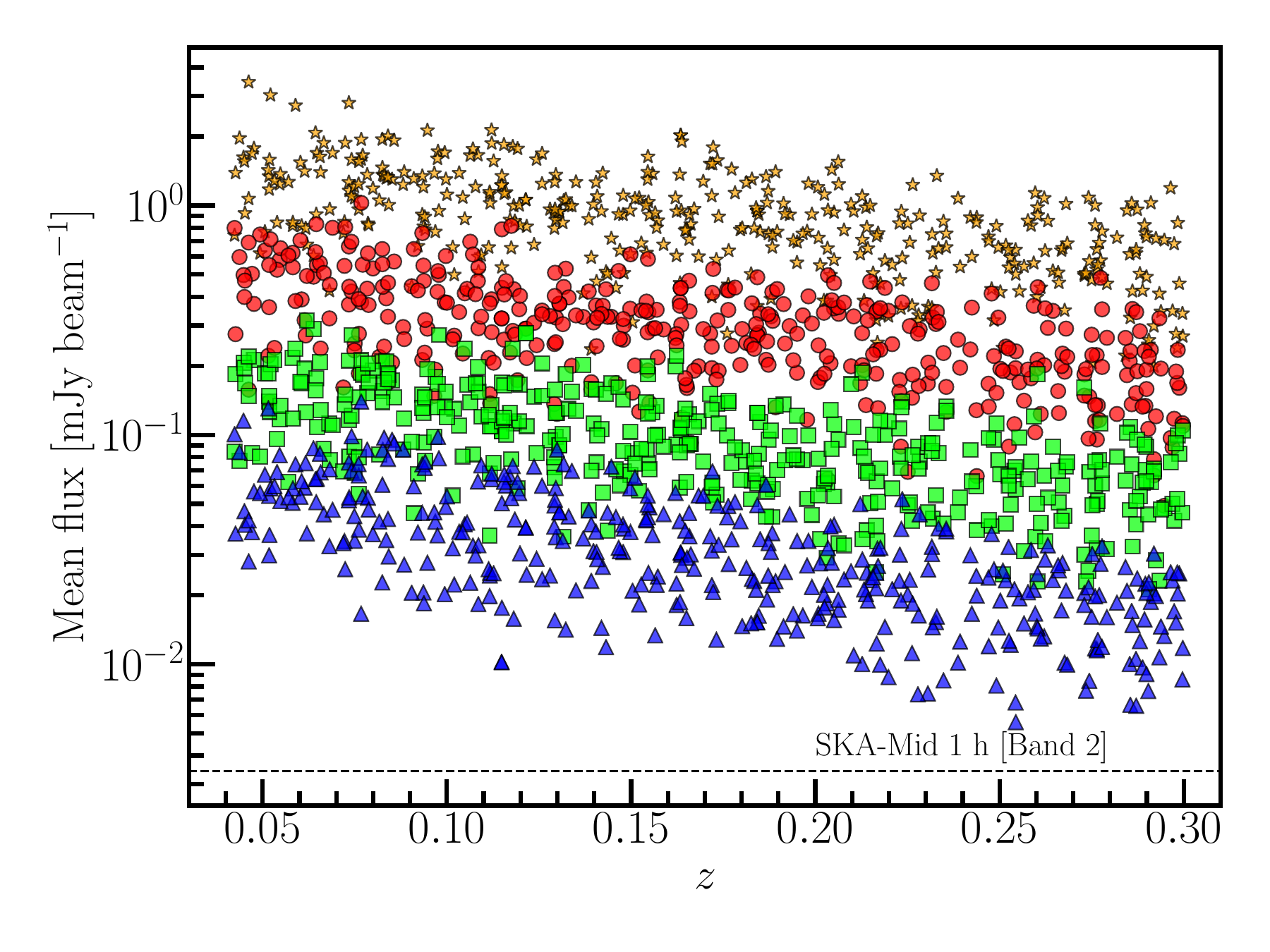} & 
    \includegraphics[width=0.49\columnwidth]{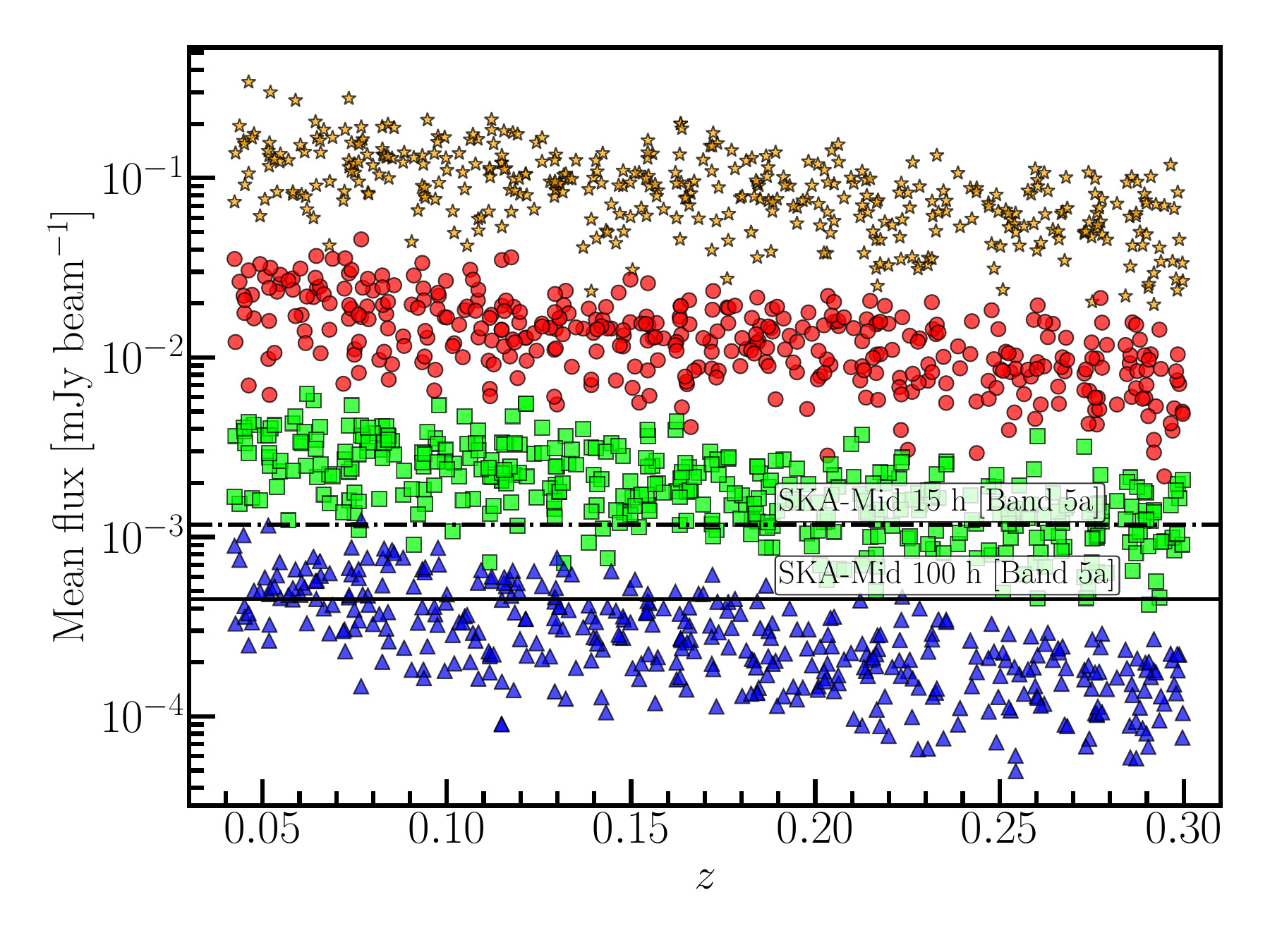}
    \end{tabular}
    \caption{Variation of bootstrap resampled mean flux density of phoenixes with redshift $z$ for different values of spectral index $\alpha$ are shown as different symbols. \textbf{\textit{Top left}} is the expectation for a 67-MHz wide sub-band centered at 0.2\,GHz of the SKA-Low at $\sim4.5\arcsec$ resolution; \textbf{\textit{top right}} is the expected mean flux density averaged over Band\,1 of SKA-Mid at a center frequency of 0.8\,GHz at $\sim 1.2\arcsec$ resolution; \textbf{\textit{bottom left}} is for SKA-Mid's Band\,2 at 1.3\,GHz with $\sim0.85\arcsec$ resolution; and \textbf{\textit{bottom right}} is for SKA-Mid's Band\,5a at 6.5\,GHz with $0.9\arcsec$ resolution. The horizontal lines show the $3\,\sigma$ sensitivity of the SKA's AA4 for the integration times mentioned in the plots.}
    \label{fig:phoenix_sensitivity}
\end{figure}

In order to establish the physical origin of radio phoenixes in galaxy clusters, and understand the properties of the synchrotron emission in them, high angular resolution observations that probe $\lesssim 1$\,kpc scales, to spatially resolve filaments, and cover a wide frequency range, to study the spectral slope and curvature, are necessary. Furthermore, a large sample is needed to establish a statistically meaningful census of the association of properties of phoenixes to the properties of clusters hosting them, e.g., any dependence of the emission/morphological properties of phoenixes on the mass and merger history of the host cluster, level of AGN activity, etc. Unfortunately, only a handful ($<20$) of `candidate' radio phoenixes have been identified in the Northern Hemisphere (based on morphological classification in \citealt{weere19}; Mandal et al., in preparation) from the LOFAR Two-metre Sky Survey (LoTSS) data at 144\,MHz \citep{LoTSSDR2}. Furthermore, their spectral properties are poorly known. 

For the purpose of this chapter, we use different values of $\alpha$ ranging between $-1.5$ and $-3$ to extrapolate the surface brightness to different frequency bands of the SKA. Here, we generate a sample of 400 candidates based on the expected total number of radio halo detectable with the SKA-Low \citep[e.g.,][]{cassano2015, Cassano01.2026.SKA}, and based on the morphological classification reported in \citet{weere19}, we consider that about 15\% of them are likely to be phoenixes. Because there are only 18 radio phoenix candidates known, we generated a bootstrap sample\footnote{Unfortunately, because of the statistically few number of phoenixes, their luminosity function and relation to host clusters are unknown, and thus the bootstrap resample of candidate phoenixes from LoTSS could be biased.} of 400 phoenixes using the distribution of radio power ($P$) at 144\,MHz in the rest frame. In order to obtain the mean surface brightness, we also generated a bootstrap sample for the projected area and assumed a uniform redshift distribution in the range 0.04--0.3, the same range as for the LoTSS sample. Finally, we scaled the mean surface brightness to different frequency bands of SKA-Low and Mid by appropriately considering the different resolutions. Note that the surface brightness scaled using LoTSS data is independent of the model, and therefore the expected detectability with the SKA is also independent of the process through which filamentary emission in phoenixes arises, provided that the small sample from LoTSS is a representative sample.

In Fig.~\ref{fig:phoenix_sensitivity}, we show the variation of the mean flux density (surface brightness) of the resampled distribution for the different bands covered by the SKA-Low and SKA-Mid in array assembly AA4. It is apparent that SKA-Low would detect all phoenixes in just 10\,min using four sub-bands. However, the relatively poor angular resolution in the range $2.5\arcsec$ to $9\arcsec$ across the band will somewhat limit the study of spatially resolved morphological properties highlighted in Sec.~\ref{sec:synthmaps} and \ref{sec:phoenix}, especially for distant phoenixes. Nevertheless, surveys with SKA-Low would be excellent for identifying phoenixes as mostly diffuse, irregularly shaped emission in galaxy clusters, and its large fractional bandwidth, covered with four sub-bands, would provide the requisite information on their in-band spectral properties. With the relatively higher resolution of a Band\,1 survey, their morphology could be further established. Then they could be followed up through deep, targeted observations in Bands\,2 and 5a to study their polarization properties.

\subsection{Detectability of polarized emission from the ICM} \label{sec:obspol}

\begin{figure}
\centering
\begin{tabular}{ccc}
{\large $l_{\rm f}=256\kpc$} & {\large $l_{\rm f}=102.4\kpc$} & {\large $l_{\rm f}=64\kpc$} \\
\multicolumn{3}{c}{\large{Total intensity ($0.9\arcsec$ resolution with AA4 in Band\,2)}}\\ 
\includegraphics[width=5.2cm, trim=10mm 15mm 12mm 15mm, clip]{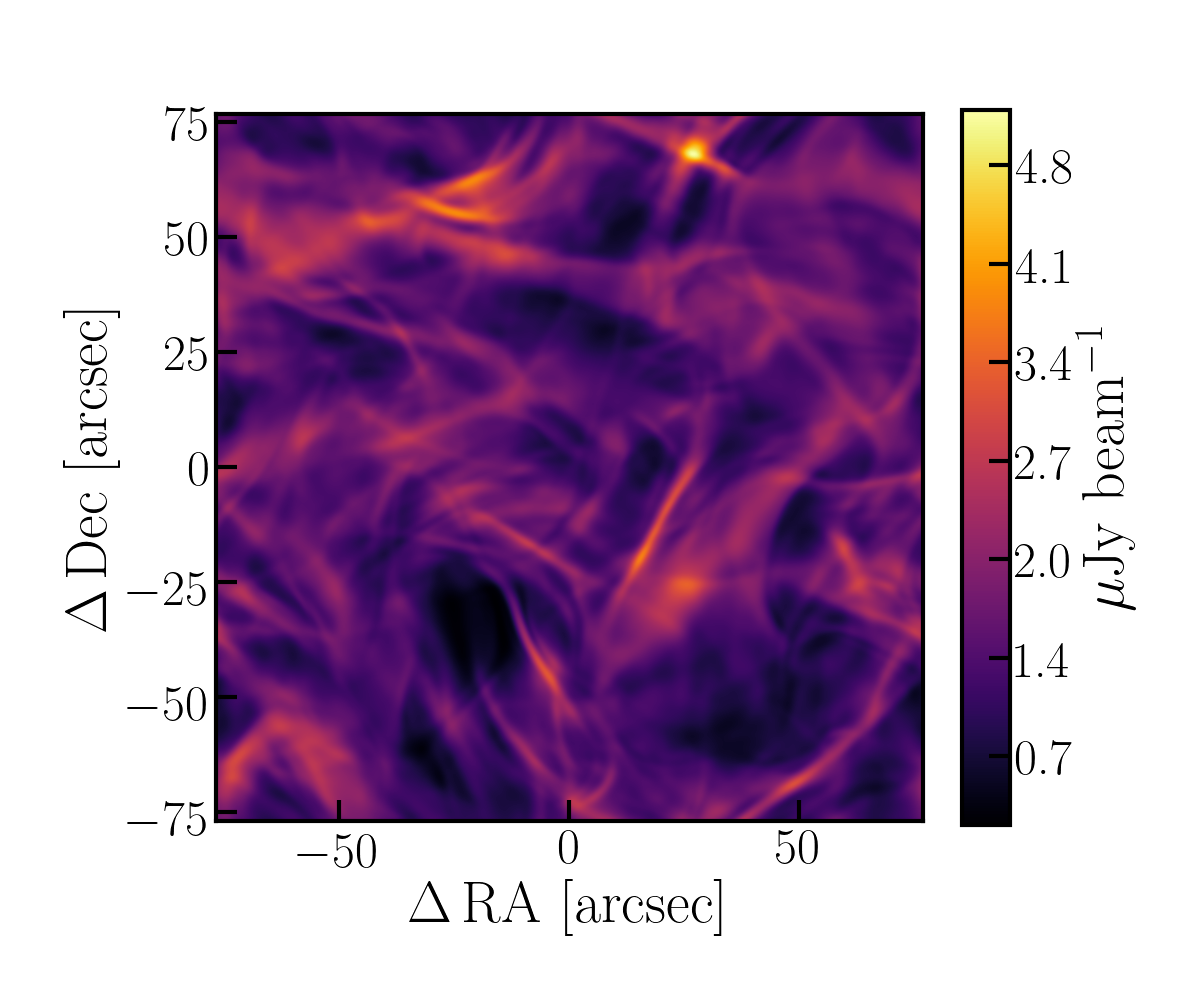} &
\includegraphics[width=5.2cm, trim=10mm 15mm 12mm 15mm, clip]{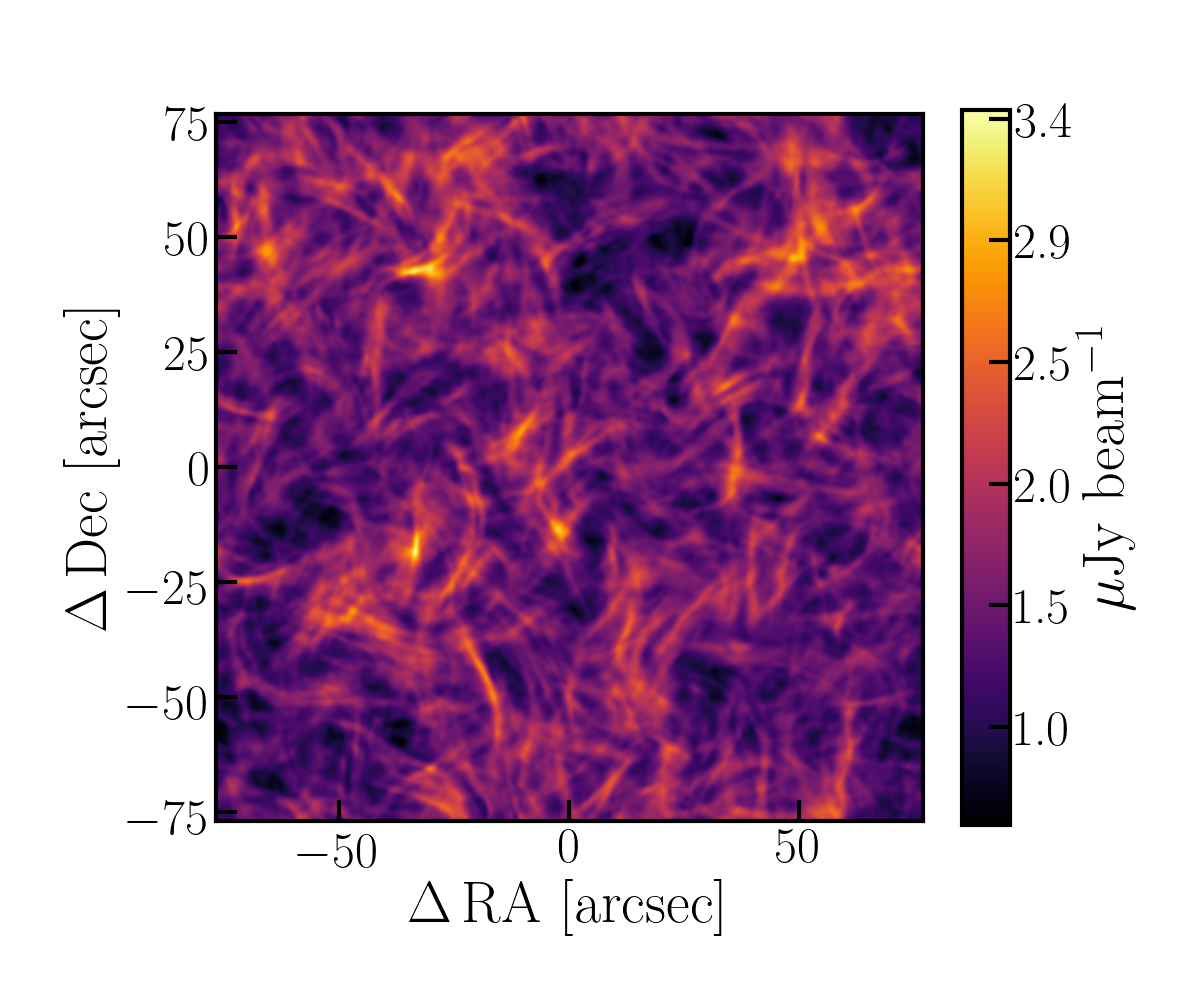} &
\includegraphics[width=5.2cm, trim=10mm 15mm 12mm 15mm, clip]{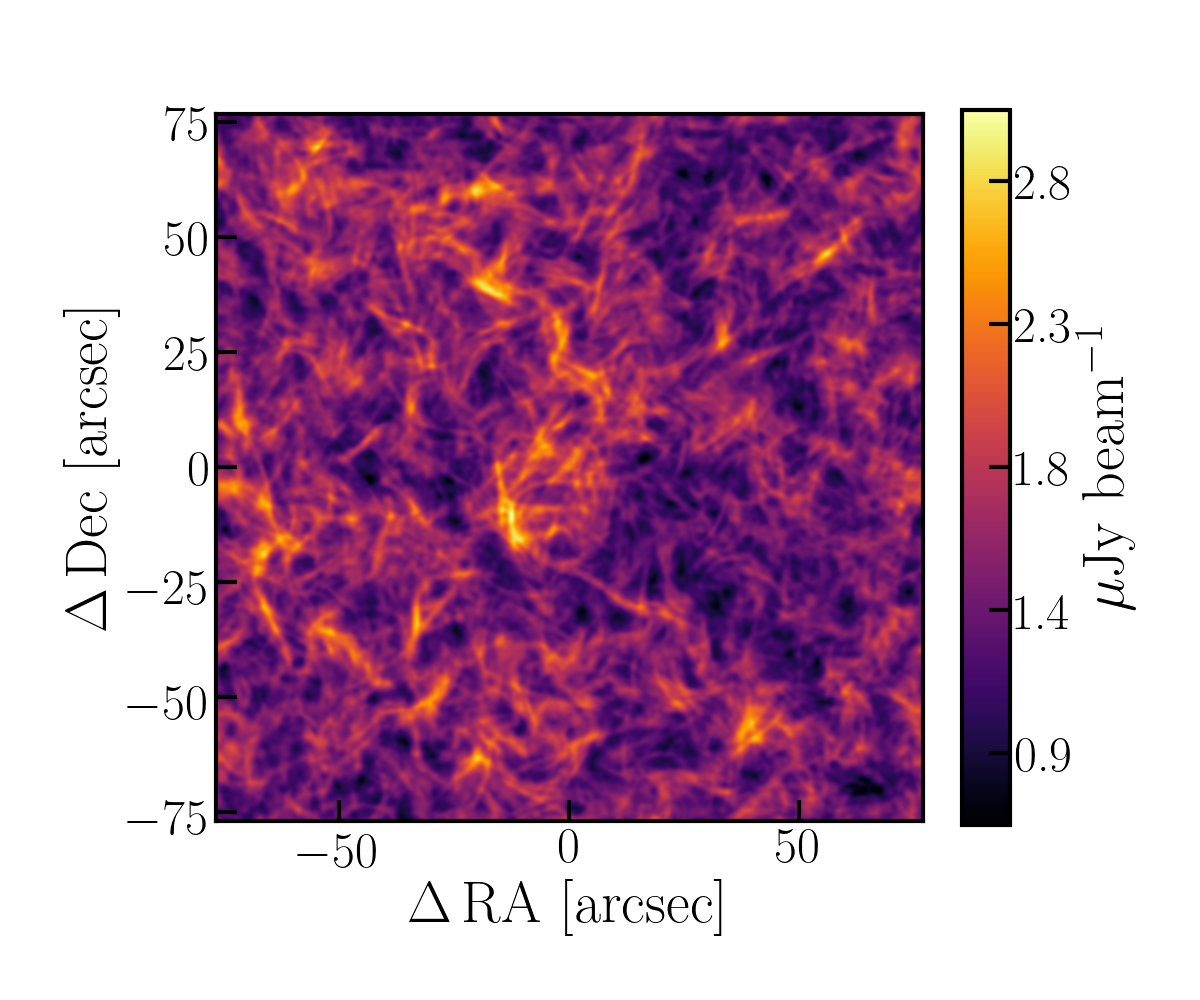} \\
\multicolumn{3}{c}{\large{Total intensity + noise ($0.9\arcsec$ resolution with AA4 in Band\,2)}}\\ 
\includegraphics[width=5.2cm, trim=10mm 15mm 12mm 15mm, clip]{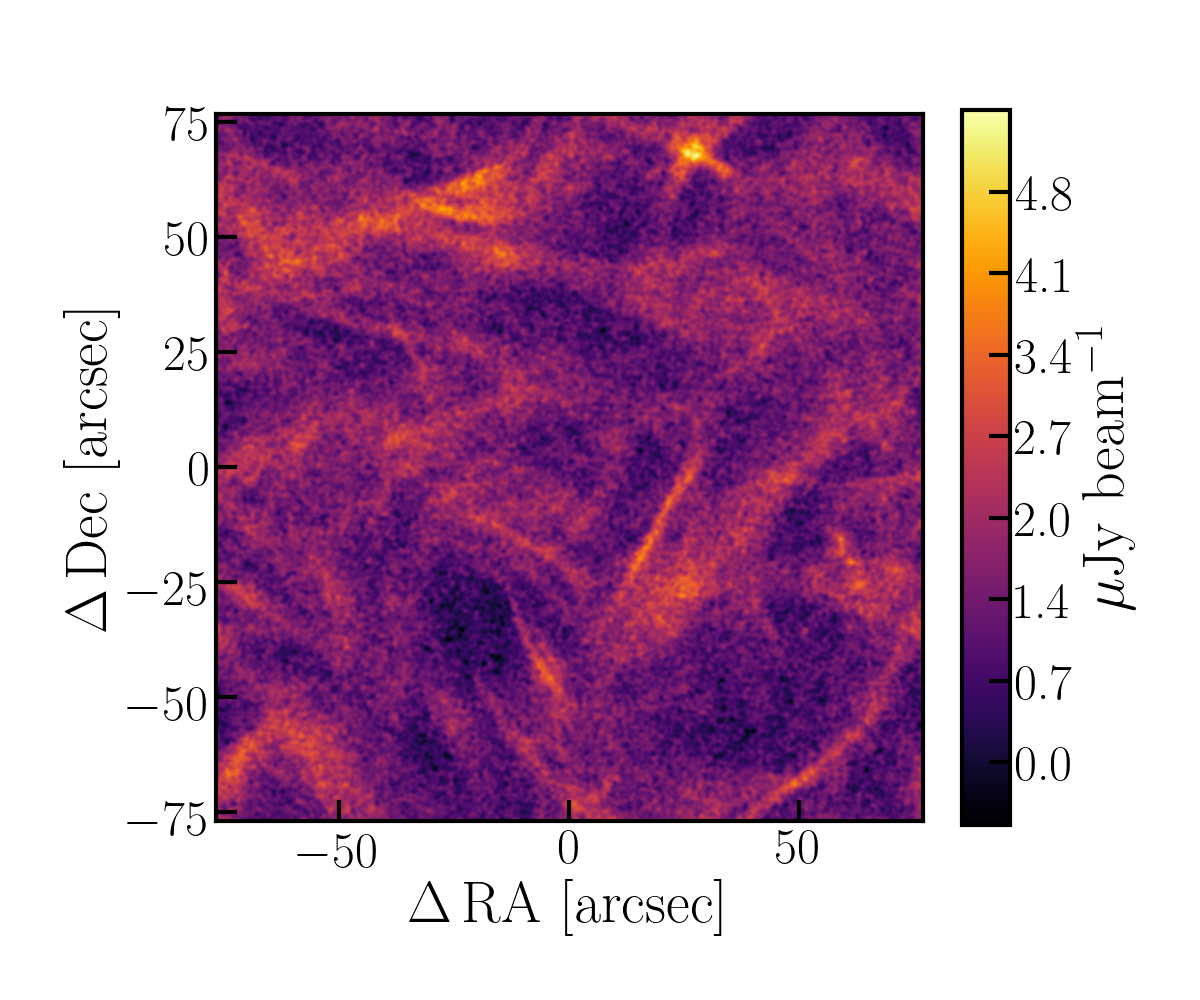} &
\includegraphics[width=5.2cm, trim=10mm 15mm 12mm 15mm, clip]{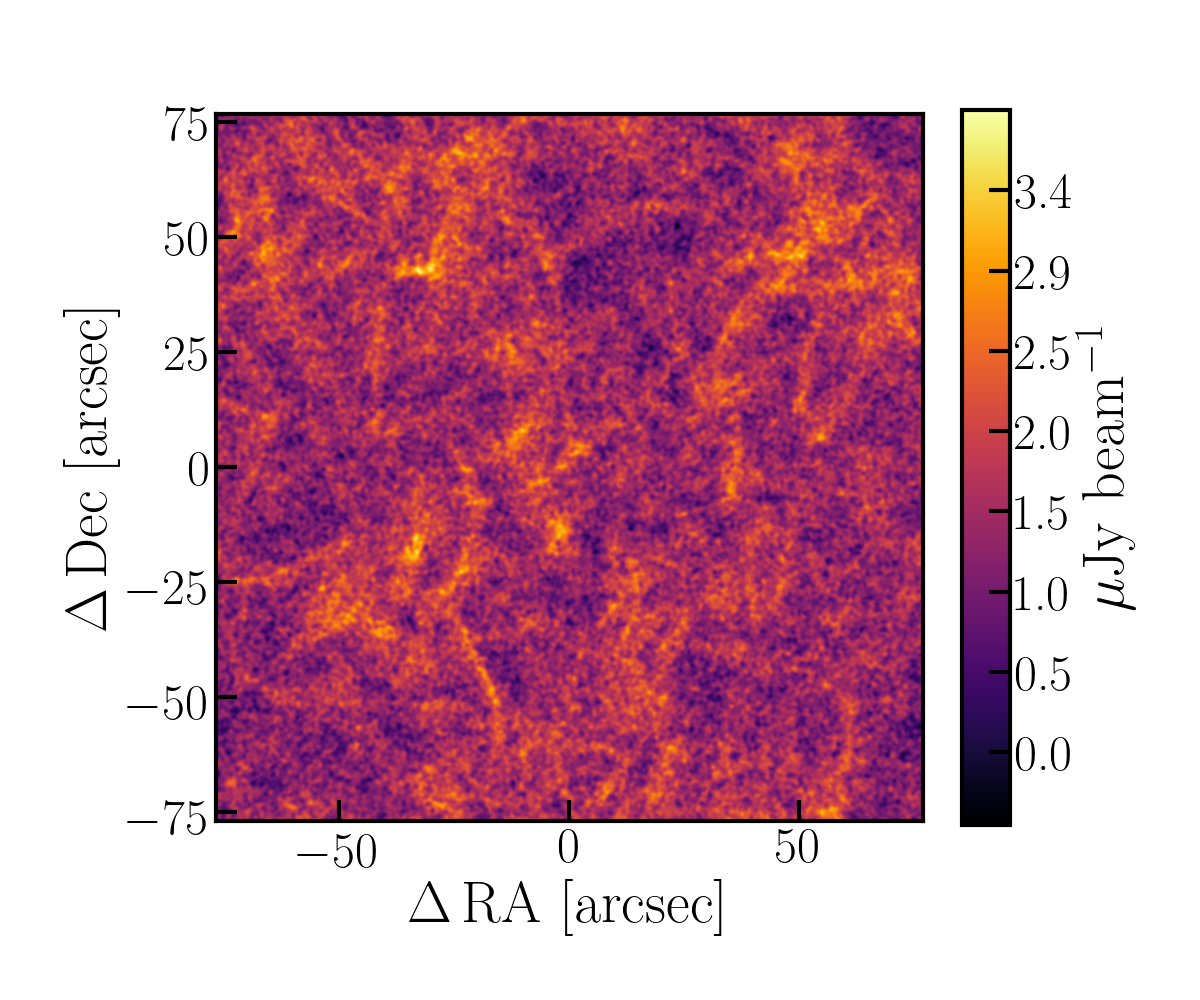} &
\includegraphics[width=5.2cm, trim=10mm 15mm 12mm 15mm, clip]{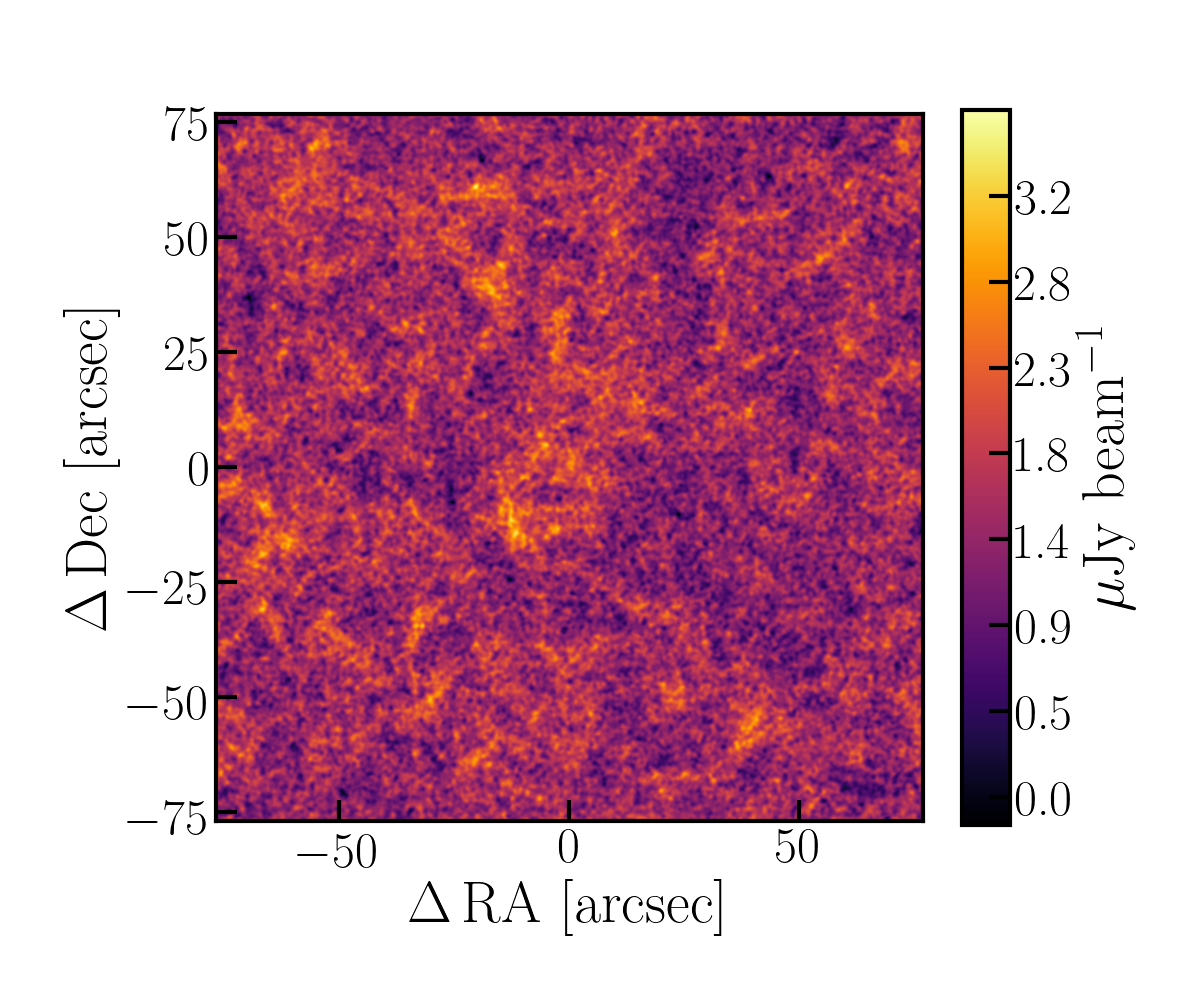} \\
\end{tabular}
\caption{Representative synthetic maps of a cluster at $z\sim0.2$ having an integrated flux density of 50\,mJy at 1.3\,GHz in the rest-frame. Here, the physical resolution is 3\,kpc, corresponding to an angular resolution of $\approx0.9\arcsec$ and $2.6^\prime$ angular size of the radio halo. \textbf{\textit{Top row}} are those obtained directly by smoothing the synthetic maps in Fig.~\ref{fig:totI} top row and rescaling them to flux density units, while the \textbf{\textit{bottom row}} shows the maps with $0.3\,\rm \upmu Jy\,beam^{-1}$ rms noise to mimic a 15\,hr on-source observations with AA4 in Band\,2.}
\label{fig:totI_noise}
\end{figure}

In the previous sections, we concentrated on the total intensity emission from the ICM, especially in the context of phoenixes. Here, we discuss the prospect of detecting polarized synchrotron emission from the ICM with the SKA-Mid. As noted in Sec.~\ref{sec:phoenix}, understanding of radio phoenixes, and using them as a stepping stone towards understanding magnetic fields in ICM in greater detail, is a possible way forward. Nevertheless, we will discuss the role that SKA will play using our MHD simulations as a general representation for total and polarized emission from the ICM. As discussed in Sec.~\ref{sec:obsreq}, it is imperative to observe galaxy clusters at $\gtrsim3\ghz$ with high spatial resolution to detect polarized emission from the ICM.

\begin{figure}
\centering
\begin{tabular}{ccc}
{\large $l_{\rm f}=256\kpc$} & {\large $l_{\rm f}=102.4\kpc$} & {\large $l_{\rm f}=64\kpc$} \\
\multicolumn{3}{c}{\large{Polarized intensity + noise ($1.5\arcsec$ resolution with AA4 in Band\,5a)}}\\ 
\includegraphics[width=5.2cm, trim=10mm 15mm 12mm 15mm, clip]{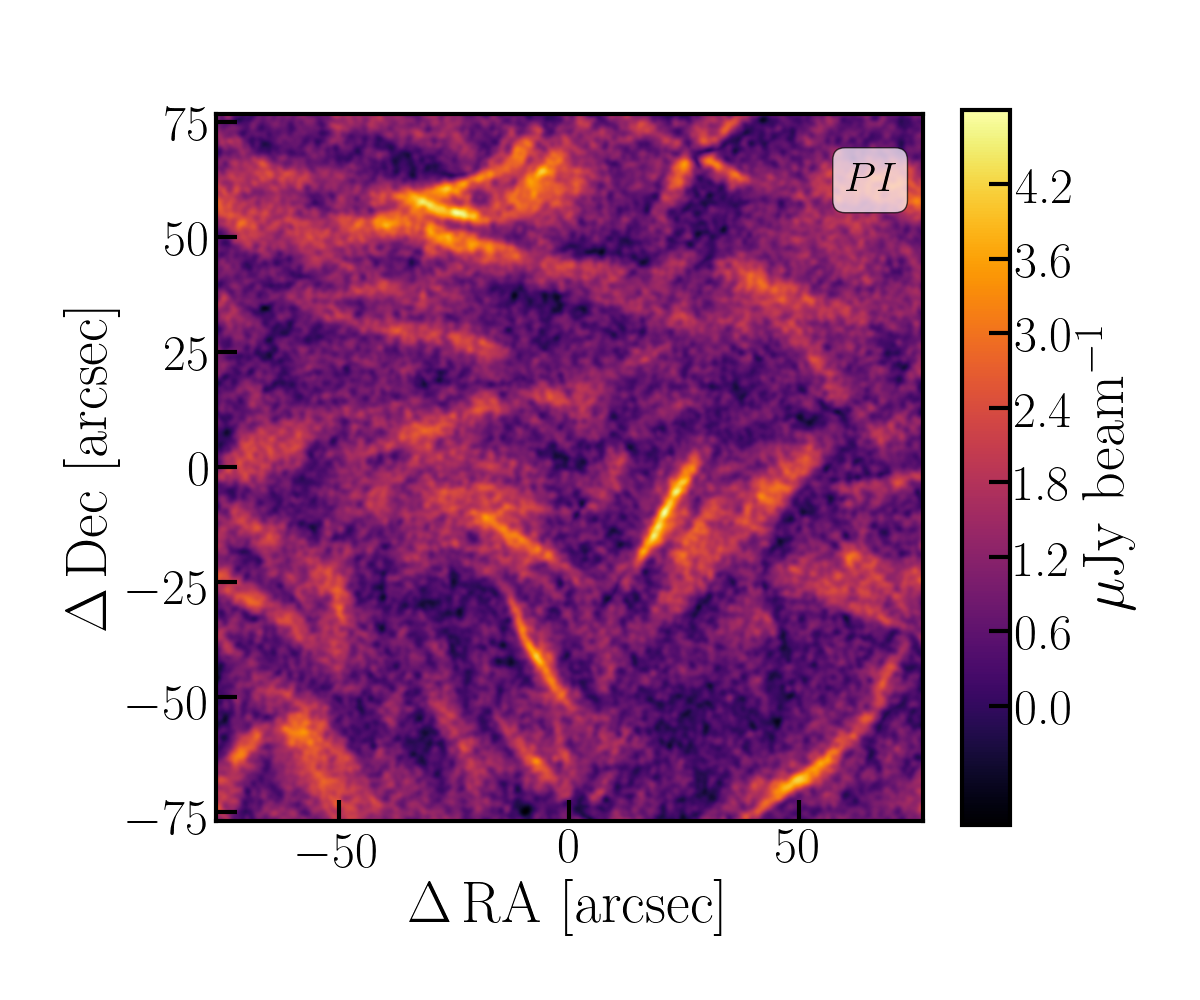} &
\includegraphics[width=5.2cm, trim=10mm 15mm 12mm 15mm, clip]{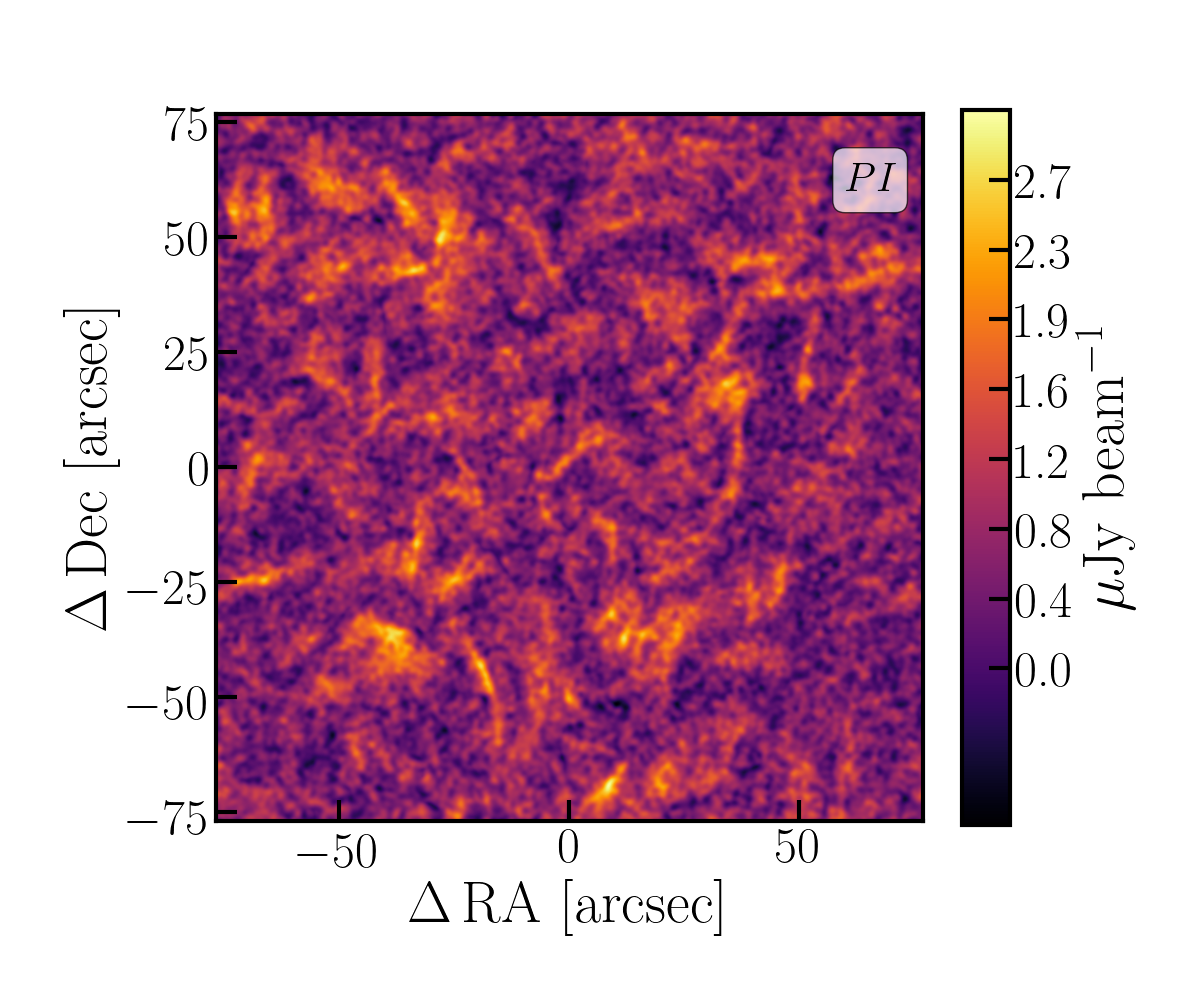} &
\includegraphics[width=5.2cm, trim=10mm 15mm 12mm 15mm, clip]{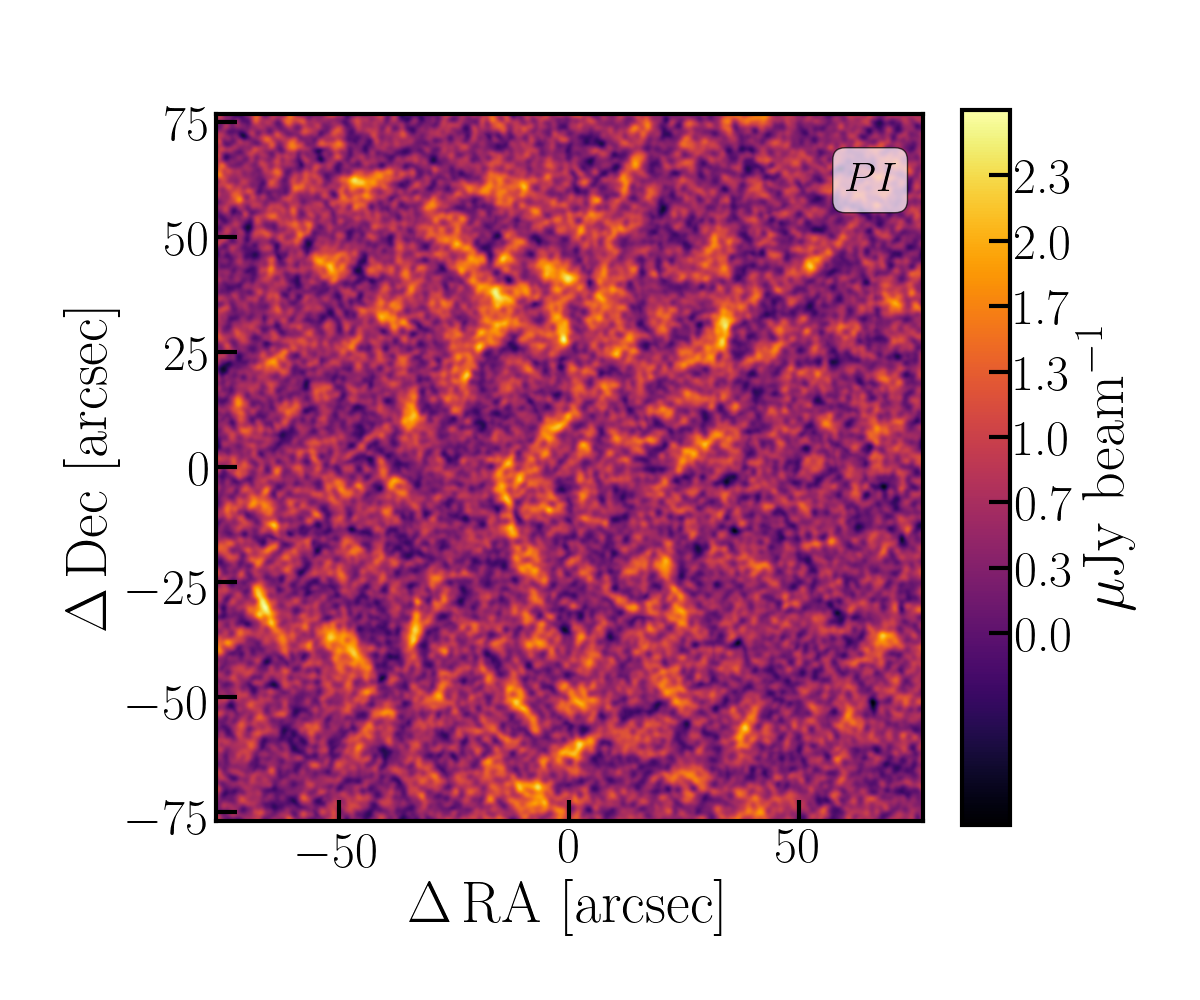} \\
\multicolumn{3}{c}{\large{Fractional polarization + noise ($1.5\arcsec$ resolution with AA4 in Band\,5a)}}\\ 
\includegraphics[width=5.2cm, trim=10mm 15mm 12mm 15mm, clip]{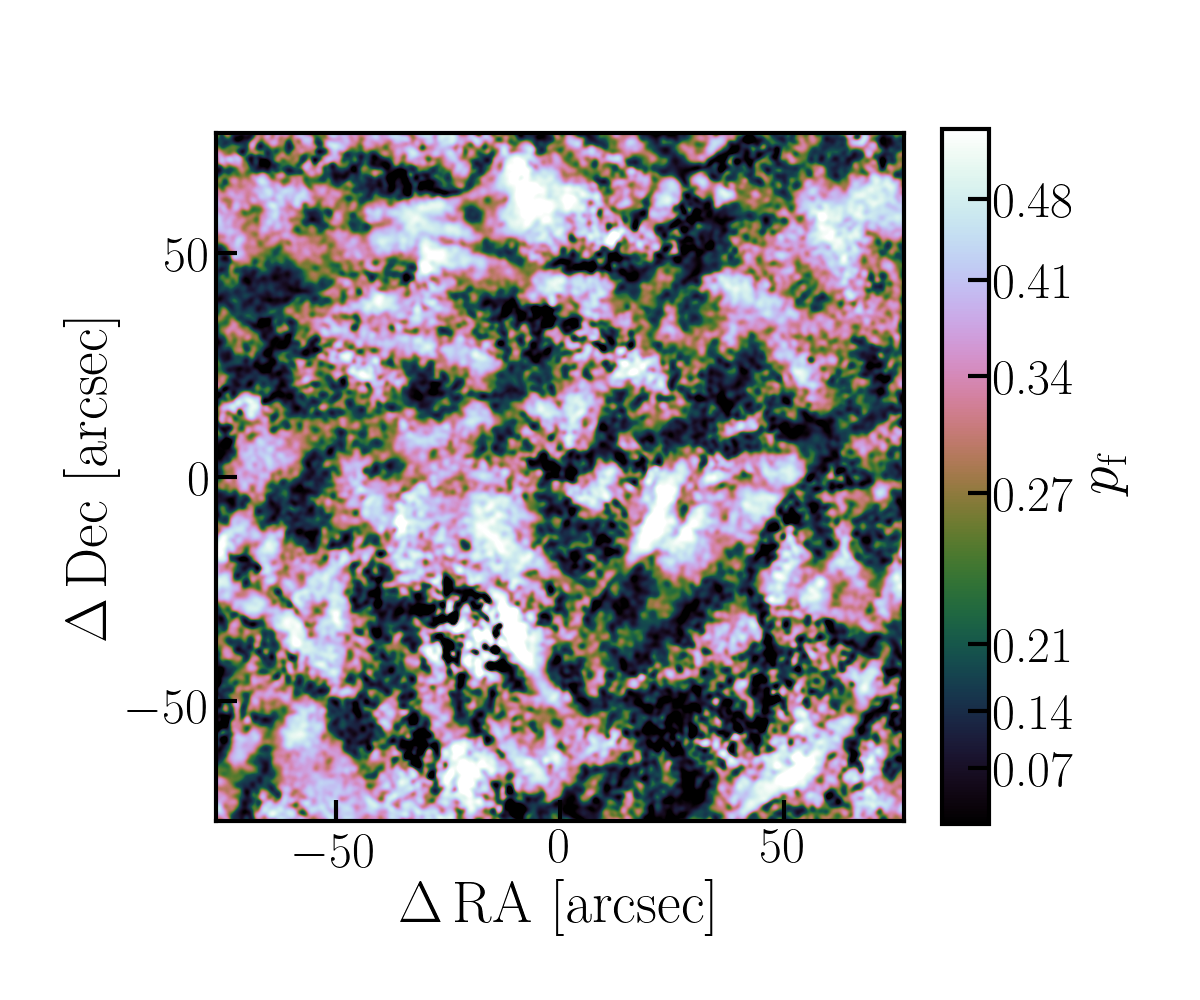} &
\includegraphics[width=5.2cm, trim=10mm 15mm 12mm 15mm, clip]{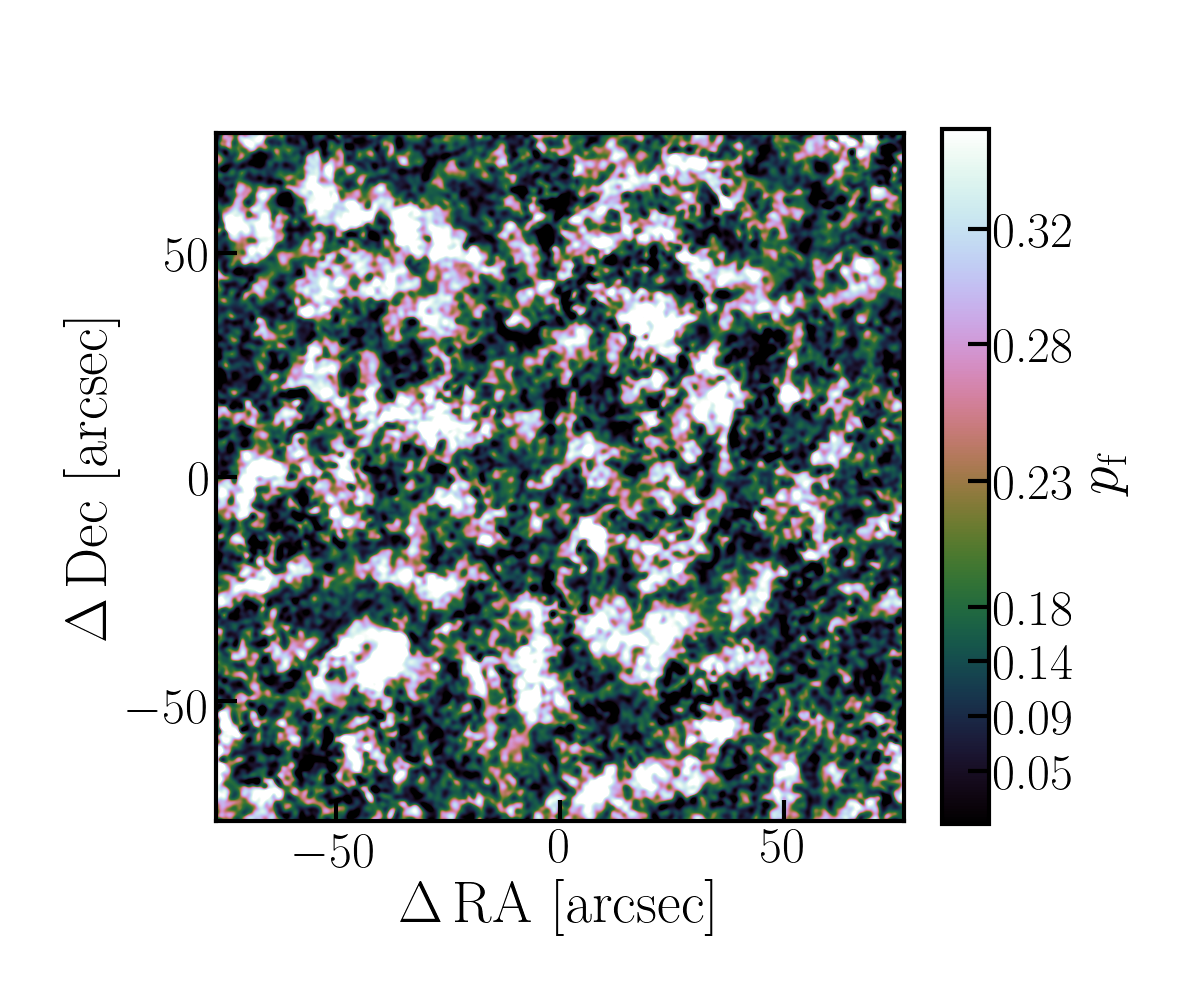} &
\includegraphics[width=5.2cm, trim=10mm 15mm 12mm 15mm, clip]{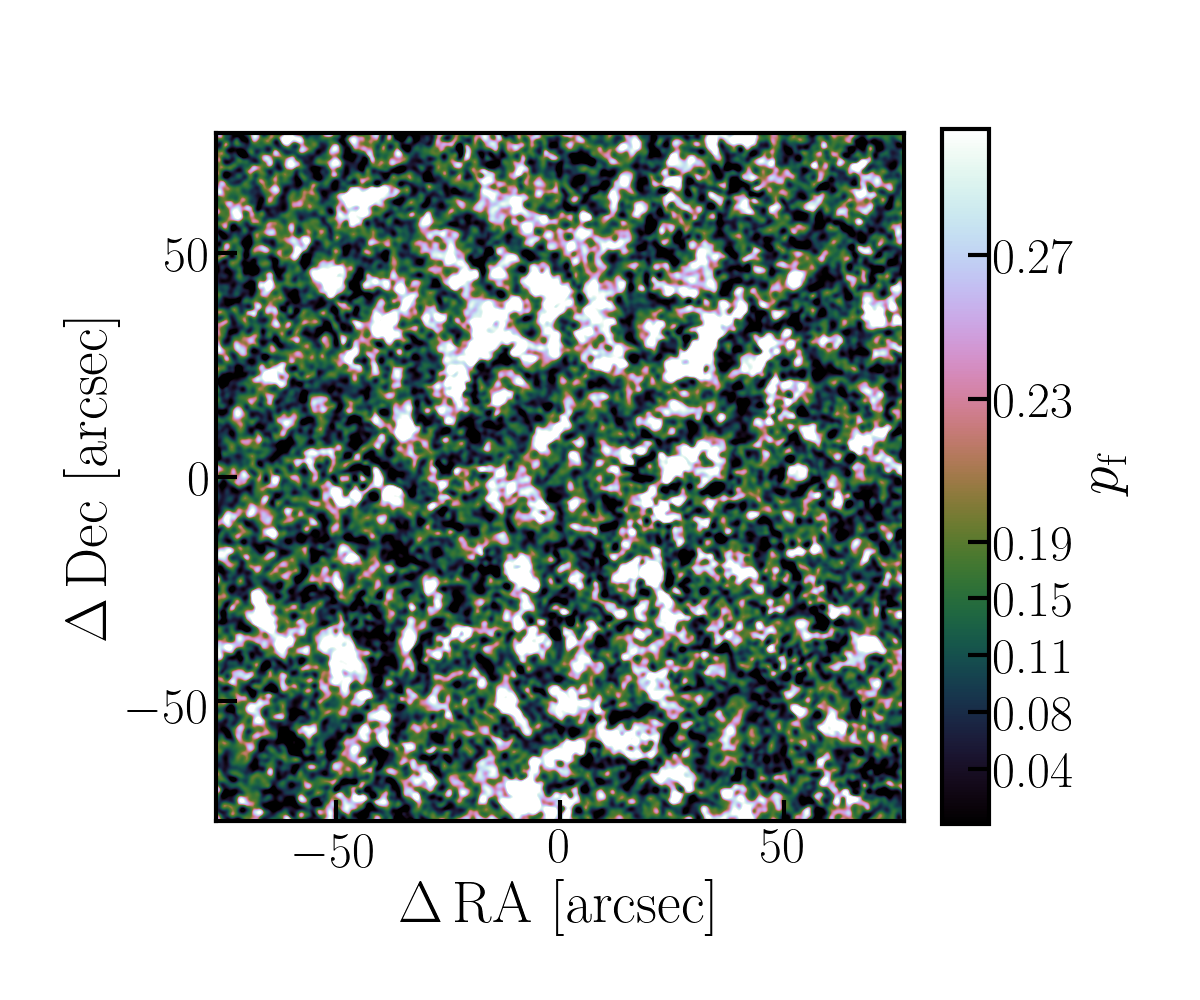} \\
\end{tabular}
\caption{Representative synthetic maps of polarized emission at 6\,GHz of a cluster at $z\sim0.2$. We assumed an integrated flux density of 500\,mJy at 1.3\,GHz in the rest-frame with $\alpha=-1.5$. Here, the physical resolution is 5\,kpc, corresponding to an angular resolution of $\approx1.5\arcsec$. The \textbf{\textit{top row}} are those obtained directly by smoothing the synthetic maps in Fig.~\ref{fig:totI} top row, while the \textbf{\textit{bottom row}} shows the maps when observed for an on-source time of 20\,hr having a rms of $0.25\,\rm \upmu Jy\,beam^{-1}$ with AA4 in Band\,5a.}
\label{fig:polI_noise}
\end{figure}

In the top panel of Fig.~\ref{fig:totI_noise} we show the synthetic maps of $I_{\rm sync}$ for a fiducial cluster at $z \sim 0.2$ at a rest-frequency of 1.3\,GHz for the three different $\lf$. We adopted the median redshift from currently available compilations of cluster halos detected at radio frequencies \citep{yuan2015, weere19}. Furthermore, unlike in our simulations, here we normalized the integrated flux density of the synthetic map to a typical value of 50\,mJy of radio halos at 1.3\,GHz. The pixel scale of 1\,kpc in our simulations corresponds to $0.3\arcsec$ at $z\sim 0.2$. We have therefore smoothed the synthetic maps using a Gaussian kernel with $3\times3$\,pixel FWHM to mimic $0.9\arcsec$ resolution with AA4 in Band\,2. Then, a noise of $\rm 0.3\,\upmu Jy\,beam^{-1}$ was added to simulate a 15-hr observation. The expected maps of the total synchrotron emission are shown in the bottom panel of Fig.~\ref{fig:totI_noise}. It is clear that for an on-source time of 15\,hr, AA4 in Band\,2 would recover the filamentary structures.

On the other hand, detecting polarized emission from the ICM is going to be arduous. As discussed in Sec.~\ref{sec:depol}, Faraday and beam depolarization at frequencies below $\sim3\ghz$ wipes out the intrinsic polarized structures, making it rather impractical to make meaningful quantitative exploration even if patches of polarized emission are detected. At higher frequencies, however, steep radio continuum spectrum in radio halos results in low surface brightness. In Fig.~\ref{fig:polI_noise}, we show the expected polarized intensity and fractional polarization at 6\,GHz rest-frame for a 20\,hr observation with AA4 in Band\,5a. Unlike for total intensity above, here we assume a brighter galaxy cluster having a total flux density of 500\,mJy at 1.3\,GHz. The flux density at 6\,GHz was determined by assuming a spectral index $\alpha = -1.5$.\footnote{Note that, in the synthetic observations discussed in Sec.~\ref{sec:simulation} we have used $\alpha=-1$. To robustly explore the efficacy of observations with the SKA we have used a steeper spectrum.} Furthermore, to recover the polarized emission over the entire 3.9\,GHz bandwidth of Band\,5a, the technique of RM-synthesis needs to be applied resulting in the angular resolution to be limited by the lower frequency end of the band. Hence, here we have used an angular resolution of $1.5\arcsec$ corresponding to a spatial resolution of 5\,kpc, and added a noise of $\rm 0.25\,\upmu Jy\,beam^{-1}$ that can be achieved with 20\,hr on-source integration. On a cursory comparison of the polarized morphology in Fig.~\ref{fig:polI_noise} with Fig.~\ref{fig:polI_6GHz}, it is seen that most of the polarized structures are expected to be well recovered above $\approx4\,\sigma$, and $\pmean$ can be determined within 10\% of $\pmean_{\rm int}$.

\subsection{Inferring turbulence scale from the observed polarization of the ICM} \label{sec:turbscale}

\begin{figure}[t]
    \centering
	\includegraphics[width=0.8\columnwidth]{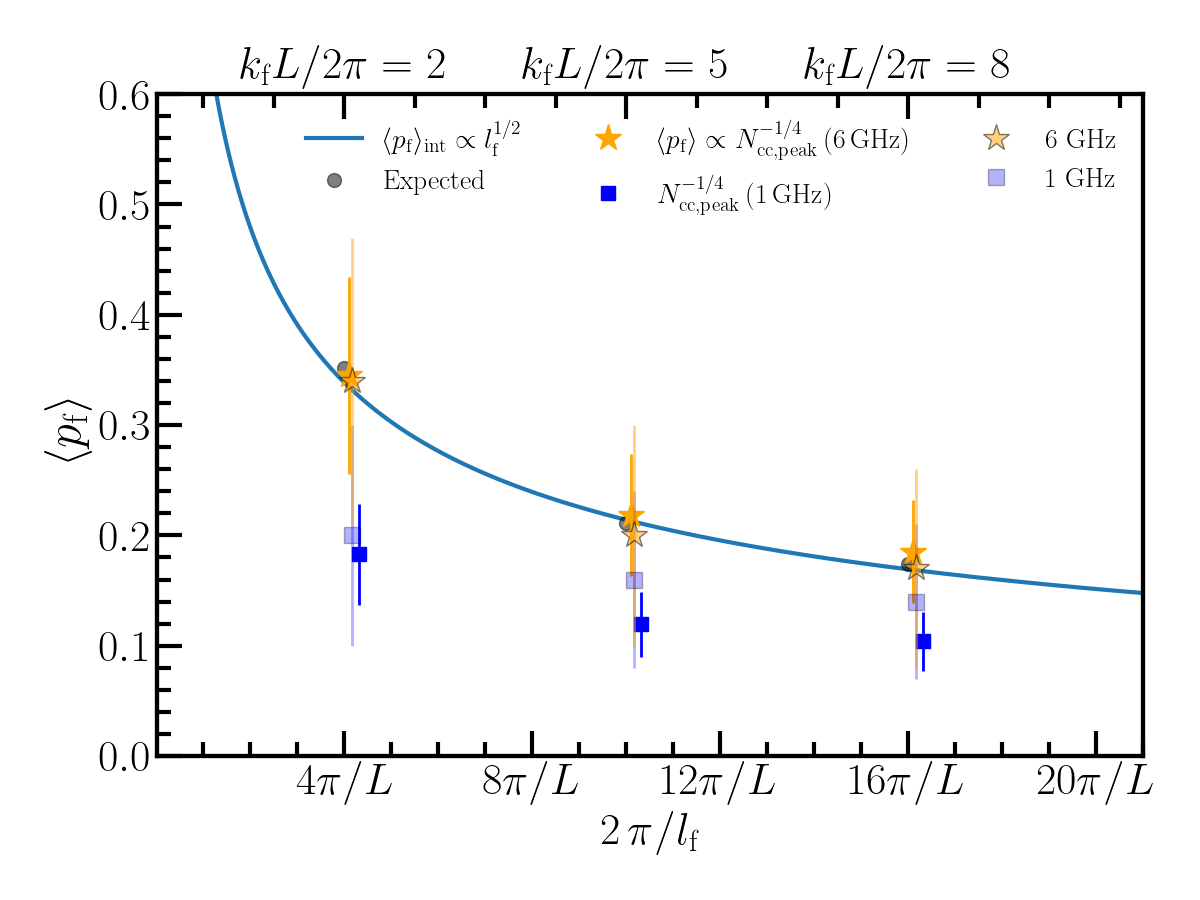}
    \caption{Variation of $\pmean$ with $\lf$. The blue curve shows the 
$\pmean \propto l_{\rm f}^{1/2}$ relation, and the grey points show the expected $\pmean$ 
computed from the magnetic integral scale, both taken from \citet{bs21}. The solid yellow stars and solid blue squares show $\pmean \propto N_{\rm cc, peak}^{-1/4}$ at 
6 and $1\ghz$ in the presence of Faraday rotation, while, the corresponding shaded symbols show $\pmean$ directly obtained from the synthetic fractional polarization maps. The data are plotted with 
slight offset in the $x$-axis to avoid overlap.}
    \label{fig:depol}
\end{figure}

As discussed above, deep observations of galaxy clusters in Band\,5a with the SKA would be instrumental in our quest to directly detect and characterize the morphological properties of polarized synchrotron emission in the ICM. However, inferring the driving scale of turbulence is paramount to understand the mechanisms that are at play in shaping the evolution of the nonthermal phase in ICM. The mean intrinsic fractional polarization ($\pmean_{\rm int}$), best measured at high frequencies where Faraday depolarization is low, is an important quantity that is directly related to the magnetic integral scale ($\ell_{\rm M}$). $\ell_{\rm M}$ is in turn linearly related to $\lf$ \citep{bs21}. Although, unlike Gaussian random fields, the polarized emission due to fluctuation dynamo is spatially intermittent, the volume averaged $\pmean_{\rm int}$ approximately follows $\pmean_{\rm int} \approx p_{\rm max}/\sqrt{L/\ell_{\rm M}} \propto \lf^{1/2}$. Here, $p_{\rm max} \approx 0.75$ is the maximum theoretical fractional polarization of synchrotron emission. Such a relation is expected because, similar to Gaussian random fields, random rotation of the plane of polarization in spatially intermittent magnetic fields also results in a random degree of polarization when integrated along the line of sight. Below, we discuss a couple of possible methods for determining $\lf$.

\paragraph{From fractional polarization maps:} In Fig.~\ref{fig:depol}, the $\pmean_{\rm int} \propto \lf^{1/2}$ relation is shown as the blue line, while the grey data points were computed from $\ell_{\rm M}$ directly estimated from the MHD simulations \citep[see][]{bs21}. This shows that mean fractional polarization can provide a direct estimate of $\lf$. As discussed earlier, $\pmean$ measured at frequencies $\gtrsim3\ghz$, e.g., in Band\,5a, is close to $\pmean_{\rm int}$. The shaded orange stars in Fig.~\ref{fig:depol} show $\pmean_{6\ghz}$ determined from synthetic maps without any noise. For comparison, the blue shaded squares show $\pmean_{1\ghz}$ expected to be measured in Band\,2. Due to severe Faraday and beam depolarization, $\pmean_{1\ghz}$ is significantly off with respect to the expected $\pmean \propto \lf^{1/2}$ relation, making it a challenge to directly connect any detectable polarized emission in Band\,2 to the turbulence driving mechanism in the ICM. It is clear that deep Band\,5a observations with the SKA will be imperative to reliably explore the structural and statistical properties of polarized emission from the ICM for the first time.

\paragraph{From connected components and `filamentarity':} The estimation of $\pmean_{6\ghz}$ from the map of $PI_{6\ghz}$ could be biased because the polarized emission at low signal-to-noise ratio levels ($\lesssim5\,\sigma$) and/or low $\pf$ are difficult to measure reliably due to the `Ricean bias' originating from positive definite background of polarized intensity maps \citep{wardl74} and instrumental leakage, respectively. For such cases, \citet{dutta2024} recently proposed an alternative method of using Minkowski functionals and connected components ($N_{\rm cc}$). They showed that the number of connected polarized components above a threshold $\pf$, $N_{\rm cc}(>\pf)$, has a characteristic peak as a function of $\pf$. The threshold $\pf$ ($p_{\rm peak}$) at which $N_{\rm cc}$ has a peak, $N_{\rm cc,peak}$, is one-to-one related to $\pmean$, and follows the relation $p_{\rm peak} \approx \pmean \propto N_{\rm cc,peak}^{-1/4}$. This is expected because, the number of emitting components $N_{\rm cc} \propto (L/\lf)^2$, and then follows from the relation $\pmean \propto \lf^{1/2}$. Thus, in other words, the surface density of connected components, $\Sigma_{\rm cc} = N_{\rm cc,peak}/L^2$, is a measure of the so-called `turbulent cells' and a direct consequence of randomly averaging polarized emission \citep{dutta2024}. Besides studying the properties of polarized emission based on Minkowski functionals, they are also useful in identifying the morphology of filaments, as shown in the context of sub-structures in radio relics \citep{WBGR23}.

In Fig.~\ref{fig:depol}, the solid orange stars show $\pmean_{6\ghz}$ estimated from $N_{\rm cc,peak}^{-1/4}$ for the three different $\lf$s, and clearly show an excellent match to the expected $\pmean_{\rm int} \propto \lf^{1/2}$ relation. An advantage of using $N_{\rm cc,peak}$ as a measure of $\pmean_{6\ghz}$ is that it is largely unaffected by non-detection of faint polarized emission and provides an estimate on $p_{\rm peak}$ within $\sim 15\%$ as long as the detectable region is larger than $\lf$ \citep{dutta2024}.

\section{Discussion}

In this chapter we have explored the efficacy of SKA in detecting, perhaps ubiquitous, filamentary synchrotron emission and associated linearly polarized emission from the ICM. Detecting characteristic filaments generated by the action of a fluctuation dynamo requires high spatial resolution ($\sim1\kpc$), corresponding to angular resolution $\lesssim 1.5\arcsec$ (see Sec.~\ref{sec:totalI}). So far, detection of filamentary synchrotron emission and polarized emission from the ICM has been limited by a combination of poor angular resolution and generically low surface brightness of radio halos. Furthermore, a robust estimate of $\pmean$ at frequencies $\gtrsim 3\ghz$ is imperative to determine the turbulence scale, $\lf$ (see Sec.~\ref{sec:turbscale}). Detection of polarized emission has been especially challenging because at low frequencies ($\lesssim 2\ghz$), severe Faraday and beam depolarization leads to a negligible degree of polarization (see Sec.~\ref{sec:depol}).

So far, only for three clusters, tentative detection of polarized emission has been reported, namely, for Abell\,2255 \citep{govoni+05}, MACS\,0717.5+3745 \citep{bonafede+09}, and Abell\,523 \citep{girardi16}. It is worth noting that all three reports of polarized halo emission are based on observations at 1.4\,GHz, a frequency at which many radio relics show polarized emission \citep{wittor19}. It seems likely that in all three cases the polarized signal is related to a radio relic seen in projection rather than the halo emission itself \citep{p+11, rajpurohit21}. \textit{To date, no clean detection of polarized halo emission has been made.} Based on cosmological MHD simulations, \citet{govoni+13} found that luminous halos may show polarized flux of 0.5--2\,$\upmu$Jy for a $3\arcsec$ beam at 1.4\,GHz, with $\pf$ in the range 10--30\%. However, our higher resolution simulations in a box indicate that, depending on $\lf$, $\pmean$ increases significantly from $\sim2\textrm{--}5\%$ at 1\,GHz to $\sim20\textrm{--}35\%$ at 6\,GHz when observed with a spatial resolution of $\sim5\textrm{--}10\kpc$. This increase of $\pmean$ compensates surface brightness decrease from 1.4 to 6\ghz. 

By adding a rms noise of $\rm 0.3\,\upmu Jy\,beam^{-1}$ for a 15-h observations with the SKA's AA4 in Band\,2 at $0.9\arcsec$ resolution, we find that a representative radio halo of 50\,mJy flux density would be comfortably detectable. Generalizing from this, all halos that have total flux density,
$$I \gtrsim 50\,{\rm mJy}\,[1.2/(1+z)]^{1-\alpha}[D_{z=0.2}/D_{z}]^2\, [2.6^\prime/\Omega_{\rm halo}]^2\, [\theta/0.9\arcsec]^2$$ 
at 1.3\ghz, should be detectable. Here, $D_z$ is the luminosity distance at redshift $z$, $\Omega_{\rm halo}$ is the angular extent of the halo emission in arcmin, and $\theta = 0.9\arcsec\, (D_{z=0.2}/D_{z})$ is the desirable angular resolution.

In order to target suitable clusters to study filamentary emission with somewhat time-inexpensive observations, radio phoenixes are ideal. 
They can be identified in surveys with the SKA-Low as extended emission of irregular morphology associated with cluster halos (see Sec.~\ref{sec:obs_phoenix}) having steep spectrum ($\alpha \lesssim -1.2$). Phoenixes are expected to be efficiently detected with SKA-Low, and their spectrum could be determined within 50--70\,MHz sub-bands. Although their origins are not yet understood, but they are potential objects that could directly trace the magnetic field morphology in the ICM. Fig.~\ref{fig:phoenix_sensitivity} suggests that all phoenixes that have properties similar to those identified in LoTSS data would be detectable by SKA-Low in about 10\,min. To study the spectral properties of filaments, Bands\,1 and 2 of SKA-Mid's AA4 would play a central role in resolving filamentary emission in $\lesssim 1$\,h of observations at $\lesssim1\arcsec$ resolution for $\alpha$ as low as $-3$ up to $z\approx 0.3$. Hence, all-sky surveys with SKA-Low, and Bands\,1 and 2 would effectively provide a large census of phoenixes.

To facilitate the study of magnetic field morphology and inferring $\lf$ using polarized emission, deep observations above 3\,GHz, e.g., at Band\,5a are necessary (see Secs.~\ref{sec:depol} and \ref{sec:obspol}). Similar to the total intensity from radio phoenixes as tracer of ICM magnetic fields, searching for polarized emission in them with the SKA-Mid is a way forward to start with. From Fig.~\ref{fig:phoenix_sensitivity}, synchrotron emission from phoenixes that have a mean spectral index $\alpha \gtrsim -2.5$ is expected to be detected at Band\,5a in 15\,h on-source observations. Detecting phoenixes with $\alpha \sim -3$ would require significantly deeper observations. However, for expected $\pmean_{6\ghz}$ in the range $\approx 0.2\textrm{--}0.35$, polarized emission from phoenixes that have mean $\alpha > -2$ should be detectable. Also, in the context of detecting polarized emission from radio halos, at least, 15--20\,hr of on-source observations at Band\,5a is required for robust characterization (see Fig.~\ref{fig:polI_noise}).

\paragraph{Prospect of a 3--5\,GHz band and synergy with XRISM:} In the scenario of strong CRE energy losses in Band\,5a, it will become a challenge to detect polarization from a meaningful sample that will facilitate a firm establishment of the connection between the origin of phoenixes and/or the driving scale of turbulence in ICM with the evolutionary state of cluster merger. This situation will get an enormous boost with the availability of Band\,4 where the surface brightness of the polarized emission will be more than a factor of 2 brighter for $\alpha < -2$. That will provide a more complete census of ICM properties in a fraction of the time than that needed at Band\,5a.

Since dynamo action relies on the conversion of kinetic energy 
in turbulent motions into magnetic energy, complementary information on the velocity fields in galaxy clusters could provide crucial insights into the magnetic field amplification mechanism in the ICM.
SKAO along with high-quality information on velocity fields from spectral-lines in X-ray band from XRISM will nail down the kinematics and role of turbulence driving mechanism in the ICM.
For this, information on the turbulent Mach number, comparison of power spectra of velocity field from X-ray emission and radio emission would be essential in understanding the dominant turbulence driving mechanism. Hence, a detailed study through non-isothermal numerical simulations to compare X-ray and radio emissions are needed to exploit the synergy.
XRISM will also provide a detailed insight into the merger state of clusters that host phoenixes \cite[e.g.,][]{heinrich2025, XRISM_A20292025, Fujita2025}. 
SKA-Low and Mid in array assembly AA4 will play a game changing role in our understanding of the magnetic fields and the dynamic role they play in evolution of ICM in galaxy clusters.

\section{Acknowledgments}

NB acknowledges support from the ERC Consolidator Grant ULU 101086378. MB acknowledges financial support from Next Generation EU funds within the National Recovery and Resilience Plan (PNRR), Mission 4 – Education and Research, Component 2 – From Research to Business (M4C2), Investment Line 3.1 – Strengthening and creation of Research Infrastructures, Project IR0000034 – "STILES – Strengthening the Italian Leadership in ELT and SKA", from INAF under the Large GO 2024 funding scheme (project "MeerKAT and Euclid Team up: Exploring the galaxy-halo connection at cosmic noon"), the Large Grant 2022 funding scheme (project "MeerKAT and LOFAR Team up: a Unique Radio Window on Galaxy/AGN co-Evolution") and the Mini Grant 2023 funding scheme (project `Low radio frequencies as a probe of AGN jet feedback at low and high redshift'). We thank Yik Ki (Jackie) Ma for helpful comments.

\bibliographystyle{abbrvnat-maxbibnames4}
\bibliography{references_main} 

\end{document}